\documentclass{aa}  
\usepackage{longtable}
\usepackage{threeparttable}
\usepackage{graphicx}
\usepackage{txfonts}
\usepackage{amsmath}
\begin{document}

   \title{Four ages of rotating stars in the rotation--activity relationship and gyrochronology}

 \author{Huiqin Yang \inst{1,2} \and Jifeng Liu\inst{1,4} \and Roberto Soria\inst{3,4,5} \and Federico Spada \inst{6} \and Song Wang\inst{1,2} \and Xiangsong Fang \inst{1} \and Xue Li\inst{1,4}
          }

   \institute{Key Laboratory of Optical Astronomy, National
              Astronomical Observatories, Chinese Academy of Sciences, Beijing 100101, China\\
              \email{yhq@nao.cas.cn}\\ 
              \and
             Institute for Frontiers in Astronomy and Astrophysics, Beijing Normal University, Beijing 102206, China\\ 
              \and
              INAF-Osservatorio Astrofisico di Torino, Strada Osservatorio 20, I-10025 Pino Torinese, Italy\\
              \and
              School of Astronomy and Space Sciences, University of Chinese Academy of Sciences, Beijing 100049, China\\
              \and
              Sydney Institute for Astronomy, School of Physics A28, The University of Sydney, Sydney, NSW 2006, Australia\\
              \and
              Università di Catania, Dipartimento di Fisica e Astronomia, Via S. Sofia 78, 95123 Catania, Italy\\
             }

 
  \abstract
   {Gyrochronology and the rotation--activity relationship are standard techniques used to determine the evolution phase and dynamo process of low-mass stars based on their slowing down. Gyrochronology identifies two tracks in color--period diagrams: the convective (C) phase (younger, faster rotating) and the interface (I) phase (older, slower rotating). These phases are separated by a transition, or  ``gap'' (g phase), and there is no precise estimate of its duration. The rotation--activity relation also identifies two stages: the saturated regime (faster, with higher activity) and unsaturated regime (slower, with declining activity). The mismatch in the definition of the evolutionary phases has so far raised many issues in physics and mathematics and has hampered the understanding of how the internal dynamo processes affect the observable properties.}
   {To address this problem, we seek a unified scheme that shows a one-to-one mapping from gyrochronology to the rotation--activity relationship.}
   {We combined LAMOST spectra, the Kepler mission, and two open clusters to obtain the chromospheric activity $R'_{\rm HK}$ of 6846 stars and their rotation periods. We used $R'_{\rm HK}$ and the rotation period to investigate the rotation--activity relationship. Instead of the traditional two-interval model, we applied a three-interval model to fit the rotation--activity relationship in the range of the Rossby number Ro$<0.7$. We also used the X-ray data to verify our new model. }
   {We find that the rotation--activity relationship is best fit by three intervals in the rang of Ro$<0.7$. We associate those intervals with the convective, gap, and interface phases of gyrochronology. The mean Ro of the C-to-g and g-to-I transition is $\approx 0.022$ and $\approx 0.15$, respectively. The g-to-I transition is on the edge of the intermediate period gap, indicating that the transition of surface brightness from being spot dominated to facula dominated can be associated with the transition from the gap to the I sequence. Furthermore, based on previous studies, we suggest an additional epoch at late times of the I phase (Ro $>0.7$; weakened magnetic breaking phase) from the perspective of activity.  We further used the three-interval models to fit the period--activity relationship in temperature bins and determine the duration of the transition phase as a function of effective temperature. By comparing the critical temperature and period of the g-to-I transition with the slowly rotating sequence of ten young open clusters whose ages range from 1 Myr to 2.5 Gyr, we conclude that our new model finds the pure I sequence without fast rotating outliers, which defines the zero-age I sequence (ZAIS). We propose that there is an ambiguous consensus on when the I sequence begins. This ambiguity is from the visually convergent sequence of the color--period diagrams in open clusters. This visually convergent sequence is younger than the ZAIS and is actually the pre-I sequence that can be associated with the stall of the spin-down. Our results unify the rotation--activity relationship and gyrochronology for the stellar evolution of low-mass stars, which we denote as the ``CgIW" scenario.}
   {}


   \maketitle
%

\section{Introduction}
Stellar structure and evolution are fundamental issues in stellar physics. Stellar activity and rotation manifest themselves in appreciable and measurable ways that reflect stellar structure and evolution. This has caused the relationships of activity, rotation, and age to be widely studied in order to reveal the evolution of stellar structure and dynamo. Quantitative relationships between stellar age, activity, and rotation have been actively investigated for more than 50 years \citep{Sku1972}, and two representative paradigms have been established: the rotation--activity relationship \citep{Noyes1984,Wright2011,Yang2017,Yang2019} and gyrochronology \citep[the rotation--age relationship;][]{Barnes2003b,Barnes2007,Barnes2010}. 

Gyrochronology is well known as a dating method for main sequence (MS) stars, while its underlying physical mechanism is based on the tracks in color--period diagrams \citep{Barnes2003b,Barnes2007}. These tracks demonstrate that a star undergoes the convective (C; younger, faster rotating) phase, the gap (a transition phase from convective to interface), and the interface (I; older, slower rotating) phase in its MS era. The rotation--activity relationship also reveals the evolutionary phase of a star as it spins down. It identifies two stages: the saturated (faster, with higher activity) and unsaturated (slower, with declining activity) regimes \citep[e.g.,][]{Noyes1984,Wright2011}. The two regimes are separated by a critical Rossby number, Ro$_\text{sat}$.

Although the twins (the rotation--activity relationship and gyrochronology) are extensively studied, respectively, the physical connections between them have barely been discussed, and efforts on unifying them have proceeded slowly \citep{Yang2019,Mama2008}. For the first time, \citet{Barnes2003a} associated the C and the I sequence of gyrochronology with the saturated and unsaturated regime of the rotation--activity relation, respectively. This scenario does not present the precise duration of the gap between the C and I sequence, partly because it is a critical point rather than a duration in the rotation--activity relationship \citep{Pizzolato2003}. \citet{Wright2011} adopted this scenario and took the Rossby number, $\rm Ro_{sat}$, where the activity reaches saturation as the transition point of the C and I phases. However, they also raised the question of why the transition between the two regimes (dynamos) occurs so smoothly (or instantaneously) given that the transition should include complicated changes in physics. By revisiting the X-ray data of \citet{Wright2011}, a slight slope of the rotation--activity relationship was found in the saturated regime \citep{Reiners2014}, indicating some remnant dependence of the activity on rotation. This implies that there may be a third regime near the transition point with a distinct activity dependence. Mathematically, a strong degeneracy between the decay rate ($\beta$) of the unsaturated regime and $\rm Ro_{sat}$ can cause a great uncertainty in $\rm Ro_{sat}$. \citet{Newton2017} and \citet{Douglas2014} used the same activity proxy H$_\alpha$ to study the rotation--activity relationship with different samples, while they obtained totally different $\rm Ro_{sat}$. This implies that the dichotomy model may not reflect the real physical properties of the rotation--activity relationship near the transition point. Moreover, recent studies have found a new epoch (the weakened magnetic braking phase) at the late time of the I phase \citep{Saders2016,Hall2021}, suggesting that a star undergoes four phases in terms of spin-down. This makes one-to-one mapping of the two paradigms more difficult and inspires us to reconsider the dichotomous model of the rotation--activity relationship.

In this study, we calculate the stellar chromospheric activity $R'_{\rm HK}$ of 6846 stars, from the Large Sky Area Multi-Object Fiber Spectroscopic Telescope \citep[LAMOST;][]{Cui2012}, and their rotation periods from the Kepler mission \citep{Borucki2010} (Sect.\ref{dm}). We use the dichotomy model to fit the rotation--chromospheric activity relationship and present several issues related to the model (Sect.\ref{sec3}). We propose a new scheme with three intervals to fit the rotation--activity relationship and remap them to gyrochronology (Sect.\ref{sec4}). We compare our results with open clusters and the evolution tracks of gyrochronology (Sect.\ref{sec5}). Finally, we give our conclusions (Sect. \ref{sec6}).  

\section{Data and method}\label{dm}
\begin{figure}[ht]%
\includegraphics[width=0.5\textwidth]{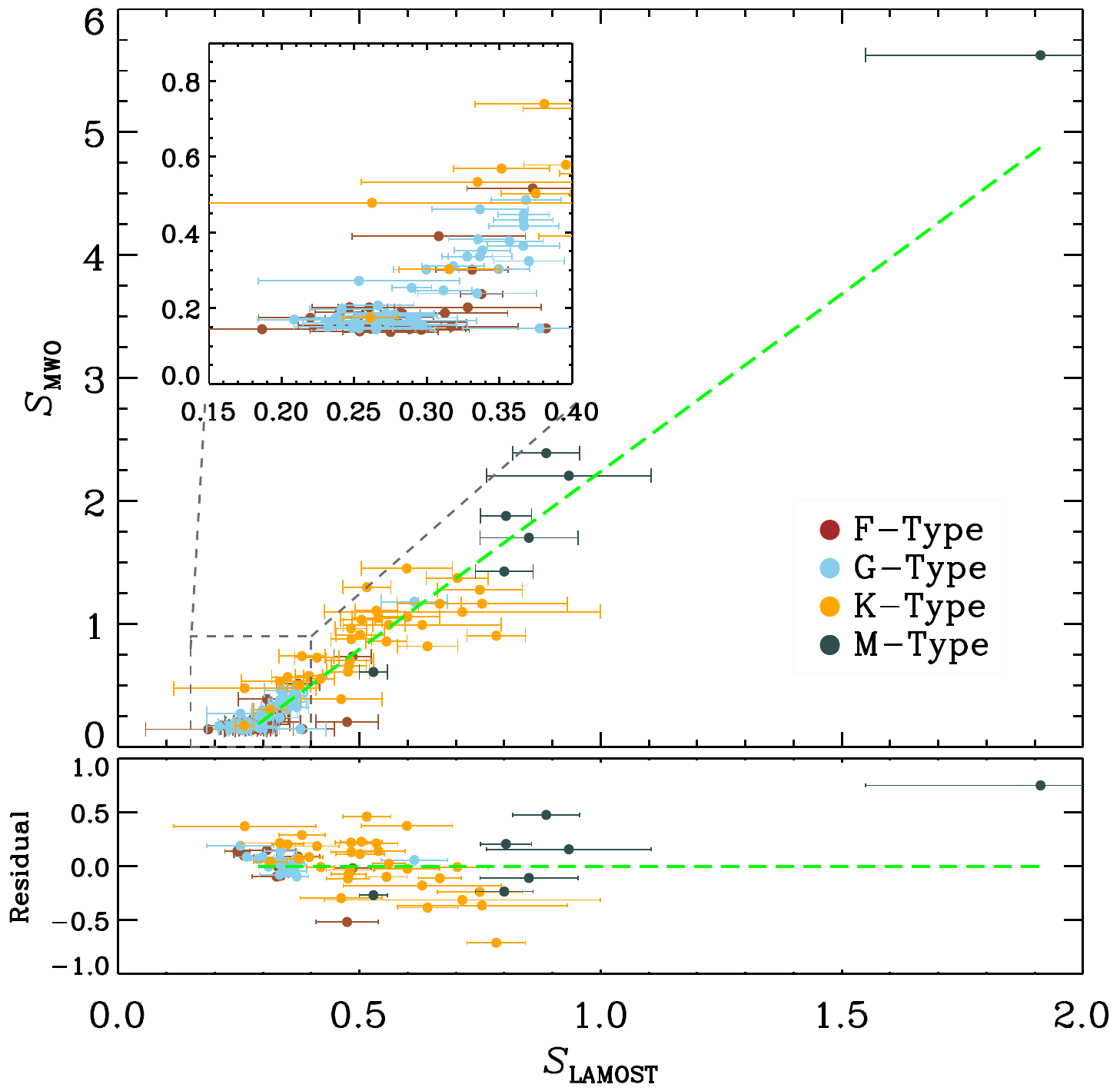}
\caption{$S_{\rm LAMOST}$ vs. $S_{\rm MWO}$. We used the 114 common stars between \cite{Boro2018} and the LAMOST DR6 to determine the calibration factors from $S_{\rm LAMOST}$ to $S_{\rm MWO}$. The green dashed line indicates the best fit given by a weighted ordinary least-square bisector \cite{Isobe1990}. The inset highlights the low-activity region, demonstrating a detection limit of $S_{\rm MWO} \approx 0.15 $ for LAMOST spectra, below which the calibration depends on extrapolation of the fit. The bottom panel shows the residual.} \label{figsc}
\end{figure}
\begin{figure}[ht]
\includegraphics[width=0.5\textwidth]{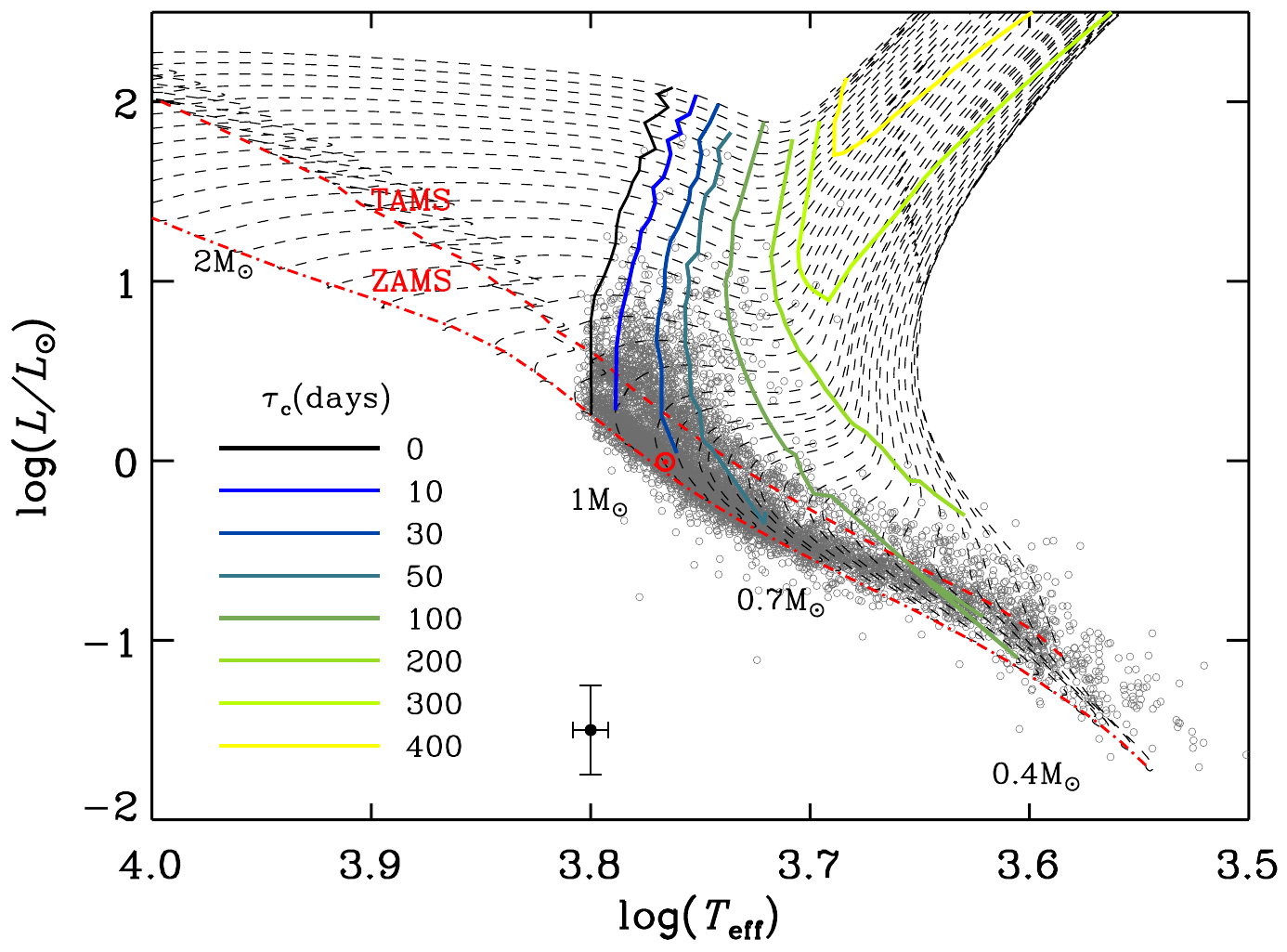}
\caption{Parameter space of $T_{\rm eff}$ and log($L/L_{\odot}$), which we used for the determination of $\tau_{\rm c}$ by matching them with the Yale-Potsdam Stellar Isochrones. The gray circles are the stars of our sample. The $\tau_{\rm c}$ isocontours are shown as solid colored lines. The red dashed lines mark the ZAMS and TAMS. The stellar mass from 0.4 $M_{\odot}$ to 2 $M_{\odot}$ are marked at the beginning of each isochrone. The typical uncertainty for the solar mass is plotted at the bottom. Note the isochrones are plotted at metallicilty [Fe/H]=0. }\label{figtaohr}
\end{figure}
\begin{figure}[ht]
\includegraphics[width=0.5\textwidth]{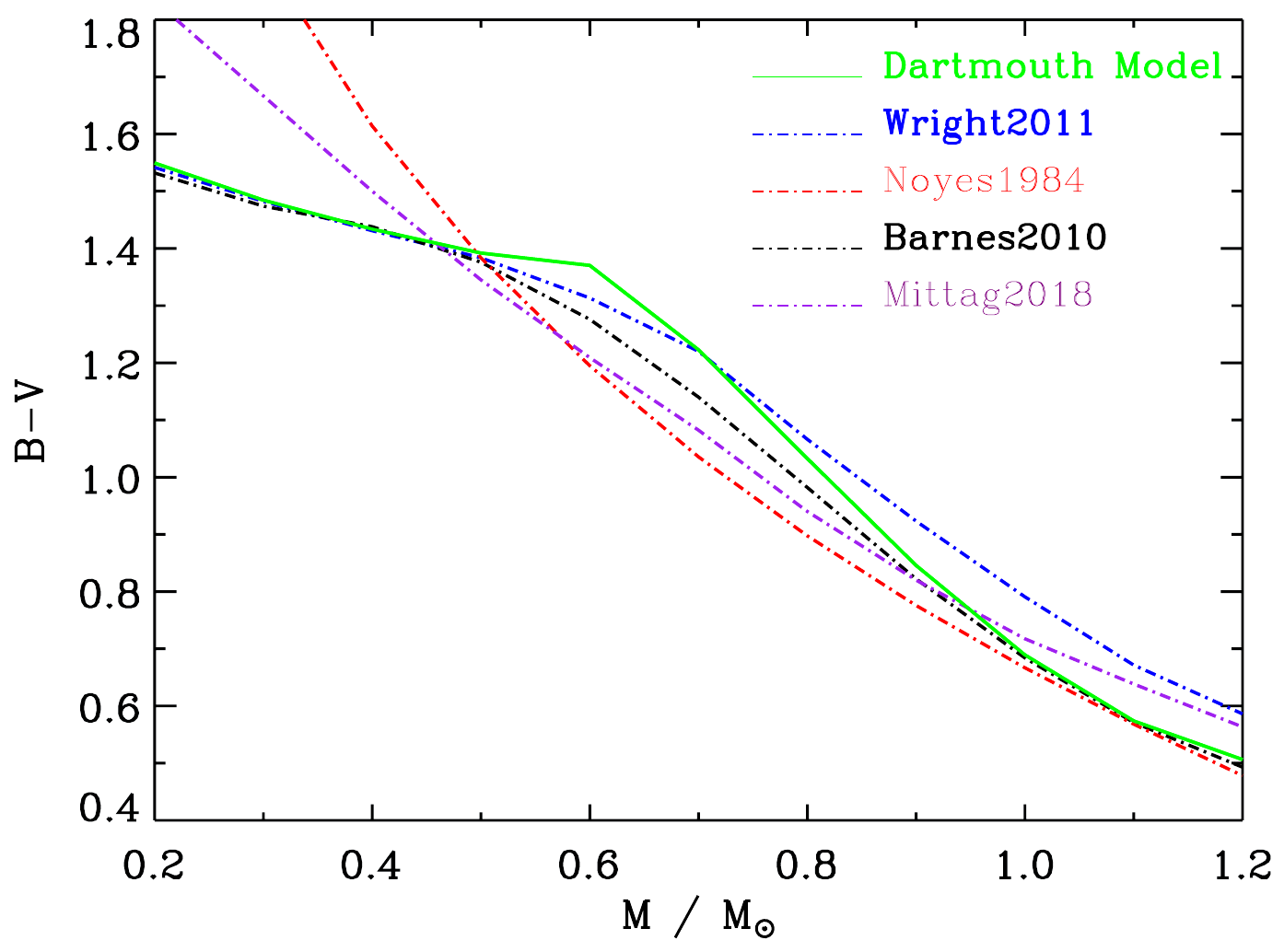}
\caption{Mass vs. $B-V$  for different color systems. The Dartmouth model is used in this study, and we calibrated the other color system to our scale to compare $\tau_c$.}
\label{figcom_bv}
\end{figure}
\begin{figure*}[ht]
    
\includegraphics[width=0.5\textwidth]{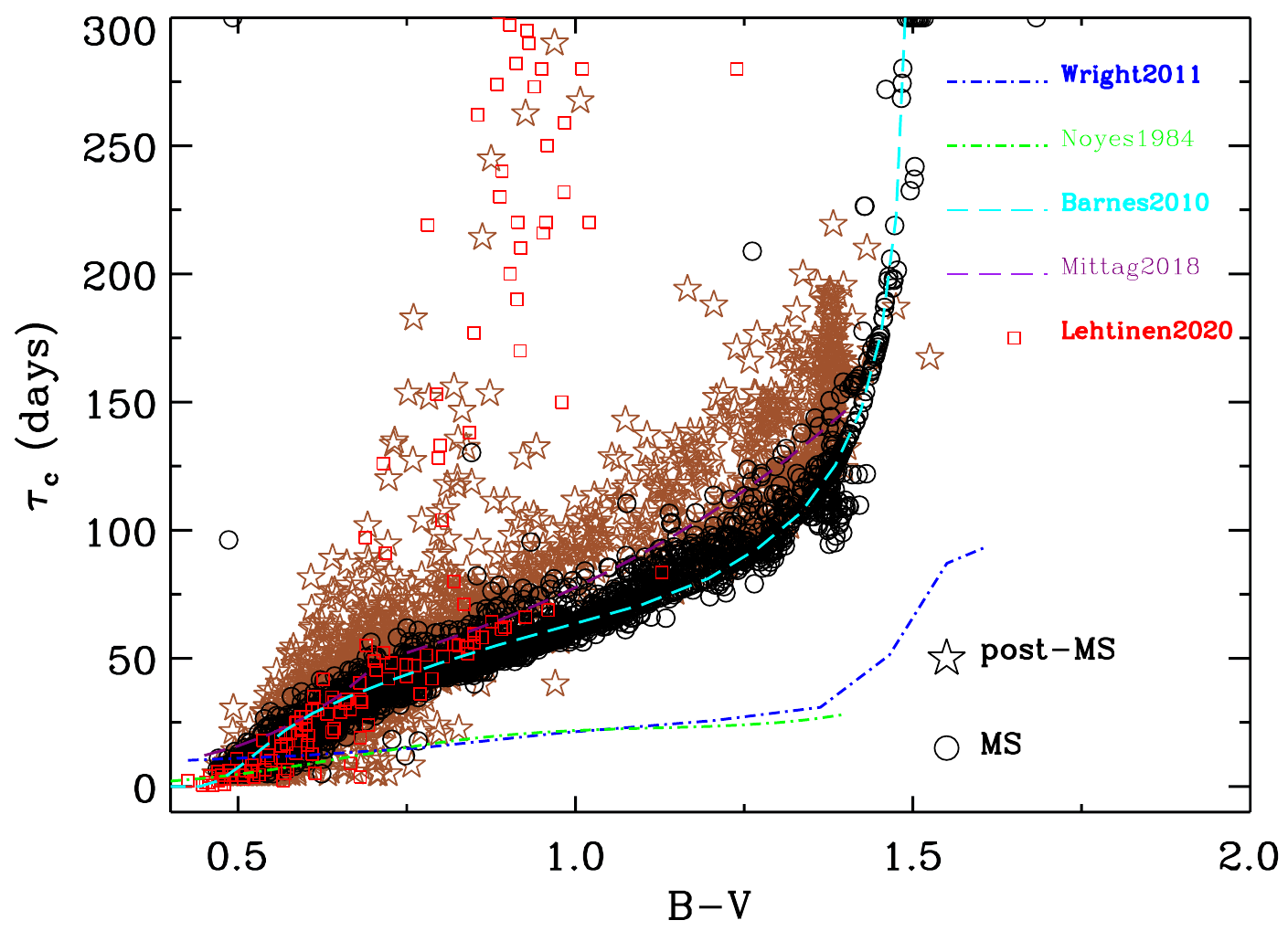}
\includegraphics[width=0.5\textwidth]{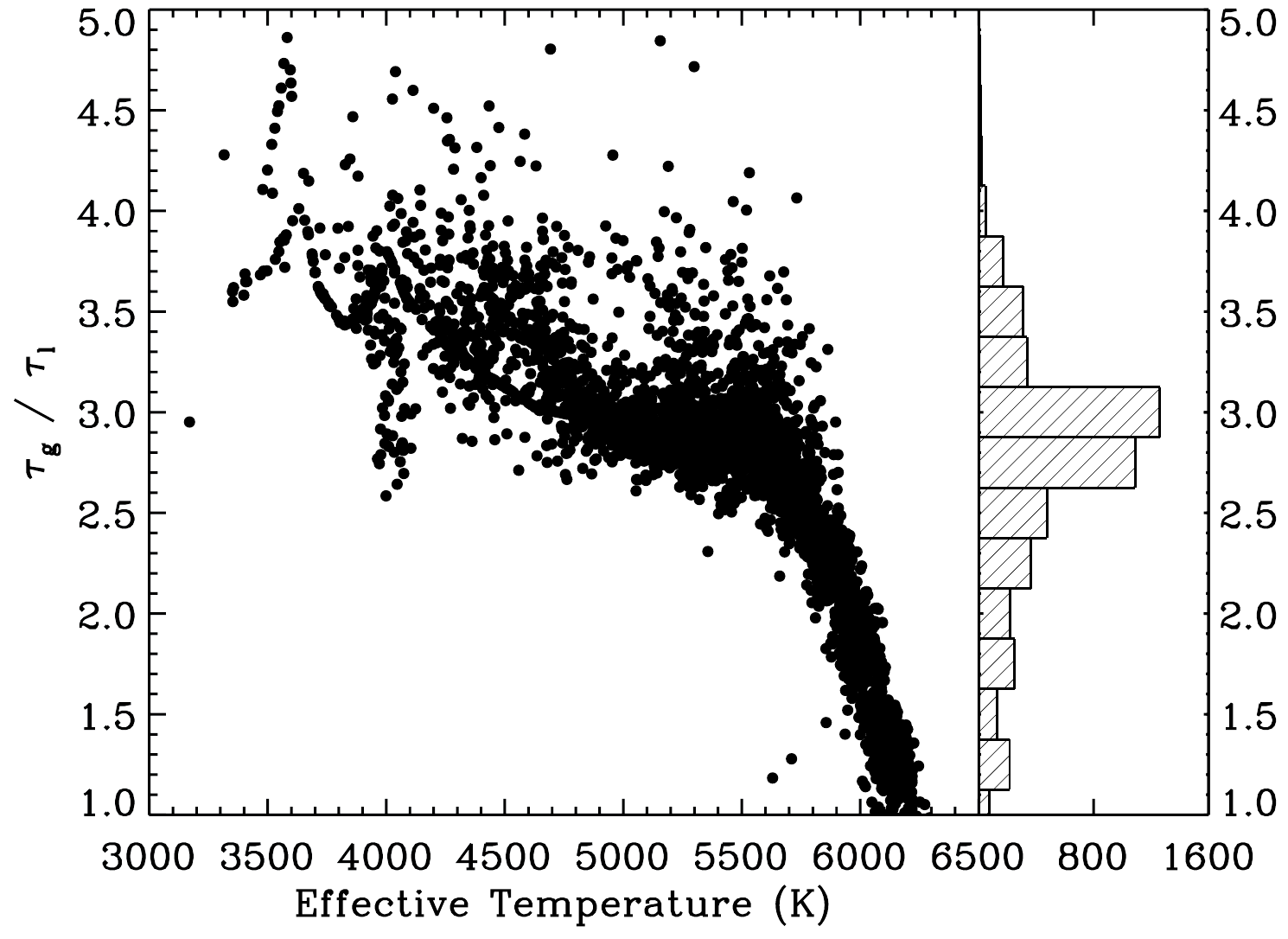}

\caption{Left panel: $B-V$ vs. $\tau_{\rm c}$. The black open circles and five-point stars are MS and post-MS stars identified in Fig.~\ref{figtaohr}. Red squares are from a MWO sample \citep{Lehtinen2020} whose $\tau_g$ is also derived from the YaPSI model. We have plotted four reference relationships (dashed lines) indicating the local $\tau_{\rm c}$ \citep{Noyes1984,Wright2011} and the global $\tau_{\rm c}$ \citep{Barnes2010b,Mittag2018}, after calibrating their $B-V$ scale to this study. Right panel: Comparison between the YaPSI $\tau_{\rm g}$ of MS stars and the empirical $\tau_l$ along with effective temperature.}\label{figtaocom}
\end{figure*}
\subsection{Sample selection}

This study required two observation quantities: the stellar chromospheric activity and the corresponding rotation period from the LAMOST spectra \citep{Zong2018} and the Kepler mission \citep{Borucki2010}, respectively. We established our sample as follows: (1) We collected rotation periods in the Kepler mission from two literature \citep{Mc2014,Santos2021}. One of them presented rotation periods of more than 34,000 stars for the entire Kepler data set by the autocorrelation function \citep[ACF;][]{Mc2014}, while the later study presented more than 55,000 stars from the same data set by an improved method that consisted of the combination of the ACF, the wavelet analysis, and the composite spectrum \citep{Santos2021}. It also argued that a small part of the catalog of \citet{Mc2014} is false-positive signals which were red giants, classical pulsators, close-in binaries and photometric pollution, while its additional more than 24,000 new rotators had higher noise.  However, with regard to the common stars between the two catalogs, there is an agreement within 15\% for 99.0\% of them \citep{Santos2021}, which is a good cross-validation. We hereby select the common stars that have an agreement within 15\% for our catalog of rotators.  In total, we picked out rotation periods of $\sim$32,000 stars from them. (2) A cross-identification between the catalog of the LAMOST DR6 low-resolution spectra and the catalog of rotators was made with a tolerance of $<2$ arcsec in coordinates. As the Ca~{\sc ii} H\&K in the spectrum were used in this work, the threshold signal-to-noise ratio (S/N) in the {\it g} band was set to 20. (3) The parameter catalog of LAMOST DR6 generated by the LAMOST stellar parameter pipeline \citep[LASP;][]{Luo2015} gives a classification for each spectrum, including a ``class" field and a ``subclass" field. Any spectrum whose class field was not ``STAR" or whose subclass field was ``Non" was discarded. Stars that had an effective temperature of $T_{\rm eff} > 6400$K were also discarded. For a star with multiple spectra, the spectrum with the highest S/N was used. This step removed targets of other types, pollution, and low quality spectra. Since most stars of the Kepler mission are field stars, there is a lack of fast rotating stars which may induce bias in the analysis of the rotation--activity relationship. We thus added about 450 stars of two open clusters ($\alpha$ Persei and Pleiades) to our sample that are also covered by the LAMOST sky survey. Both of $\alpha$ Persei (age $\sim$ 85 Myr) and Pleiades (age $\sim$ 125 Myr) have enough ultra-fast rotators with measured rotation periods in the C sequence and gap. The rotation periods of $\alpha$ Persei are from \citet{Boyle2023}, and the rotation periods of Pleiades are from \citet{Hartman2010},\citet{Covey2016}, and \citet{Rebull2016}. The number of the final sample is 6846 stars.

\subsection{The physical parameters of the sample}

Surface gravity $g$ is difficult to measure accurately. Its typical uncertainties are $25\% \sim 50\%$ through spectroscopy and $90\% \sim 150\%$ through photometry. We selected values of ${\rm log} g$ from the so-called flicker gravity \citep{Bastien2016}, the LASP and the Kepler Input Catalog \citep{Mathur2017}. We preferentially chose flicker gravity, whose uncertainty approaches the accuracy of asteroseismic gravity. We adopted values of ${\rm log} g$ from the LASP for the rest of the F-, G-, and K-type stars and values from the KIC for M-type stars.

The spectral types are given by LASP. The other basic parameters, including effective temperature, metallicity and radial velocity, are given by the LASP for F-, G-, K-type stars. We calibrated LAMOST spectra to the rest-frame on wavelength. The typical external errors of the LASP are -47$\pm$ 95 K, 0.03$\pm$0.25 dex, -0.02$\pm$0.1 dex and -3.75$\pm$6.65 km$\rm s^{-1}$ for $T_{\rm eff}$, ${\rm log} g$, [Fe/H] and radial velocity, respectively \citep{Luo2015}. We note that the influence of those uncertainties are much small compared to internal errors of signal noises, transformation errors and fitting errors when we calculate the S-index. Nevertheless, parameters of M-type stars are not presented by the LASP mainly due to their coupled relation with each other and the poor performance of template spectra on molecular bands. We use the spectral index of several molecular bands to determine effective temperature and [Fe/H] of M-type stars in our sample \citep{Fang2016}, and we calculate their RVs through cross-matching method with template spectra \citep{Boch2007}.

\subsection{The Mount Wilson S-index and the scaling from LAMOST observations}\label{sect_mwo_lamost}
The definition of the S-index traces back to the Mount Wilson Observatory (MWO) stellar cycle program \citep{Wilson1968,Baliunas1995}, which is the ratio between the flux through two 1.09{\AA} full-width at half maximum triangular bandpasses centered on the Ca~{\sc ii} H\&K  lines and two 20{\AA} rectangular pseudo-continuum bandpasses located some distance blueward and redward of the K and H lines, respectively (V is centered at 3901{\AA}, and R is centered at 4001{\AA}). The MWO used two spectrophotometers in its program successively, dubbed the HKP-1(from 1966 to 1977) and HKP-2 (post-1977). The MWO S-index ($S_{\rm MWO}$) for HKP-2 is given as

\begin{center}
\begin{equation}
 S_{\rm MWO}=\alpha \cdot \frac{f_{\rm H} + f_{\rm K}}{f_{\rm V} + f_{\rm R}},
 \end{equation}
\end{center}

where $f_{\rm H}, f_{\rm K}, f_{\rm V}, f_{\rm R}$ are the total counts in four bandpasses. The correction factor, $\alpha$, is used to calibrate the HKP-2 data to the HKP-1 data since HKP-2 used a different telescope and reference bandpasses (i.e., V and R bandpasses) compared to the HKP-1 \citep{Vaughan1978}, and it was found to be 2.4 \citep{Duncan1991}.

As LAMOST observations are done with a spectrograph and not with a spectrophotometer, we followed the previous studies by rewriting the LAMOST S-index ($S_{\rm LAMOST}$) in terms of mean flux per wavelength interval $\tilde{f}_{\rm H} =f_{\rm H}/\Delta\lambda_{\rm H}, \tilde{f}_{\rm K} =f_{\rm K}/\Delta\lambda_{\rm K}, \tilde{f}_{\rm V} =f_{\rm V}/\Delta\lambda_{\rm V},\tilde{f}_{\rm R} =f_{\rm R}/\Delta\lambda_{\rm R}$
rather than integrated flux \citep{Lovis2011,Karoff2016,Def2017}:

\begin{equation}\label{eq_sl}
\begin{aligned}
 S_\text{LAMOST}&=8\alpha \cdot \frac{\Delta \lambda_\text{H}}{\Delta \lambda_\text{R}} \cdot \frac{\tilde{f}_\text{H} +\tilde{f}_\text{K}}{\tilde{f}_\text{V} +\tilde{f}_\text{R}}\\ &= 8\alpha \cdot \frac{1.09  \text{\AA}}{20 \text{\AA}} \cdot \frac{\tilde{f}_{\text{H}} +\tilde{f}_\text{K}}{\tilde{f}_\text{V} +\tilde{f}_\text{R}}
 \end{aligned}.
 \end{equation}

The factor of eight accounts for the HKP-2 duty cycle (with an 8:1 ratio of core-to-continuum exposure time), and the bandpass width is $\Delta\lambda_\text{H}=\Delta \lambda_\text{K} = 1.09 \text{\AA},  \Delta \lambda_\text{R}= \Delta \lambda_\text{V} = 20 \text{\AA}$.

We note that LAMOST has no target in common with the MWO targets to directly verify the consistency of its S-index values with the MWO scale because the MWO program focuses on relatively bright stars (V mag brighter than 10), which saturate the LAMOST charge-coupled devices (CCDs). Nevertheless, several subsequent surveys on chromospheric activity have targets in common with LAMOST, which have been calibrated to the MWO scale. A MWO-scaled S-index catalog has been compiled from several surveys \citep{Boro2018}. It consists of 4454 unique cool stars, and some of them have multiple measurements from different surveys (in total 6962 S-index values). LAMOST has 114 targets in common with it, which were used to perform the calibration in Fig.~\ref{figsc}. Fig.~\ref{figsc} shows a systematic deviation from $S_{\rm LAMOST}$ to $S_{\rm MWO}$ and a detection limit on very low activity stars, which could be due to the low resolution of the LAMOST spectra. The residuals are mainly from the systematic uncertainties between the LAMOST and the MWO telescope. The variation of $S_{\rm MWO}$ in a star is much smaller than the uncertainties. We added the uncertainty of the fitting result and the uncertainty of $S_{\rm LAMOST}$ in quadrature as the uncertainty of $S_{\rm MWO}$, which is larger than the intrinsic uncertainty of a star.

As shown in Fig.~\ref{figsc}, we calibrated $S_{\rm LAMOST}$ to the MWO scale by performing a linear regression on the common stars \citep{Def2017,Boro2018,Hall2007} whose $S_{\rm MWO} > 0.15$, such that
\begin{equation}\label{eq_sc}
\begin{aligned}
 S_{\rm MWO}&=a \cdot S_{\rm LAMOST} + b\\
 a&= 2.89^{+0.21}_{-0.17}\\
 b&= -0.65^{+0.18}_{-0.20},
 \end{aligned}
 \end{equation}
where $a$ and $b$ are given by a weighted ordinary least-square bisector \citep{Isobe1990}. The weights were derived from the uncertainty $\sigma_{S}$ of each $S_{\rm LAMOST}$. This fit has an rms dispersion in $S_{\rm MWO}$ of 0.22 dex. The uncertainty of $a$ and $b$ was calculated by separately changing each parameter until the deviation of $\chi^2$ from its best-fit value reaches $\Delta\chi^2$ = 1.

The lower limit of this calibration is $S_{\rm MWO} > 0.15$, under which the calibration depends on the extrapolation. Equation~\ref{eq_sc} could therefore result in a very small or even negative $S_{\rm MWO}$. This issue need us to set a threshold to verify the reliability on the value of $S_{\rm MWO}$. According to previous surveys, we found the minimum $S_{\rm MWO}$ is $\sim$ 0.09. We thus took 0.1 as the lower limit of $S_{\rm MWO}$ and discarded stars whose $S_{\rm MWO} < 0.1$ .

\begin{figure*}[ht]

\includegraphics[width=0.9\textwidth]{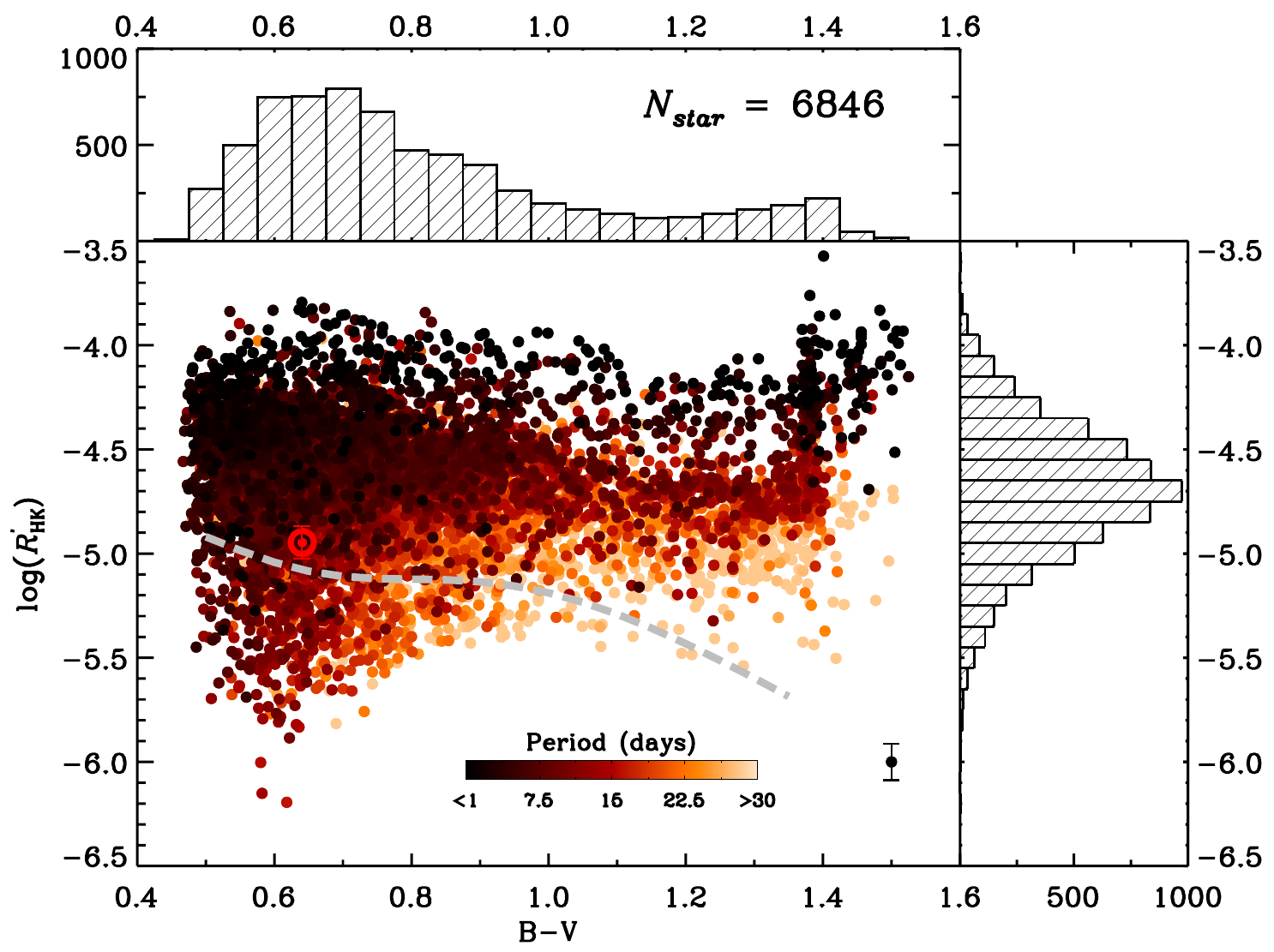}
\caption{Chromospheric activity $R'_{\rm HK}$ as a function of $B-V$~ in our sample. The color bar indicates the rotation period. The Sun is marked as $\odot$ symbol with $B-V =0.64$, ${\rm log} R'_{\rm HK} =$ -4.94 \citep{Egeland2017}.  The gray dashed line denotes $S_{\rm MWO} = 0.15$, below which $S_{\rm MWO}$ were obtained by extrapolation that could have large uncertainties. The typical error of ${\rm log} R'_{\rm HK}$ = -4.5 is plotted as a black circle at the bottom right. The top and right histograms show the number distribution along $R'_{\rm HK}$ and $B-V$, respectively.}\label{figrhkhr}
\end{figure*}

\begin{table*}
\setlength{\tabcolsep}{1pt}
\centering
\caption{\label{table_stat}Example of the parameters and information of stars in this study.}

\begin{tabular}{ccccccccccccccccccccccc}
\hline\hline
Gaia DR3&KIC&Spectral&$T_{\rm eff}$&log$g$&[Fe/H]&S-index&${\rm log}(R'_{\rm HK})$&Period&$\tau_{g}$&$B-V$&log($L/L_{\odot}$)& Phase&Cluster\\[1ex]
ID&&Type&(K)&($\rm cm/s^2$)&&&&(day)&(day)&&&&\\[1ex]

\hline\\
2051694572946664960&1028018& F9& 5783&  4.28&  0.168&  0.6708 & -4.124& 0.621&33.55 &0.691&0.016&MS&Field Star\\
2050241710075252352&1162715& F9 &5707&4.56& 0.082 & 0.2416& -4.718&6.625 &37.50 &0.706 &-0.078& MS& Field Star\\
2051694989565637760&1163579& F9 &5741&4.50&-0.025 &0.4193 & -4.361&5.429& 40.82 & 0.681 &0.241& post-MS& Field Star\\
2051692962341150976&1164363&G8 &5296& 4.66& 0.097& 0.5033& -4.446 &8.092 &48.51& 0.866&-999.000& MS& Field Star\\
2050254457537860608&1292666&G6&5611&4.39 & 0.356 & 0.1252 & -5.304 & 18.437& 36.65&0.775& 0.018& MS& Field Star\\
2050245970677463168&1293907&G9& 5129 & 4.73 & 0.011&0.1760&-5.041&26.300 &59.08 & 0.910& -999.000 &MS&Field Star\\
2051744914267116544&1295195& K4 &4870 &4.64  &0.203 &0.9155 &-4.415 &21.960&67.79& 1.057& -0.459& MS& Field Star\\
70286319462343808&--& G9& 5193 & 4.66 & 0.060 & 0.7165 &  -4.243 &   6.438  & 51.20 &0.836&    -0.846& MS& Pleiades\\
70310027681371648&--& K7& 3905& 4.53 &-0.380 & 6.1897 & -4.152 &  7.148&  131.43 & 1.396  &  -1.547 &MS& Pleiades\\
70941383577307392&--& G7 &5409 & 4.55 & 0.050 & 0.3522 &  -4.525 &   6.531 &  44.04 &  0.754  &  -0.624& MS &Pleiades\\\\
\hline

\hline

\end{tabular}
\tablefoot{The entire table is available online at the CDS as coordinates and uncertainties are not shown.}
\end{table*}
\subsection{The determination of the global convective turnover time}\label{ms_pms}

The convective turnover time cannot be observed directly and instead has to be estimated using either empirical fits \citep{Noyes1984,Wright2011} or computations of theoretical models \citep{Spada2017,Kim1996}. The difference between the local convective turnover time, $\tau_l$, and the global convective turnover time, $\tau_g$, could influence the calculation of Ro and the partition of the relationship. The local convective turnover time is proportional to the ratio between the pressure scale height and the local convective velocity. Its value is entirely determined at a specific location, which is usually half a mixing length above the bottom of the convection zone \citep[hence the name ``local";][]{Noyes1984}. The global convective turnover time is an integral of the ratio over the convection zone, which essentially takes into account the local contributions of all layers in the convection zone \citep[hence the name ``global";][]{Spada2017,Kim1996}.

The traditional rotation--activity relationship usually uses the local convective turnover time, but the global convective turnover time should be preferred with regard to this study because: (i)The global convective turnover time is adopted by the gyrochronology that need us to keep consistent with. (ii) The local convective turnover time is obtained by empirical fits \citep{Noyes1984,Wright2011} that have hardly taken into account subgiants and giants. Our sample comprises a substantial portion of subgiants and giants which also conform to the rotation--activity relationship in terms of the global convective turnover time \citep{Lehtinen2020}.

We use a grid of stellar evolutionary tracks from the Yale-Potsdam Stellar Isochrones \citep[YaPSI;][]{Spada2017} \footnote{Available online: URL http://www.astro.yale.edu/yapsi/} to derive $\tau_g$ by finding the best match of them to physical parameters of each star. The YaPSI release is intended as an update and an extension of the Yonsei--Yale isochrones, improving the accuracy of physical parameters, in particular for low-mass stars ($M \le 0.6 M_\odot$). All tracks are constructed using a solar-calibrated value of the mixing length parameter, $\alpha_{\rm MLT} = 1.918$ or 1.821 to guarantee a smooth transition between low-mass stars ($M \le 0.7 M_\odot$) and massive stars ($M > 0.7 M_\odot$). Rather than the local convective turnover time which is location and local variables dependent, the YaPSI models computes the global convective turnover time to parameterize the characteristic timescale of convective turnover \citep{Kim1996}.

First, we remapped the YaPSI tracks as a function of uniformly spaced equivalent evolutionary points\citep[EEPs;][]{Spada2017}. We then constructed a series of synthetic tracks by linearly interpolating EEPs with steps of 0.02 $M_{\odot}$. We compared $T_{\rm eff}$ and log($L/L_{\odot}$) of each star with tracks from $0.2 M_{\odot}$ to $3M_{\odot}$ and obtain $\tau_g$ from the closest point of all the tracks to the star (see Appendix~\ref{appendix_lum} for the estimate of the bolometric luminosity). The uncertainties of parameters are according to LASP with 100 K in $T_{\rm eff}$, and we set the uncertainty of log($L/L_{\odot}$) to be 0.25 dex. The standard deviation of the values of $\tau_g$ within the uncertainties is computed as the uncertainty of $\tau_g$ for each star. In this way, we could repeat the above procedures to match the $\tau_g$ for different metallicities from -1.5 to 0.3. The final value of $\tau_g$ is obtained by linear interpolation in metallicity. Fig.~\ref{figtaohr} shows a matching process at metallicity [Fe/H]=0. Some stars are beyond the scope of the tracks, which could be due to the uncertainty of the parameters or the boundary of the tracks. We match the closest points in the tracks to them.  We take 5 days as the lower limit of $\tau_g$ as some F-type stars are near the boundary of $\tau_g = 0 $. We take 300 days as the upper limit of $\tau_g$ to reduce the influence of a few outliers. Very few stars in our sample are without a reliable extinction or distance. We estimated their $\tau_g$ by interpolating the tabulated data in \citet{Barnes2010b}, which has the same scale as the YaPSI model (see Fig.~\ref{figtaocom} for the comparison of the YaPSI and \citet{Barnes2010b}).

It should be noted that in matching the observed parameters of the stars in our sample with the YaPSI tracks, we only considered the portion of the track after the zero-age MS (ZAMS). In other words, we exclude the (significantly less likely) solutions on the pre-MS, in favor of those on the MS or post-MS (post-MS). The ZAMS is defined as the point on the evolutionary track at which the central hydrogen content has dropped below 0.999 times its initial value. The termination-age MS (TAMS) is determined by that when the hydrogen mass fraction in the core of a star reaches below $10^{-4}$, after which the stars enter their subgiant phase. The classification of MS and post-MS stars in this study is identified by the two tracks of ZAMS and TAMS in Fig.\ref{figtaohr}.

\subsection{ The comparison of global and local convective turnover time}\label{app_com_tau}

The convective turnover time is often expressed as a function of $B-V$. In order to compare our $\tau_g$ with previous studies \citep{Noyes1984,Wright2011,Barnes2010b,Mittag2018}, all of the references should have the same scale as ours on $B-V$. We obtained the ($B-V$)--mass--$\tau_c$ relation from the tabulated data \citep{Wright2011,Barnes2010b} and functions \citep{Noyes1984} from previous studies. We remapped their $B-V$ values to our scale (the Dartmouth model) through stellar mass as shown in Fig.~\ref{figcom_bv}, and perform the calibration. After the calibration, we establish their new ($B-V$)--$\tau_c$ relations with the same $B-V$ scale. We plot these new relations in Fig.~\ref{figtaocom}.

The left panel of Fig.~\ref{figtaocom} shows the ($B-V$)--$\tau_g$ relation of the YaPSI model overplotted with local and global relations of previous studies. The $\tau_g$ of a MS star is consistent with that of \citet{Barnes2010b}, which is independent of its age, while the $\tau_g$ of a post-MS star (sub-giant or giant) varies as it is aging. In particular, the bifurcation near $B-V =0.7$ reveals the different dependence between dwarfs and giants \citep[e.g.,][]{Lehtinen2020}. The left panel of Fig.~\ref{figtaocom} also shows the asymptotic line near the fully convective stars ($B-V \approx 1.55, T_{\rm eff} \approx 3300 K$, M4). We note that to date, there has been no unified treatment for the convective turnover time of fully convective stars. For example, the empirical relation treated fully convective stars the same as partially convective stars \citep{Wright2018}, while theoretical models treated them in a totally different way\citep[see e.g.,][and references therein]{Irving2023}. However, since our sample has very few stars with $T_{\rm eff} < 3500$ K, we take all of them as partially convective stars and match them with the closest points in the tracks (the right bottom region of Fig.~\ref{figtaohr}).

The right panel of Fig.~\ref{figtaocom} shows a comparison between our theoretical $\tau_g$ of MS stars and the empirical $\tau_l$ \citep{Wright2011}. The ratio of $\tau_g/\tau_l$ varies along with the effective temperature. The histogram shows that most G-type and K-type stars are near $\tau_g/\tau_l \sim 3 $.  As $\tau_l$ \citep{Wright2011} have little constraint in $ T_{\rm eff} > 5700 $  (nearly a constant of $\sim$ 10 days), and G-type and K-type stars account for $\sim 80\%$ stars of our sample, we conclude that $\tau_g \approx 3 \tau_l$ in our sample, which is roughly in agreement with the previous comparison \citep{Lehtinen2020,Mittag2018}. This ratio is also supported by the theoretical calculation showing $\tau_g/\tau_l \approx 2.5$ \citep{Kim1996,Landin2010}. We should note that in the empirical fitting of \citet{Wright2011}, they set the $\tau_l$ of the Sun as $\approx 12$ days, and take it as a benchmark to rescale their initial convective turnover time. Given that the $\tau_g$ of the Sun in the YaPSI model is 34.9 days, which is also by a factor of 3 to $\tau_l$ \citep{Wright2011}, we can infer that the initial fitting result of $\tau_l$ \citep{Wright2011} is the global convective turnover time.

We also compared our method with others \citep{Lehtinen2020}, which construct a new EEPs track by estimating the stellar mass. We found an agreement within 30\% for 82\% of stars and an agreement within 50\% for 93\% of stars. The uncertainty is mainly caused by two kinds of stars: (1) stars out of the boundary of the tracks, where the method of \citep{Lehtinen2020} has a high uncertainty on the mass estimate (2) very low mass stars which has a dramatic variation on the convective turnover time. Our method could obtain a more robust result in those areas.
\begin{figure*}[ht]

\includegraphics[width=1\textwidth]{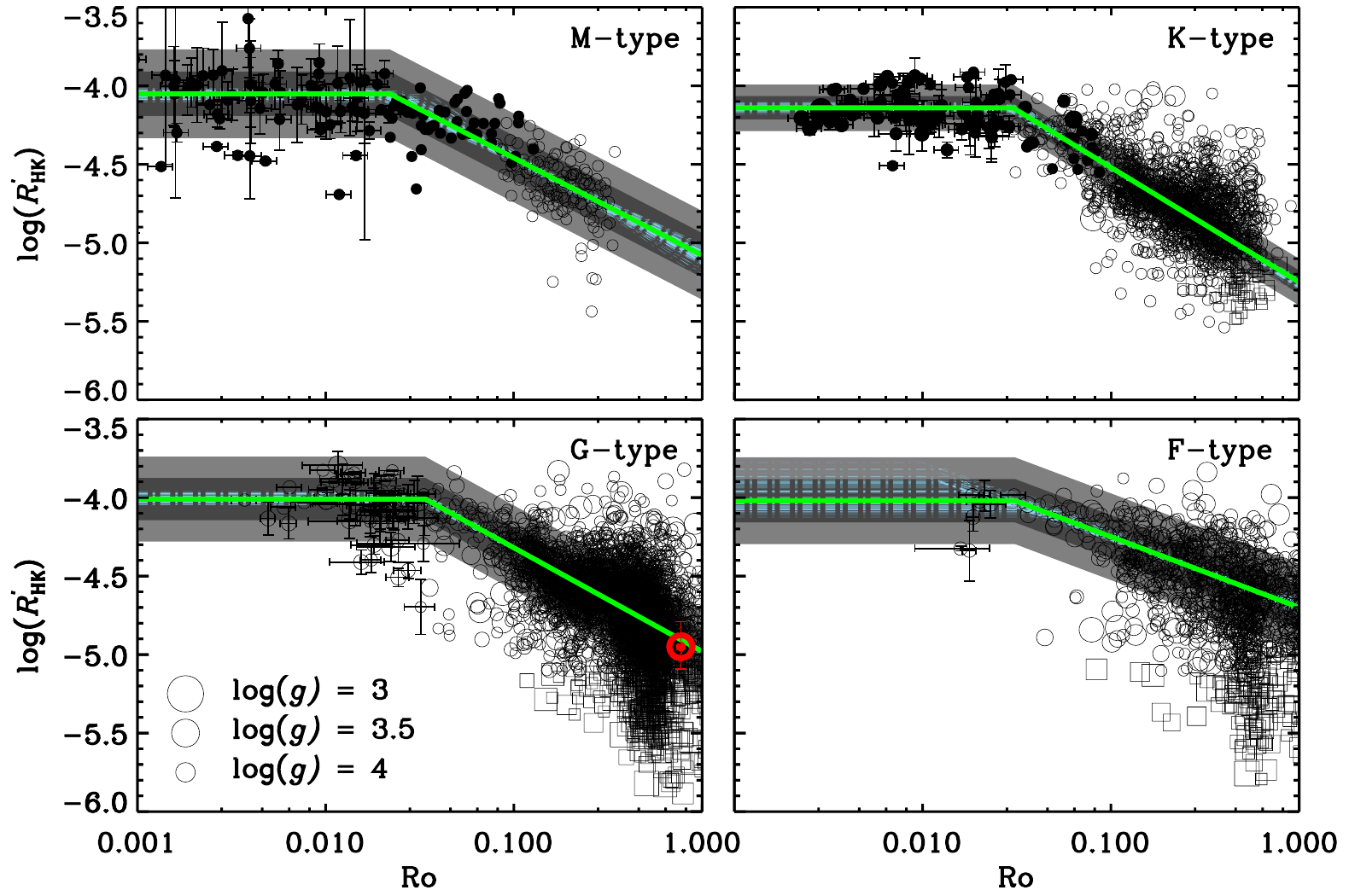}
\caption{Chromospheric activity $R'_{\rm HK}$ vs. the Rossby numer $R_{o}$ for F-,G-,K-, and M-type stars.  The filled circles represent emission line stars, while the open circles represent absorption line stars. The green line indicates the best fit. The squares represent stars whose $S_{\rm MWO} < 0.15$ which may have large uncertainties. The blue dotted lines are 100 random draws from the posterior distribution of the fit. The 1$V$ and 2$V$ scatter of the fit are indicated by the shaded gray regions (see Appendix~\ref{rare} for the meaning of $V$).The Sun is marked with an $\odot$ symbol. Error bars are only plotted in the saturated regime for clarity.}\label{fig_fgkm_dic}
\end{figure*}
\begin{figure}[ht]
    
\includegraphics[width=0.5\textwidth]{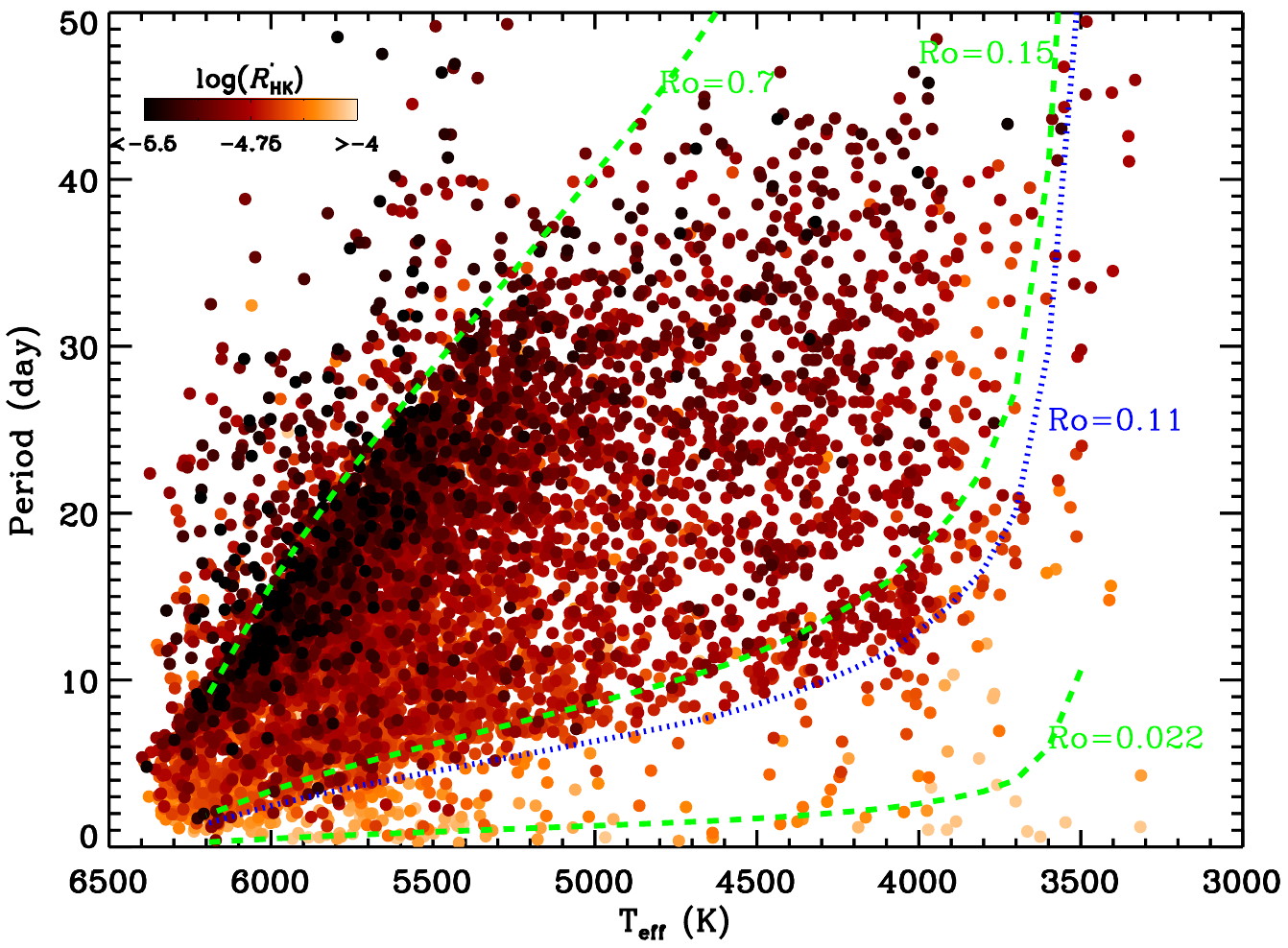}

\caption{Effective temperature vs. rotation period. The color bar indicates the chromospheric activity $R'_{\rm HK}$. The green dashed lines denote the three critical Rossby number that separate the convective region, gap, the interface region, and the weakened magnetic braking region. The blue dotted line (Ro=0.11) denotes the lower envelope of the Kepler slow rotator distribution, where fast rotators begin to converge. In order to clarify this lower envelope, we only plot the Kepler star of our sample (i.e., do not include stars of the two clusters).} \label{fig_teff_period}
\end{figure}
\section{The ${\rm log}(R'_{\rm HK})$ and the canonical rotation--activity relationship}\label{sec3}
The S-index includes contribution from chromospheric and photospheric radiation. It also includes contribution from the spectral continuum that varies with spectral types. In order to calculate the true contribution from chromosphere and normalize to the bolometric luminosity, we converted the S-index to the quantity of $R'_{\rm HK}$ (see Appendix~\ref{strhk} for details of the conversion).

In total, we have obtained 6846 stars with ${\rm log}(R'_{\rm HK})$ and Rossby number. Their $B-V$ range is from 0.3 to 1.5 including 5199 MS stars and 1647 post-MS stars. Table~\ref{table_stat} presents an example of our catalog. Fig.~\ref{figrhkhr} shows the chromospheric activity proxy ${\rm log}(R'_{\rm HK})$ as a function of $B-V$. It is similar to Fig.~3 of \citet{Boro2018}, while our sample has much more stars at $B-V >1$. The majority of stars lie in the range $-5.2 \le {\rm log}(R'_{\rm HK}) \le -4.4$ which is consistent to \citet{Boro2018}. However, the median value of ${\rm log}(R'_{\rm HK})$ is 0.5 dex larger than that of \citet{Boro2018}, which probably because the detection limit of the LAMOST spectra excludes the low-activity stars, and our sample has more low-mass stars that are more active.

As noticed by \citet{Boro2018}, rapidly rotational velocities could cause that the wings around Ca~{\sc ii} H\&K lines would be filled in, mimicking an active chromosphere \citep{Schro2009}. As a result, one can see much more artificially active stars in the range of $B-V < 0.5$, where rotational velocity increases dramatically. We note that this effect is more serious for the low-resolution spectrum, since it cannot separate the H1 and K1 minima from the wings at all. However, this effect has a smaller influence on K-type and M-type stars, because most of those fast rotators are emission line stars, which have a very high intrinsic H and K flux. The conversion equation from the S-index to $R'_{\rm HK}$ has fewer reference points on K- and M-type stars. This may also account for the dip in the upper envelope of $R'_{\rm HK}$ to some extent. As we have used the improved conversion equation for M-type stars (see Appendix~\ref{strhk}), their saturation activities are in line with F- and G-type stars.

The gray dashed line indicates the detection limit of the LAMOST spectra ($S_{\rm MWO} \sim 0.15$), below which the calibration of $S_{\rm MWO}$ is by extrapolation and subsequently has a larger uncertainty. As mentioned in Section~\ref{sect_mwo_lamost}, we have set 0.1 as the lower limit of $S_{\rm MWO}$, which depicts the lower boundary of Fig.~\ref{figrhkhr}. Therefore, this artificial boundary is not the real low-activity boundary, which is the so-called basal flux \citep[e.g.,][]{Rut1984,Mittag2013}. However, for the regime of $B-V >1$ the basal flux is larger than the artificial boundary and has a rapid increase along with increasing $B-V$. This phenomenon is also reported by \citet{Mittag2013} and \citet{Boro2018}, which is within our expectation because the increase of the convective turnover time could strengthen the dynamo action.

It should be noted that in our sample, there are a few slow rotators with anomalous activities that are much larger than the average level. This phenomenon can also be found in the X-ray luminosity \citep{Wright2011,Pizz2019} and studies on the light curve modulation \citep[e.g.,][]{Mc2014,Santos2021}. \citet{Cao2022} studied the filling factor of stars in Pleiades and found that a few slow rotators have very large filling factors, which are far from the rotation--filling factor relationship. They further found that those stars also have high chromospheric and coronal activity, implying the activities of those stars may be true. The physical nature of those stars are not clear, as they cannot be explained by the stellar cycle.

Figure~\ref{fig_fgkm_dic} shows the canonical rotation--activity relationship that identifies two stages of stellar evolution: the saturated (faster, with higher activity) and unsaturated (slower, with declining activity) regimes. This dichotomy requires a piecewise function (Eq.~\ref{eq_cf}) with two segments to fit it \citep[e.g.,][]{Wright2011,Newton2017}. Here, we fit the relationship for F-,G-,K-, and M-type stars, respectively (see Appendix \ref{rare} for details of the fit).

\begin{equation}\label{eq_cf}
{\rm log} R'_{\rm HK} = \begin{cases} {\rm log} R'_{\rm HK,sat}, & {\rm Ro} \le {\rm Ro_{sat}}  \cr \beta{\rm log} {\rm Ro}+C, &{\rm Ro} > {\rm Ro_{sat}} \end{cases}.
\end{equation}

The parameter $R'_{\rm HK,sat}$ is the value of $R'_{\rm HK}$  at saturation, $\beta$ is the slope of the decay in the unsaturated regime, and $\rm Ro_{sat}$ is the critical Rossby number that separates the two regimes. Table~\ref{table_cra} presents the results of the fitting. 

The dichotomy of the relationship has been widely used by various activity proxies \citep[e.g.,][]{Noyes1984,Wright2011,Newton2017,Yang2017,Yang2019}. The physics interpretation of its mathematical form is that a star goes through two dynamos (convective and solar-like) as it ages, which is revealed by the activity dependence on rotation. The transition of the two dynamos occurs at $\rm Ro_{sat}$.  In the following subsections, we raise issues regarding the dichotomy by discussing its relation to gyrochronology.

\begin{table}
\caption{\label{table_cra}Fitting parameters of the rotation--activity relationship in terms of the dichotomy model for F-,G-,K-, and M-type stars.}
\centering
\begin{tabular}{cccc}
\hline\hline\\
Spectral type&$\beta$&$\rm Ro_{sat}$&${\rm log} R'_{\rm HK,sat}$\\[1ex]
\hline\\
F           &$-0.45^{+0.03}_{-0.03}$     &$0.031^{+0.011}_{-0.012}$ & $-4.02 \pm 0.14$\\[1ex]
G           &$-0.66^{+0.02}_{-0.02}$   &$0.034^{+0.003}_{-0.003}$& $-4.01 \pm 0.13$\\[1ex]
K          &$-0.73^{+0.02}_{-0.03}$   &$0.030^{+0.003}_{-0.002}$&$-4.14 \pm 0.10$\\[1ex]
M         & $-0.62^{+0.04}_{-0.05}$    &$0.022^{+0.004}_{-0.003}$ &$-4.05 \pm 0.14$\\[1ex]
\hline\\
Mean value      &$-0.62^{+0.03}_{-0.03}$    &$0.029^{+0.005}_{-0.005}$&$-4.06 \pm 0.13$\\[1ex]
\hline
\end{tabular}
\tablefoot{The uncertainty of ${\rm log} R'_{\rm HK,sat}$ is calculated using the  1$V$ scatter of the fit (see Appendix~\ref{rare} for the meaning of $V$).}
\end{table}

\begin{table*}[ht]
\setlength{\tabcolsep}{1pt}
\caption{\label{table_para_act_ro}Parameters of the activity--rotation relationship for different proxies.}
\begin{tabular}{cccccccccc}
 \hline \hline
 Reference & Spectra & Sample  & Activity &
Rotation&\multicolumn{2}{c}{$\beta$} &\multicolumn{2}{c}{$\rm Ro_{sat} \times 10^{2}$}& ${\rm log} R_{\rm act,sat}$ \\
                    &  Type & Number & Proxy&
Proxy&Local $\tau_{\rm c}$&Global $\tau_{\rm c}$ &Local $\tau_{\rm c}$& Global $\tau_{\rm c}$ &\\ 

\hline
\citet{Soder1993}&F,G,K& $\sim$110 & H$\alpha$,Ca~{\sc ii} IRT & $v$sin$i$ &$--$&$--$ &$--$ &$--$& $-$(3.5--4)\\
\citet{Jackson2010}&K,M& $\sim$240 & Ca~{\sc ii} IRT & $v$sin$i$ & $\sim-1$&$--$ & $\sim 10$&$--$&$\sim-4.4$ \\
\citet{Mohanty2003}&M,L& $\sim$100 & H$\alpha$ & $v$sin$i$ & $--$ &$--$&$--$ &$--$& $-$(3.9--4.2)\\
\citet{Delfosse1998}&M& $\sim$120 & H$\alpha$ & $v$sin$i$ & $--$ &$--$ &$--$&$--$&$-$(3.5--4)\\
\citet{Newton2017}&M& $\sim$350 & H$\alpha$ & Period & $-1.7 \pm 0.1$ & $-2.0\pm 0.2$& 21$\pm 2$&6$\pm 1$&$-3.8$ \\
\citet{Douglas2014}&M&$\sim$100 & H$\alpha$ & Period & $-0.73 \pm 0.1$& $-0.83 \pm 0.3$ & $11\pm 3$&$3.5\pm 0.6$& $ -3.9$ \\
\citet{Mclean2012}&M,L& $\sim$160 & Radio & $v$sin$i$ & $-1.1$ & $--$& $--$ &$--$& $-7.5$\\
\citet{Pizzolato2003}&F,G,K,M& $\sim$250 & X-ray & Period & $-2$ & $--$& $\sim 13$&$--$& $-3.2\pm 0.3$\\
\citet{Wright2011,Wright2018}&F,G,K,M & $\sim$820 & X-ray & Period & $-2.3\pm 0.5$ & $-2.2\pm 0.1$ &$14 \pm 6$&$5 \pm 0.2$& $-3.1\pm 0.2$\\
\citet{Mittag2018}&F,G,K,M& $\sim$170 & $R^{+}_{\rm HK}$ & Period & $--$& $-0.93 \pm 0.1$ & $--$&$3.3 \pm 0.7$ &$-3.6 \pm 0.2$\\
\citet{Yang2019}&K,M& $\sim$1480 & Flare & Period & $--$& $-3.9 \pm 0.2$ & $--$&$6.6 \pm 0.5$ &$-4.1 \pm 0.1$\\
\citet{Lehtinen2020}&F,G,K& $\sim$150 & $R'_{\rm HK}$ & Period &$--$& $-1.0\pm 0.1$ & $--$& $--$&$\sim -4.0$\\
This study&F,G,K,M & $\sim$6850 & $R'_{\rm HK}$ & Period &$--$& $-0.62 \pm 0.05$ & $--$&$3.0\pm0.3$& $-4.0 \pm 0.1$\\
\hline
\end{tabular}

\tablefoot{The studies on the rotation--activity relationship whose sample number $N_{\rm star} >100$. The column $R_{\rm act,sat}$ indicate the activity values in the saturation regime for different proxies. Parameters of this study are mean values of F-,G-,K-,M-type stars. Parameters derived by global $\tau_c$ for $R_{\rm flare}$ \citep{Yang2019} and $R^{+}_{\rm HK}$ \citep{Mittag2018}, $R_{\rm X}$ \citep{Wright2011} and $R_{\rm H\alpha}$ \citep{Douglas2014,Newton2017} were fit by the emcee method used in this study, see Appendix \ref{rare} for details. The sample of \citet{Lehtinen2020} is lack of fast rotators, which may lead to a biased $\beta$. The sample of \citet{Soder1993} is from open cluster the Pleiades ($t_{\rm age} \sim 125$ Myr). The sample of \citet{Douglas2014} is from open cluster the Praesepe and the Hyades ($t_{\rm age} \sim 750$ Myr). The sample of \citet{Jackson2010} is from open cluster NGC2516 ($t_{\rm age} \sim 150$ Myr). Parameters derived by local $\tau_c$ for X-ray are from \citet{Wright2018}, which includes fully convective stars. About 45\% stars of \citet{Wright2011} are from open clusters whose ages are from 40 Myr to 700 Myr.}
\end{table*}
\begin{figure}[ht]
\centering
\includegraphics[width=0.5\textwidth]{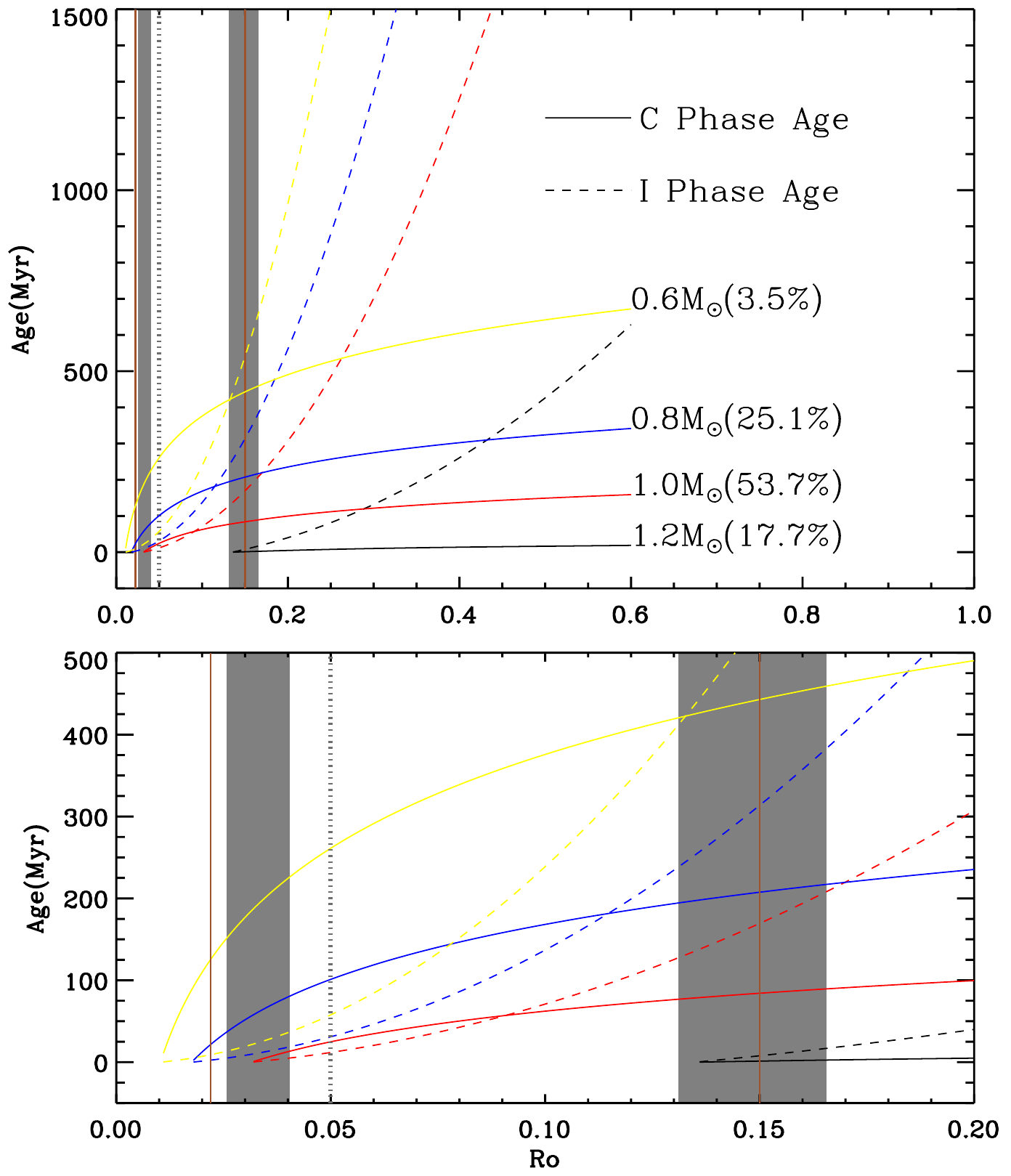}
\caption{Top panel: Rossby number--age relations for 1.2, 1.0,0.8, and 0.6 solar mass stars. The solid and dashed curves represent the C sequence age and the I sequence age (the first term and the second term of Eq.~\ref{eq_age}) respectively. The same color indicates the same stellar mass and the fraction of the sample number is also presented. The vertical brown lines are Ro$_\text{sat}=0.022$ and Ro$_\text{g-to-i}=0.15$ respectively, separating the C, g and I region. The dotted vertical line is the separating line given by the dichotomy. The shaded region indicates the variation range of the critical Rossby number of the C-to-g and g-to-I transition (see Fig.~\ref{fig_isp_r}), which depends on stellar mass. Bottom panel: Lower-left region in the top panel. }\label{fig_cls}
\end{figure}
\subsection{The mapping from gyrochronology to the rotation--activity relationship}

Fig.~\ref{fig_teff_period} shows the Kepler stars of our data in the color--period diagram. There is a sparse region of stars at low rotation periods, indicating that the angular momentum loss rate has a drastic variation near Ro$ = 0.11$. \citet{Curtis2020} proposed that this lower envelope separates the dense region and spares region of rotation periods, indicating that fast rotators begin to converge near this envelope. Inspired by the morphology of stars of young open clusters in the color--period diagram, \citet{Barnes2003b} proposed that a star went through three phases (convective, gap, and interface) in its MS era that was called the CgI scenario \citep{Barnes2007,Barnes2003b}.  Then he associated the convective phase and interface phase with the saturated and unsaturated regime of the rotation–-activity relationship, respectively \citep{Barnes2003a}. The two phases are supposed to have the convective dynamo and interface dynamo (or the solar-like dynamo), respectively, and thus result in different activity dependences on rotation and the angular momentum loss rate. In order to clarify the difference of the three phases in physics, \citet{Barnes2003b} pointed out that the convective dynamo might have a decoupled stellar core and envelope, the interface dynamo had a coupled state, and the gap corresponded to the recoupling process. 

The CgI scenario forms the physical basis of gyrochronology, while \citet{Barnes2003b} neglected the gap (or took it as a transition point rather than a duration) when mapping it to the rotation--activity relationship. \citet{Wright2011} 
accepted this mapping and found the transition point $\rm Ro_{sat} \sim 0.05$  in terms of our Rossby number scale. This value is close to the middle of the gap Ro $\sim 0.06$ that is given by \citet{Barnes2003b}.

\subsection{The issues of the canonical rotation--activity relationship}\label{issue}

From a physical point of view, one of the biggest issues of the dichotomy is the transition of the two regimes.\citet{Wright2011} raised this question: Why is the transition between the two regimes so smooth? In other words, since the dynamo transition (from convective to solar-like) should involve complex physical changes in magnetism and stellar structure, it is unlikely to occur instantaneously. Moreover, \citet{Reiners2014} found that the activity in the saturated regime has remnant dependence on rotation. This implies that some regions near the transition point could have an independent activity dependence rather than that of the saturated regime or the unsaturated regime. Other activity proxies from light curve modulation found the so-called intermediate period gap in the unsaturated regime, which can be associated with the transition of surface brightness from spot-dominated to facula-dominated \citep[e.g.,][]{Mc2014,Reinhold2020,Mathur2025}. This evidence implies that stars in the unsaturated regime may have different physical properties on stellar structure and dynamos, which cannot be depicted by one mechanism. Emission lines are common for fast rotating K- and M-type stars. They are markers for distinguishing whether a star is active or not. However, $\rm Ro_{sat}$ is not at the place that separates emission line stars and absorption line stars (top panels of Fig.~\ref{fig_fgkm_dic}).

From a mathematical point of view, the value of $\rm Ro_{sat}$ has not been well determined yet. 
We collect studies on the rotation--activity relationship whose sample number $N_{\rm star} >100$ and summarize the fitting parameters in Table~\ref{table_para_act_ro}. Since some studies used a local $\tau_c$, we supplement the parameters fit by global $\tau_c$. The fitting details are the same as Appendix~\ref{rare}. Table~\ref{table_para_act_ro} shows that the choice between local $\tau_c$ and global $\tau_c$ has little influence on $\beta$ and saturation activity, but it will change $\rm Ro_{sat}$ by a factor of about three. It is in agreement with $\tau_g \sim 3\tau_l$, which we have discussed in Section~\ref{app_com_tau}. Different activity proxies account for the difference in $\beta$ and $R_{\rm act,sat}$. Appendix~\ref{app_cda} presents the conversion of these activity proxies, which shows that the parameters $\beta$ and $R_{\rm act,sat}$ are basically the same after the conversion, while $\rm Ro_{\rm sat}$, which is supposed to be independent of the activity proxy, has a large uncertainty among these proxies (Table~\ref{table_para_act_ro}). Even for the same proxy, \citet{Douglas2014} and \cite{Newton2017} found totally different $\rm Ro_{sat}$ based on different samples of M-type stars. Although the samples of \citet{Douglas2014} and \citet{Newton2017} focused on young and old M-type stars, respectively, the discrepancy in the samples cannot fully explain the uncertainty (see Appendix~\ref{app_cda} for more discussion on the uncertainty of $\rm Ro_{sat}$).

From the point of view of gyrochronology, the value of $\rm Ro_{sat}$ of the dichotomy could raise several conflicting issues on interpreting the C and I sequence. \citet{Barnes2010} propose that the stellar age could be expressed as follows: 
\begin{equation}
t=t_{C}+t_{I}=\frac{\tau_g}{k_C}{\rm ln}(\frac{\tau_g Ro}{P_0})+\frac{k_I}{2\tau_g}((\tau_g Ro)^2-P_0^2) \label{eq_age},
\end{equation}
where $t_C$ is the C sequence age and $t_I$ is the I sequence age, $k_C$ and $k_I$ are constant, $P_0$ is the initial rotation period. The mean initial period is suggested to be 1.1 days \citep{Barnes2010}, which corresponds to Ro = 0.03 for solar-like stars and Ro=0.01 for early M-type stars. The rotation of pre-MS stars will speed up as they evolve towards MS stars \citep{Spada2020} and the angular momentum redistribution are still an open question once a star evolves into the MS \citep[e.g.,][]{Bouv2014}, which may result in dispersion in rotations of ZAMS stars. Given that the recent observations of very young clusters indicate that the initial rotation period of pre-MS and ZAMS stars varies from the breakup velocity (0.3 day) to 30 days, and its distribution peak depends on stellar mass \citep{Rebull2018,Rebull2022,Douglas2024}, we note that Eq.~\ref{eq_age} is a semi-empirical relation and its initial period of 1.1 days should be taken as a parameter that makes gyrochronology self-consistent and the lower limit where gyrochronology makes sense in mathematics. \citet{Barnes2010} argued that on the condition that the first term $>>$ the second term in Eq.~\ref{eq_age}, a star was on the C sequence and vice versa. The equation indicates that the evolutionary phase (C,g and I) depends on which term ($t_C$ and $t_I$) is dominated at a stellar age. We plot $t_C$ and $t_I$ as a function of the Rossby number for different stellar mass in Fig.~\ref{fig_cls}, so that the C, g and I phase can be inferred from the comparison of $t_C$ and $t_I$. As shown in Fig.~\ref{fig_cls}, the dichotomy scheme defines a direct transition from C to I phases at Ro $\approx 0.05$\citep{Wright2011,Wright2018}, where $t_C$ is much larger than $t_I$. This Ro corresponds to stellar ages of $\approx$40 Myr, $\approx$130 Myr and $\approx$320 Myr for 1.0-M$_{\odot}$, 0.8-M$_{\odot}$ and 0.6-M$_{\odot}$ stars respectively. \citet{Barnes2003b} and \citet{Meibom2009} investigated the fraction of C and I sequence stars along stellar age and found that the fraction of I sequence stars at 40 Myr, 130 Myr and 320 Myr is less than $30\%$, $40\%$, and $60\%$, respectively. This means that most stars are not on the I sequence at Ro=0.05, while the dichotomy take it as the start point of the I sequence.

\section{The new rotation--activity relationship}\label{sec4}
Besides the dichotomy, there are various expressions on the rotation--activity relationship such as in a linear-log plane \citep[e.g.,][]{Mama2008}, piecewise functions with different slopes \citep[e.g.,][]{Reiners2014,Fang2018} and a high-order piecewise function \citep{Mittag2018}. All of them are based on a mathematical point of view, aiming at minimizing the scatter of the fit. However, in order to address the issues of the canonical rotation--activity relationship, a more appealing scheme on both physics and mathematics is necessary.

\begin{figure*}[ht]%
\centering
\includegraphics[width=1.0\textwidth]{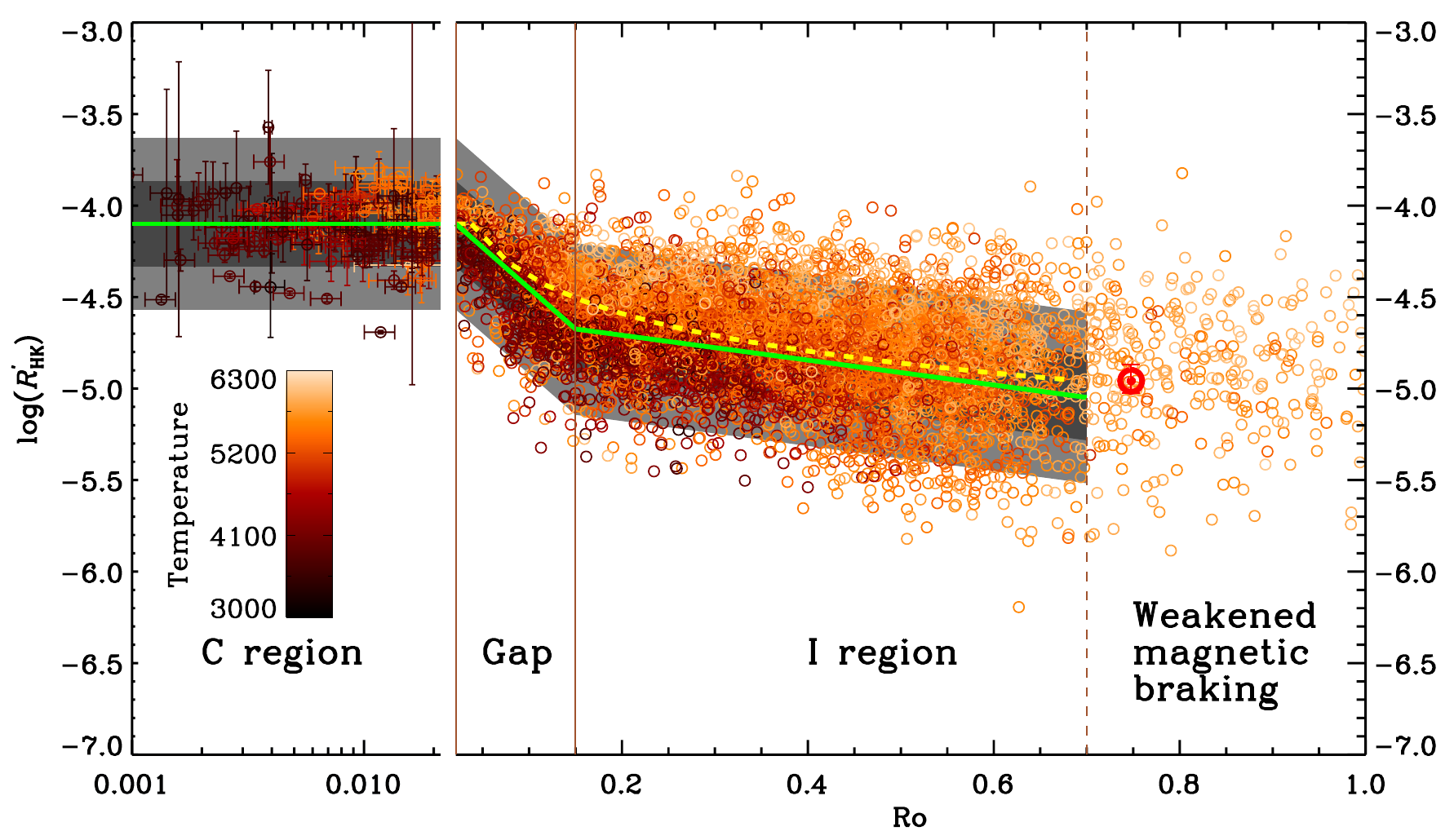}
\caption{Rotation--chromospheric activity relation divided into four parts according to the ``CgIW" scenario. The solid vertical lines indicate the critical points between the C, gap, and I regions (Ro $\approx 0.022$ and Ro $\approx 0.15$, respectively). The dashed vertical line marks the transition to the weakened-magnetic-braking phase at $\rm Ro \approx 0.7$. The green lines represent the best fit of our proposed model and yellow lines represent the best fit of the dichotomy model. The 1$\sigma$ scatter and 2$\sigma$ scatter of the fit are indicated by the shaded gray region. The location of the Sun is marked with an $\odot$ symbol. Representative error bars are plotted for the datapoints in the C region.}\label{fig_act_ro_color}
\end{figure*}

\begin{figure*}[ht]%
\centering
\includegraphics[width=1.0\textwidth]{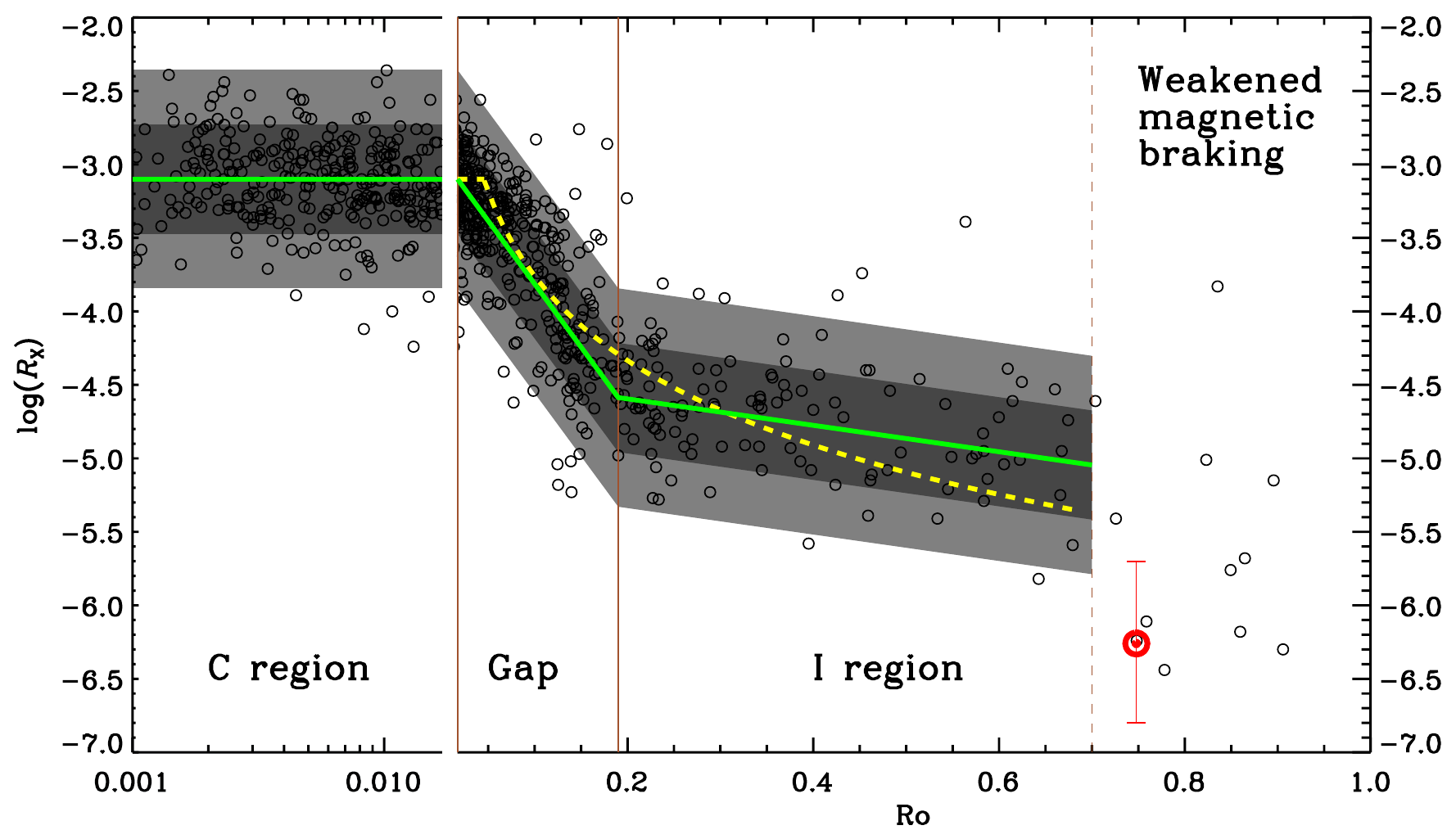}
\caption{Same as Figure~\ref{fig_act_ro_color} but for X-ray data. Solid vertical lines indicate the critical points between the C, gap, and I regions (Ro $\approx 0.017$ and Ro $\approx 0.19$, respectively.
}\label{fig_ro_rx}
\end{figure*}
\begin{table*}
\setlength{\tabcolsep}{1pt}
\centering
\caption{\label{table_pfit}Fitting parameters of the period--activity relationship in terms of the CgIW scenario along with the effective temperature.}

\begin{tabular}{cccccccccccc}
\hline\hline
Temperature range&$\beta_g$&$\beta_i$&$P_{\rm sat}$&$P_{\rm g-to-I}$&$\rm Ro_{sat}$&$\rm Ro_{g-to-I}$&${\rm log} R'_{\rm HK,sat}$&Age$_{\rm C-to-g}$&Age$_{\rm g-to-I}$&$P_{\rm I-to-W}$&Age$_{\rm I-to-W}$\\[1ex]
(K)&&&(day)&(day)&&&&(Myr)&(Myr)&(day)&(Gyr)\\[1ex]

\hline\\
$3500--4000$         &$-0.032^{+0.003}_{-0.003}$ &$-0.000^{+0.01}_{-0.01}$ & $5.13^{+1.79}_{-1.50}$& $22.92^{+1.86}_{-2.55}$&$0.031^{+0.011}_{-0.009}$&$0.138^{+0.011}_{-0.015}$&$-4.15^{+0.20}_{-0.20}$&$429^{+106}_{-107}$&$1494^{+141}_{-181}$&$116$&19.5\\[1ex]
$3750--4250$         &$-0.041^{+0.003}_{-0.003}$ &$-0.005^{+0.01}_{-0.01}$ & $4.57^{+0.56}_{-0.86}$& $15.93^{+0.94}_{-0.99}$&$0.039^{+0.005}_{-0.007}$&$0.135^{+0.008}_{-0.008}$&$-4.15^{+0.15}_{-0.15}$&$297^{+31}_{-51}$&$972^{+70}_{-70}$&$82$&13.8\\[1ex]
$4000--4500$         &$-0.038^{+0.004}_{-0.004}$ &$-0.008^{+0.01}_{-0.01}$ & $2.69^{+0.47}_{-0.45}$& $13.57^{+0.86}_{-0.97}$&$0.029^{+0.005}_{-0.005}$&$0.146^{+0.009}_{-0.010}$&$-4.19^{+0.11}_{-0.11}$&$144^{+30}_{-32}$&$806^{+67}_{-72}$&$65$&10.9\\[1ex]
$4250--4750$         &$-0.052^{+0.005}_{-0.005}$ &$-0.008^{+0.01}_{-0.01}$ & $2.00^{+0.45}_{-0.51}$& $11.65^{+1.11}_{-1.01}$&$0.026^{+0.006}_{-0.006}$&$0.150^{+0.014}_{-0.013}$&$-4.15^{+0.14}_{-0.14}$&$80^{+31}_{-36}$&$675^{+89}_{-76}$&$54$&9.1\\[1ex]
$4500--5000$         &$-0.058^{+0.005}_{-0.005}$ &$-0.012^{+0.01}_{-0.01}$ & $1.86^{+0.43}_{-0.45}$& $10.41^{+0.68}_{-0.63}$&$0.028^{+0.006}_{-0.006}$&$0.157^{+0.010}_{-0.009}$&$-4.08^{+0.15}_{-0.15}$&$62^{+27}_{-31}$&$596^{+56}_{-50}$&$47$&7.7\\[1ex]
$4750--5250$         &$-0.065^{+0.005}_{-0.005}$ &$-0.015^{+0.01}_{-0.01}$ & $1.51^{+0.35}_{-0.31}$& $9.21^{+0.75}_{-0.69}$&$0.026^{+0.006}_{-0.006}$&$0.160^{+0.013}_{-0.012}$&$-4.03^{+0.15}_{-0.15}$&$33^{+23}_{-24}$&$517^{+63}_{-55}$&$40$&6.7\\[1ex]
$5000--5500$         &$-0.069^{+0.005}_{-0.005}$ &$-0.018^{+0.01}_{-0.01}$ & $1.51^{+0.26}_{-0.29}$& $8.16^{+0.49}_{-0.44}$&$0.031^{+0.005}_{-0.005}$&$0.165^{+0.010}_{-0.009}$&$-4.05^{+0.19}_{-0.19}$&$29^{+16}_{-20}$&$453^{+42}_{-36}$&$35$&5.7\\[1ex]
$5250--5750$         &$-0.085^{+0.007}_{-0.007}$ &$-0.019^{+0.01}_{-0.01}$ & $1.66^{+0.25}_{-0.28}$& $5.89^{+0.57}_{-0.68}$&$0.040^{+0.006}_{-0.007}$&$0.144^{+0.014}_{-0.017}$&$-4.07^{+0.22}_{-0.22}$&$34^{+14}_{-16}$&$291^{+45}_{-49}$&$29$&4.7\\[1ex]
$5500--6000$         &$-0.145^{+0.010}_{-0.010}$ &$-0.023^{+0.01}_{-0.01}$ & $0.79^{+0.21}_{-0.29}$& $4.24^{+0.58}_{-0.71}$&$0.035^{+0.009}_{-0.013}$&$0.131^{+0.018}_{-0.021}$&$-4.00^{+0.25}_{-0.25}$&$0^{+11}_{-0}$&$185^{+43}_{-48}$&$22$&3.7\\[1ex]
$5750--6250$         &$-0.192^{+0.010}_{-0.010}$ &$-0.045^{+0.01}_{-0.01}$ & $0$& $2.02^{+0.62}_{-0.35}$&$0$&$0.090^{+0.028}_{-0.016}$&$-4.00^{+0.27}_{-0.27}$&$0$&$50^{+38}_{-20}$&$16$&2.6\\[1ex]
$6000--6500$         &-- &$-0.089^{+0.02}_{-0.02}$ & 0& 0&0&0&--&0&0&7&1.1\\[1ex]
\hline

\hline

\end{tabular}
\tablefoot{The period--activity relationship in terms of the CgIW scenario along with the effective temperature is shown in Fig.~\ref{fig_ro_act_teffbin1},Fig.~\ref{fig_ro_act_teffbin2},and Fig.~\ref{fig_ro_act_teffbin3}. This table present the fitting results. The parameters $\beta$ represents the slope in each evolution phase. The parameters $P$ and Ro represent the critical rotation periods and Rossby number that separate two neighboring phases, respectively. The parameter ${\rm log} R'_{\rm HK,sat}$ represents the saturated activity. The parameter Age represents the corresponding age of the critical rotation period, which is derived from Eq.\ref{eq_age}. The critical Rossby number of the transition from the I phase to the W phase is 0.7 \citep{Saders2016}, which is used to calculate $P_{\rm I-to-W}$ and Age$_{\rm I-to-W}$. The median of the temperature range is used to calculate Rossby number and the critical ages. The uncertainty of ${\rm log} R'_{\rm HK,sat}$ is calculated using the  1$\sigma$ scatter of the fit.}
\end{table*}
 Here, we propose a tentative scheme to redefine the phase of the rotation--activity relationship (Fig.~\ref{fig_act_ro_color}). The scheme is based on the CgI scenario and the weakened magnetic braking scenario \citep{Saders2016,Hall2021} (see Sect.~\ref{wmb} for the discussion of the weakened magnetic braking phase), for which we coin the ``CgIW" scenario. It divides the relationship into four parts in a linear-log plane instead of a log-log plane. We found that the relationship with Ro $\lesssim$0.7 could be well fit by three intervals (green curve in Fig.~\ref{fig_act_ro_color}), instead of the traditional two. The first interval at low Rossby numbers is consistent with a constant activity level; the second and third interval are fit by linear functions with different decay timescales (see Appendix~\ref{app_newra} for details of the fit). The fitting results are
 \begin{equation}\label{eq_3interval_fit_result}
\begin{aligned}               
                 {\rm log} R'_{\rm HK,sat}&= -4.10^{+0.1}_{-0.1}\\
                 {\rm Ro}_{\rm sat}&= -0.022^{+0.005}_{-0.005}\\
                 {\rm Ro_{g-to-I}}&=0.15^{0.01}_{-0.01}. \\               
 \end{aligned}
 \end{equation}
Where the parameter ${\rm Ro_{g-to-I}}$ is the critical point that separates the second and third intervals.

 The reduced $\chi^2$ of our three-interval model fit is 0.3021, compared with the dichotomy model $\chi^2 = 0.3050$.  We did a Monte Carlo simulation to test whether the small improvement may simply be caused by random noise. We started from the best-fitting of the dichotomy model to the observed dataset, assuming that this is the true physical form. Then, we created sets of mock datapoints with their $R'_{\rm HK}$ values equal to the best-fitting of the dichotomy plus random noise. The noise followed a Gaussian distribution with $\sigma$ equal to the observed $R'_{\rm HK}$ scatter of our real data. As the noise increases along Rossby number, we calculated the $\sigma$ in a bin of 0.1 dex along Rossby number. We built 1000 simulated datasets, and we refit each of them with both the traditional dichotomy model and our proposed three-interval model. In all cases, the dichotomy model has a lower $\chi^2$. Thus, the probability that the lower $\chi^2$ of our new model for the observed dataset is caused by noise is less than 1/1000. We conclude that a three-phase model more likely reflects a real physical property. We note that our method on evaluating the goodness-of-fit of the two models is suitable (Vinay Kashyap private communication) rather than the Bayesian information criterion or F-test, which requires one model to be nested in the other \citep{Prota2002}.

\subsection{The mapping from gyrochronology to the new rotation--activity relationship}

We propose that those three intervals correspond to the three evolutionary phases (CgI) identified by gyrochronology. Specifically, we argue that the first interval (faster rotators) in our scheme corresponds to the convective dynamo regime (``C phase''), a decoupled state of the radiative core and convective envelope. The third interval (slower rotators) corresponds to the interface dynamo regime (``I phase''), when core and envelope are fully coupled. The transition between the two phases (``g phase'' or gap) in gyrochronology was conventionally located at Ro $\approx 0.06$ \citep{Barnes2007}, but the actual finite duration of the transition process and its dependence on stellar mass have remained a matter of debate. Ro $\approx 0.06$ falls inside the intermediate interval in our new rotation--activity scheme. Thus, we identified that interval as the gap phase. With our newly derived functional form of the rotation--activity relation, we can not only confirm the existence of a finite transitional phase but also quantify its duration. This enables a one-to-one correlation between optical color--rotation tracks and activity levels as proxies for stellar aging on the MS, in studies of open clusters.

From our three-interval fit, we find that the average transition points for the whole sample are at the critical point Ro$_{\rm sat} \approx 0.022$ for the C-to-g transition and Ro$_{\rm g-to-I} \approx 0.15$ for the g-to-I transition. At Ro$_{\rm sat} \approx 0.022$, \citet{Meibom2009} and \citet{Barnes2003b} shows that the fraction of C sequence stars is dominated, and Fig.~\ref{fig_cls} shows $t_C >> t_I$. Equally, there is an opposite relation at Ro$_{\rm g-to-I} \approx 0.15$. This makes the classification of the rotation--activity relationship and gyrochronology self-consistent. It should be noted that the average critical point can be used to interpret the mapping between the rotation--activity relationship and gyrochronology, while the precise periods and ages of the transition point depend on stellar mass (or effective temperature). We fit the relationship in temperature bins and compare the fitting results of critical periods with those found by open clusters in the next section.


We propose that the new model can resolve the issues of the dichotomy: (1) It ensures a ``pure" saturated regime (see Appendix~\ref{app_newra}), avoiding the remnant dependence. (2) It takes the gap as an independent transition phase rather than a critical point, which makes sense in physics. (3) The critical point of the g-to-I transition separates the unsaturated regime near the intermediate period gap (Fig.\ref{fig_teff_period}). This also makes sense in physics as it can be associated with the transition from spot-dominated to facula-dominated. (4) The new model makes the transition from emission line stars to absorption line stars be within the gap (Fig.~\ref{fig_ro_act_teffbin1} and Fig.~\ref{fig_ro_act_teffbin2} ). (5) We argue that the transition point of the dichotomy could vary within the gap, and the variation can explain the large uncertainty of Ro$_{\rm sat}$ of the dichotomy.

It should be noted that since the C,g,and I sequences have different angular momentum loss rates, their counterparts in the rotation--activity relationship certainly should have three distinct activity dependences on rotation (Fig.~\ref{fig_act_ro_color}). We recently found that the flaring activities of fast rotating stars have a solar-like latitude distribution \citep{Yang2025}. This implies that the activity dependence on rotation may not be a dynamo proxy. Moreover, recent numerical simulation shows that convection zone without a tachocline can also produce a solar-like dynamo \citep[e.g.,][]{Charb2020,Nelson2013,Fan2014,Zhang2022}. Therefore, we suggest that the three phases reflect changes in geometry, configuration, and strength of the magnetic field instead of the dynamo transition. 

\subsection{The weakened magnetic braking phase}\label{wmb}
The beginning of the weakened magnetic braking phase (W phase) is identified at Ro $\approx 0.7$ in terms of our Rossby number scale \citep{Saders2016,Hall2021}\footnote{The critical Rossby number of the weakened magnetic braking phase is $\approx 2.1$ in the scale of $\tau_{l}$. As $\tau_{g} \approx 3 \tau_{l}$, we get Ro $\approx 0.7$ in our Rossby number scale.}. Due to the change in the angular momentum loss rate, \citet{Saders2016} suggested that it is in a change of dynamo. In the rotation--activity relationship, \citet{Mama2008} studied the high accuracy $R'_{\rm HK}$ of the Mount Wilson system and reported that the stars with ${\rm log}R'_{\rm HK} < -5$ are independent of rotation, which are mainly at Ro $>$ 0.7.\citet{Chahal2023} also reported this trend by using LAMOST spectra. The study on stellar cycle indicates that the relation between the cycle period and the rotation period has a turnoff point near the Sun \citep{Irving2023}, which occurs at Ro $\approx 0.7$ in our Rossby number scale. Studies on the light curve modulation of the Kepler mission showed the independence at the tail of the relationship \citep{Mc2014,Santos2021,Reinhold2020,Mathur2025}. However, in this study, the uncertainty of $R'_{\rm HK} < -5$ is large because of the low-resolution spectra (Fig.\ref{figrhkhr}). We cannot directly identify whether the stars with Ro $>$ 0.7 have a weak dependence or are independent of rotation. Given that previous studies present the range of the activity independence, which is consistent with the W phase, we associate the last part of the rotation--activity relationship with the W phase.

Physically speaking, both the weakened magnetic braking and the activity independence can be attributed to a very weak magnetic field. In this case, the influence of rotation on the strength of the magnetic field may be a marginal factor, so that the geometry and configuration of the magnetic field become more important in the angular momentum loss and activity (see Sect.~\ref{x-ray} for the influence of the small- and large-scale field lines on the angular momentum loss and chromospheric activity). 

As the Rossby number of the Sun places it shortly (a few hundred megayears) after the I-to-W phase transition (Fig.~\ref{fig_act_ro_color}), it may produce some observational evidence on the Sun that indicates the transition and in turn proves the existence of the W phase in the rotation--activity relationship. For example, the study on the stellar cycle \citep{Metcalfe2016} and the brightness variation \citep{Reinhold2020b} indicate that the Sun might be undergoing a transition phase. If it is correct, both \citet{Metcalfe2016} and \citet{Reinhold2020b} predict that the Sun is heading to a phase of higher activity and variability. Combing their results and the framework of the CgIW model, the I-to-W transition represents absolute minimum of chromospheric activity during stellar evolution. It is interesting that the ``Snowball Earth" \citep{Hoffman1998} occurred in 750 Myr to 550 Myr ago (the corresponding Ro $\approx 0.69$), which also has the temperature minimum on the Earth. Then, the Earth started the ``Cambrian explosion" \citep{Smith2013} (525 Myr ago, Ro $\approx 0.7$), when the diversification and evolution rate of life significantly accelerated. The two eras imply that there might be a rapid rise of temperature on the Earth, which could be associated with the activity rise after the I-to-W phase transition. It may not be a coincidence that the I-to-W transition roughly corresponded to the emergence of life out of the oceans 400 million years ago.

\subsection{The new relationship for X-ray data}\label{x-ray}

We tested our scheme further by using a different activity proxy: the X-ray luminosity. Stellar X-ray emission is caused by magnetic reconnections in the corona. It is therefore a function of the magnetic activity and driven by differential rotation inside the star \citep{Parker1955}. In lower mass (G to M) MS stars, the X-ray over bolometric luminosity ratio $R_{\rm X} = L_{\rm X}/L_{\rm bol}$ traces the spin-down of a star over time \citep{Pallavicini1981}. We used publicly available values \citep{Wright2011} of $R_{\rm X}$ for 817 stars with reliable rotational periods. We fit the data with the traditional dichotomy model (reduced $\chi^2 = 0.344$) as well as with our revised CgI scheme (reduced $\chi^2 = 0.320$; see Appendix~\ref{app_newra} for details of the fit). In our proposed three-phase scheme, we find that the average transition values are Ro$_{\rm sat} \approx 0.017$ for the C-to-g transition and Ro$_{\rm g-to-I} \approx 0.19$ for the g-to-I transition (Fig.~\ref{fig_ro_rx}). As Ro$_{\rm g-to-I}$ is higher than that of the chromospheric activity, we note that the rarity of X-ray stars at high Rossby number (especially near the g-to-I transition region ) may result in the discrepancy. It also prevents us from fitting the relationship in temperature bins for X-ray data in the next section.

During the first three ages of low-mass MS stars, chromospheric and coronal properties evolve in lockstep. However, the fourth age of coronal activity has a significant drop, although its sample is small.
Physically, the W phase was explained as the weakening of the open magnetic field lines and the large-scale magnetic field, responsible for the spin-down. X-ray activity is a coronal phenomenon caused by large-scale fields \citep{Holzwarth2007}. Thus, the observed strong decline of $R_{\rm X}$ is in line with the theoretical expectations. Instead, the re-kindling of chromospheric activity $R'_{\rm HK}$ is also positively correlated with the mean intensity of the large-scale field \citep{Petit2008,Brown2022}, which is monotonically decreasing with age and Rossby number. However, stellar aging is also a story of redistribution of magnetic energy between field components. In the I phase, the large-scale toroidal component gradually disappears while the large-scale poloidal field (responsible for the loss of angular momentum) is only moderately reduced \citep{Petit2008}. We suggest that in the W phase, while the remaining large-scale poloidal field decreases, more energy is concentrated in small-scale fields (active regions and spots). Chromospheric lines are formed much closer to the photosphere, and are likely to be more strongly correlated with small-scale fields. We note that in G and K stars, for mean surface-averaged magnetic field strengths lower than $\approx$3 G, as may be the case in the W phase, the dependence of $R'_{\rm HK}$ on the large-scale field flattens \citep{Brown2022}; this supports our suggestion that the increase in chromospheric activity caused by small-scale fields is enough to offset the decrease of the large scale field component.
\begin{figure*}
\includegraphics[width=0.5\textwidth]{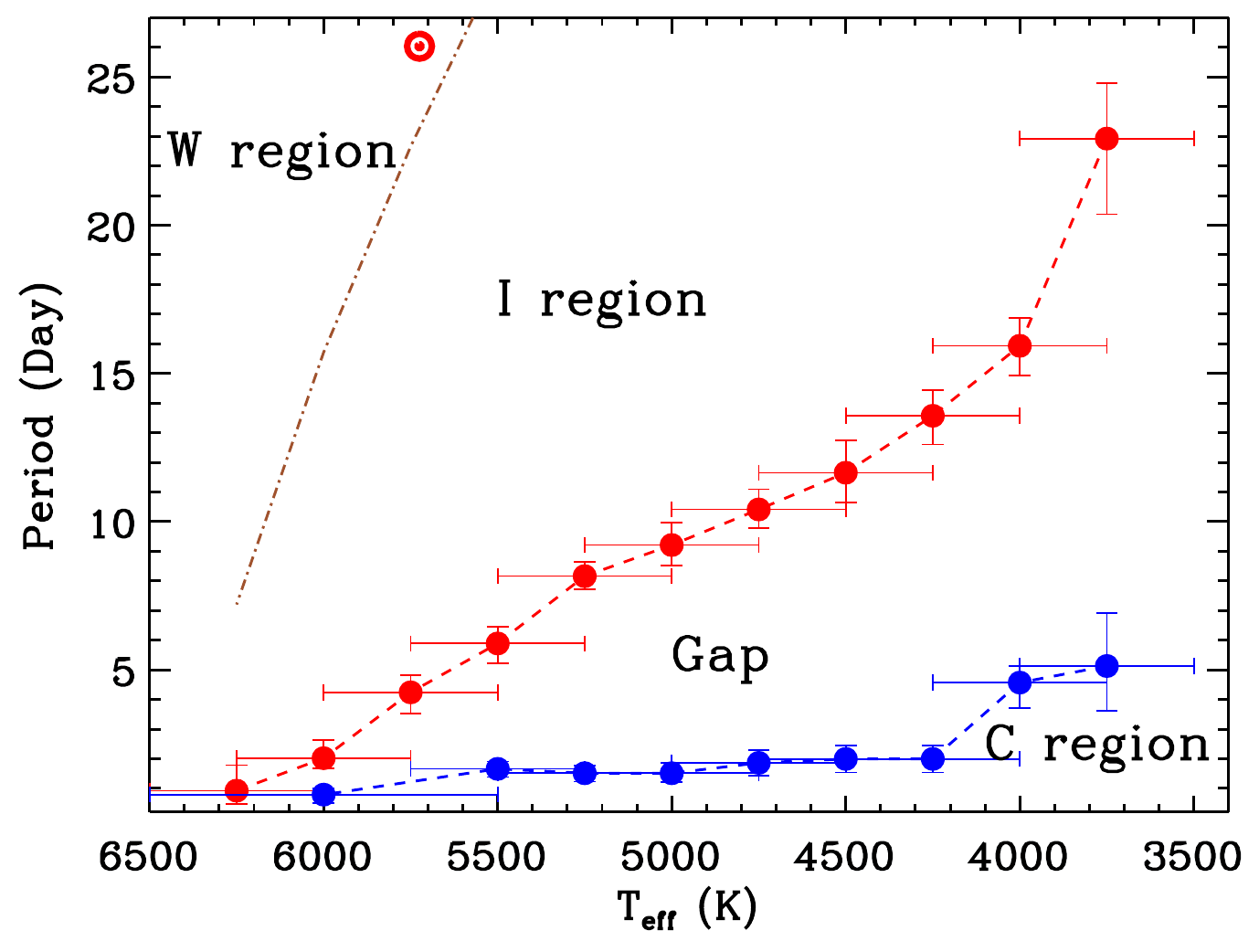}
\includegraphics[width=0.5\textwidth]{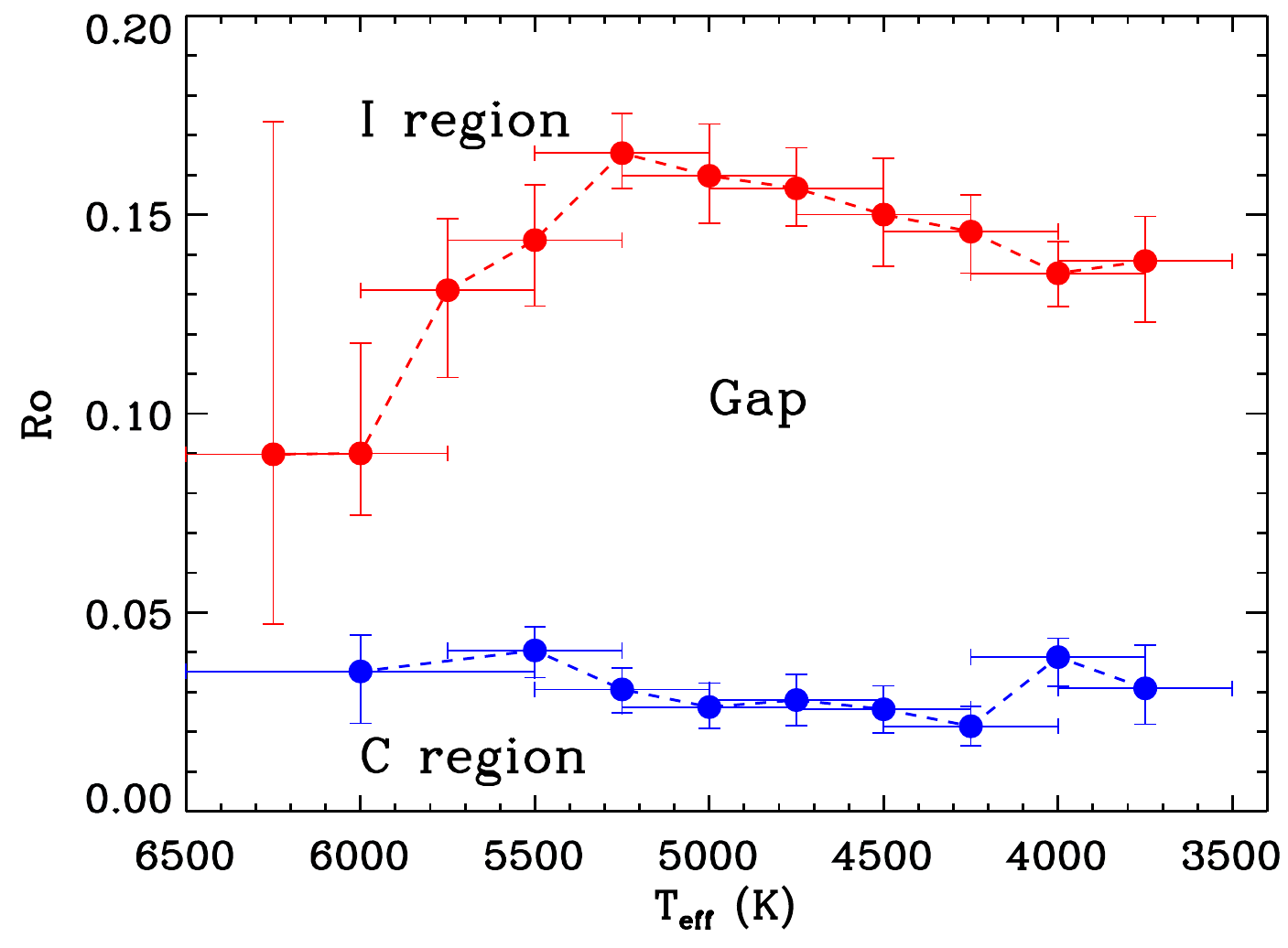}
\caption{Four regions in the color--period diagram. The red dashed lines denote the critical periods or Rossby number that separate the gap and I region in each temperature bin (the ZAIS line). The blue dashed lines denote the critical periods or Rossby number that separate the C and gap region in each temperature bin. The brown dashed line denotes the critical periods that separate the I region and the W region. The Sun is marked with an $\odot$ symbol.}
\label{fig_isp_r}
\end{figure*}

\begin{figure}
\includegraphics[width=0.5\textwidth]{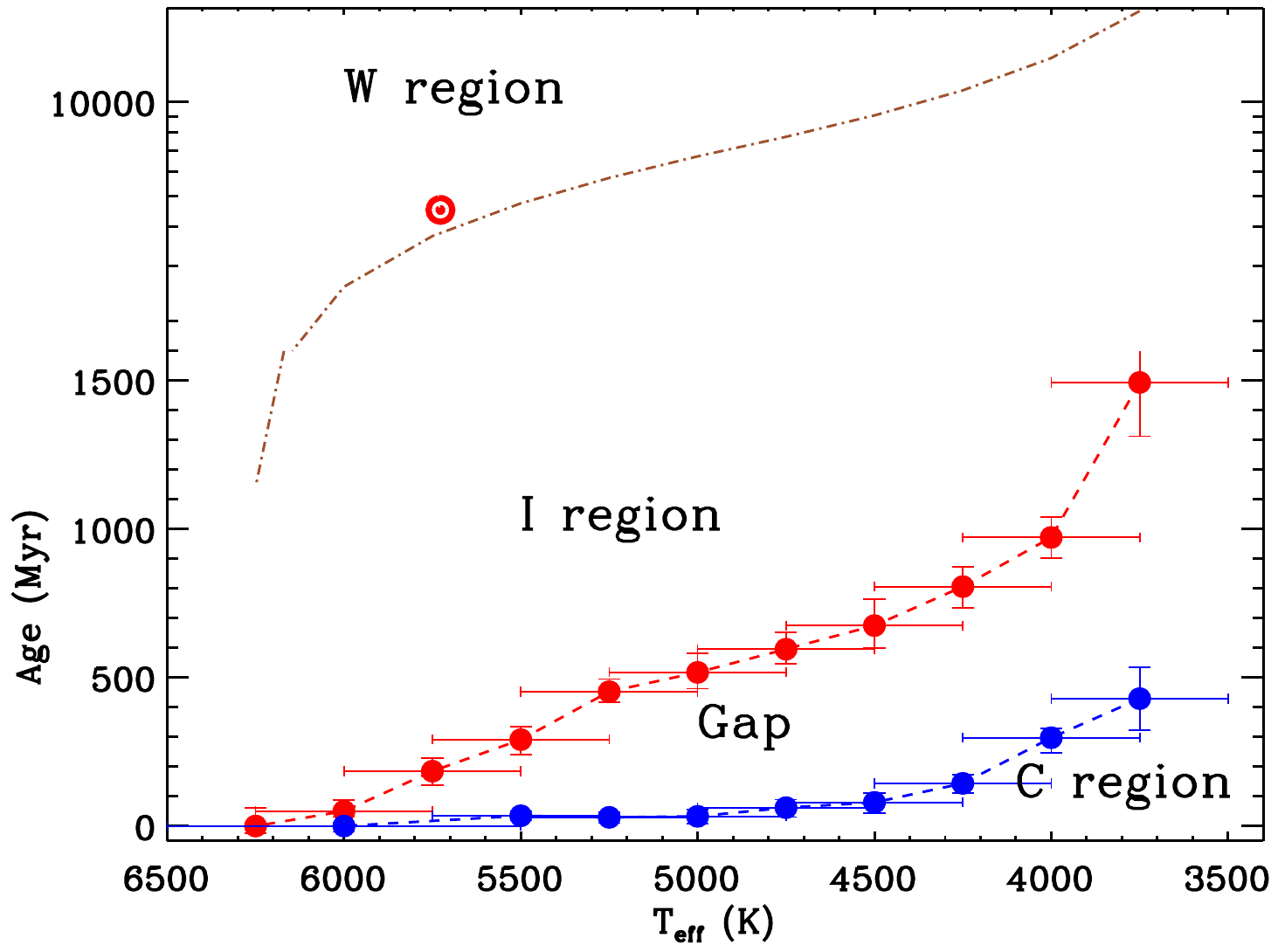}
\caption{Same as Fig.~\ref{fig_isp_r} but for the stellar age.}
\label{fig_isp_age}
\end{figure}
\section{The comparison with the color--period diagrams in open clusters}\label{sec5}
As the new rotation--activity relationship is from the framework of the CgIW scenario, it is important to compare the critical points of the transitions with those of the color--period diagrams in open clusters to validate the evolution phases. Meanwhile, the rotation--activity relationship, as a counterpart of gyrochronology, can be an independent proxy to constrain the evolutionary phase, given that gyrochronology could also be biased by other factors such as metallicity. For example, the current rotation--age relationship is based on establishing several benchmarks through open clusters, in which stars have the same metallicity. However, model prediction and observation from the Kepler field show that stars with higher metallicities have slower rotations \citep{Amard2020,Amard2020b,See2024}. Although the ages of those stars are not as precise as open clusters and there is an outlier near the solar age \citep{Karoff2018}, it indicates that the metallicity could have an impact on stellar spin-down that is not negligible. 

\subsection{The new rotation--relationship along with effective temperature}

\begin{table*}
\centering
\caption{\label{table_cluster} Information of clusters used in this study and results of the comparison.}

\begin{tabular}{cccccccccccc}
\hline\hline

Open Cluster&Age&Rotation Period&$T_{\rm eff,ZAIS}$&$P_{\rm ZAIS}$&$T_{\rm eff,CI}$&$P_{\rm CI}$\\[1ex]
&(Myr)&Reference&(K)&(day)&(K)&(day)\\[1ex]
\hline\\

$\rho$ Oph.$\&$Upper Sco.& $5\pm 3$ &\citet{Rebull2018}&--&--&--&--&\\[1ex]
$\alpha$ Persei&$85 \pm 20$&\citet{Boyle2023}&$5900^{+250}_{-250}$&$2.91^{+0.60}_{-0.49}$& 5940&$2.93^{+1.18}_{-1.18}$\\[1ex]
Pleiades&$125 \pm 30$&\citet{Rebull2016}& $5940^{+250}_{-250}$&$2.55^{+0.61}_{-0.44}$& 5741& $3.04^{+0.89}_{-0.89}$ \\[1ex]
M34&$220\pm 30$& \citet{Meibom2011}&$5650^{+250}_{-250}$&$4.90^{+0.58}_{-0.70}$&5485& $5.43^{+0.53}_{-0.53}$\\[1ex]
NGC 3532&$300\pm 50$&\citet{Fritz2021}& $5450^{+250}_{-250}$ &$6.34^{+0.55}_{-0.63}$ &5286& $7.15^{+0.74}_{-0.74}$\\[1ex]
NGC 2548 (M48)& $450\pm 50$&\citet{Healy2023}&$5300^{+250}_{-250}$&$7.71^{+0.51}_{-0.49}$& 4936& $9.06^{+2.88}_{-2.88}$\\[1ex]
Praesepe &$700 \pm 50$&\citet{Douglas2019}&$4500^{+250}_{-250}$&$11.65^{+1.10}_{-1.01}$& 4528& $12.03^{+0.86}_{-0.86}$\\[1ex]
Hyades&$750 \pm 50$&\citet{Douglas2019}& $4450^{+250}_{-250}$&$12.03^{+1.05}_{-1.00}$& 4481& $11.44^{+0.44}_{-0.44}$\\[1ex]
NGC 6811&$1000 \pm 100$&\citet{Curtis2019}& $4350^{+250}_{-250}$&$12.80^{+0.96}_{-0.99}$& 4341& $12.97^{+1.52}_{-1.52}$\\[1ex]
Ruprecht 147$\&$NGC 6819&$2500\pm200$&\citet{Curtis2020}& $3900^{+250}_{-250}$&$18.73^{+2.55}_{-2.55}$&3840&$21.83^{+1.98}_{-1.98}$\\[1ex]


\hline

\hline

\end{tabular}
\tablefoot{ The parameters $T_{\rm eff,ZAIS}$ and $P_{\rm ZAIS}$ represents the effective temperature and rotation period of the crossing point between the slowly rotating sequence and the ZAIS line, respectively (the red five-point stars in Fig.~\ref{fig_comp_cluster1} and Fig.~\ref{fig_comp_cluster2}). The parameters $T_{\rm eff,CI}$ and $P_{\rm CI}$ represents the effective temperature and rotation period of the convergent points between the slowly rotating sequence and the semi-empirical evolution tracks of the C sequence of gyrochronology, respectively (the green five-point stars in Fig.~\ref{fig_comp_cluster1} and Fig.~\ref{fig_comp_cluster2}). The rotation periods of $\alpha$ Persei are from the table of \citet{Boyle2023} whose quality flag ``Manual = g". The rotation periods of Ruprecht 147 are from the table of \citet{Curtis2020} whose quality flag `` benchmark = Yes or Yes-Rapid Outlier". }
\end{table*}

\begin{figure*}
\includegraphics[width=0.5\textwidth]{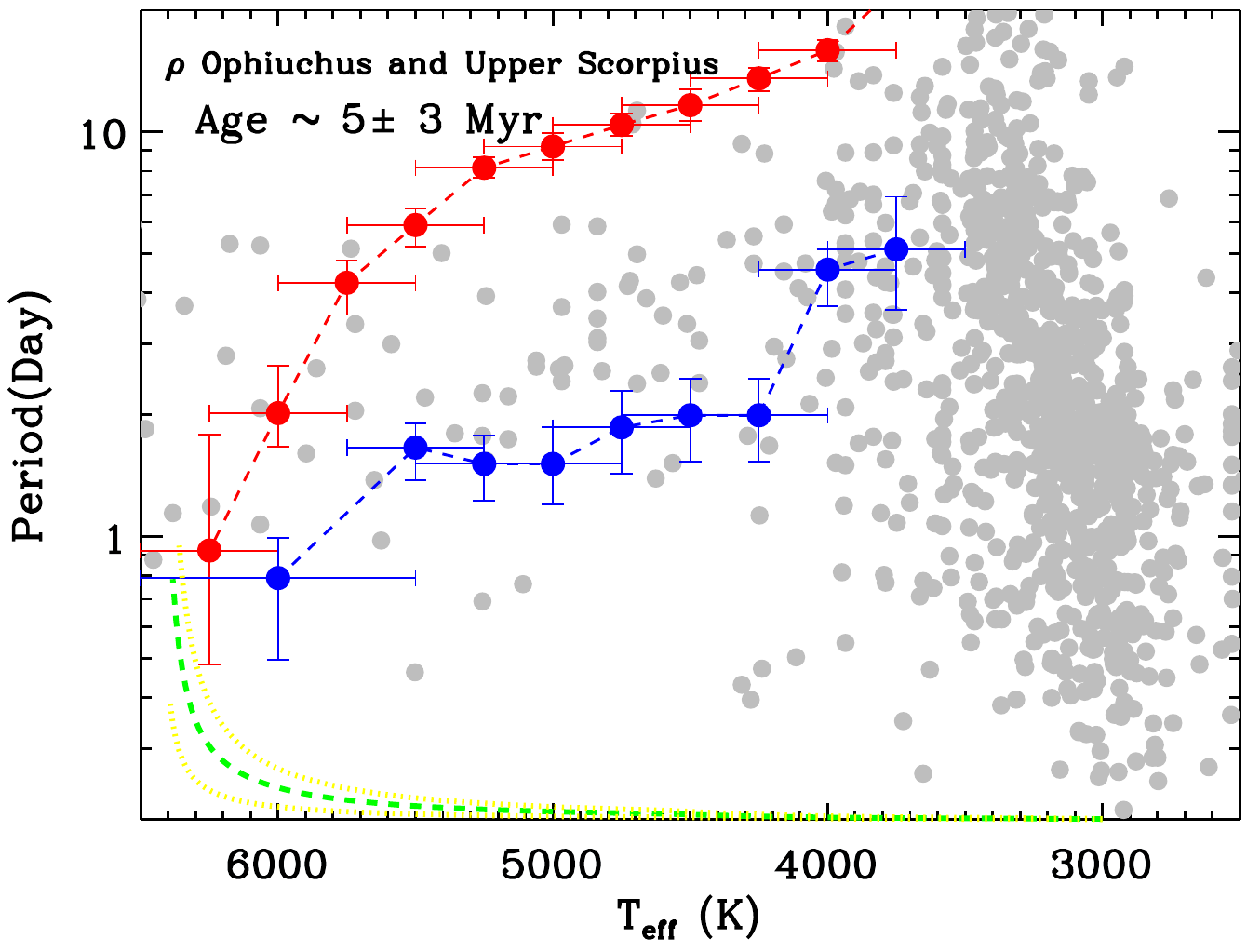}
\includegraphics[width=0.5\textwidth]{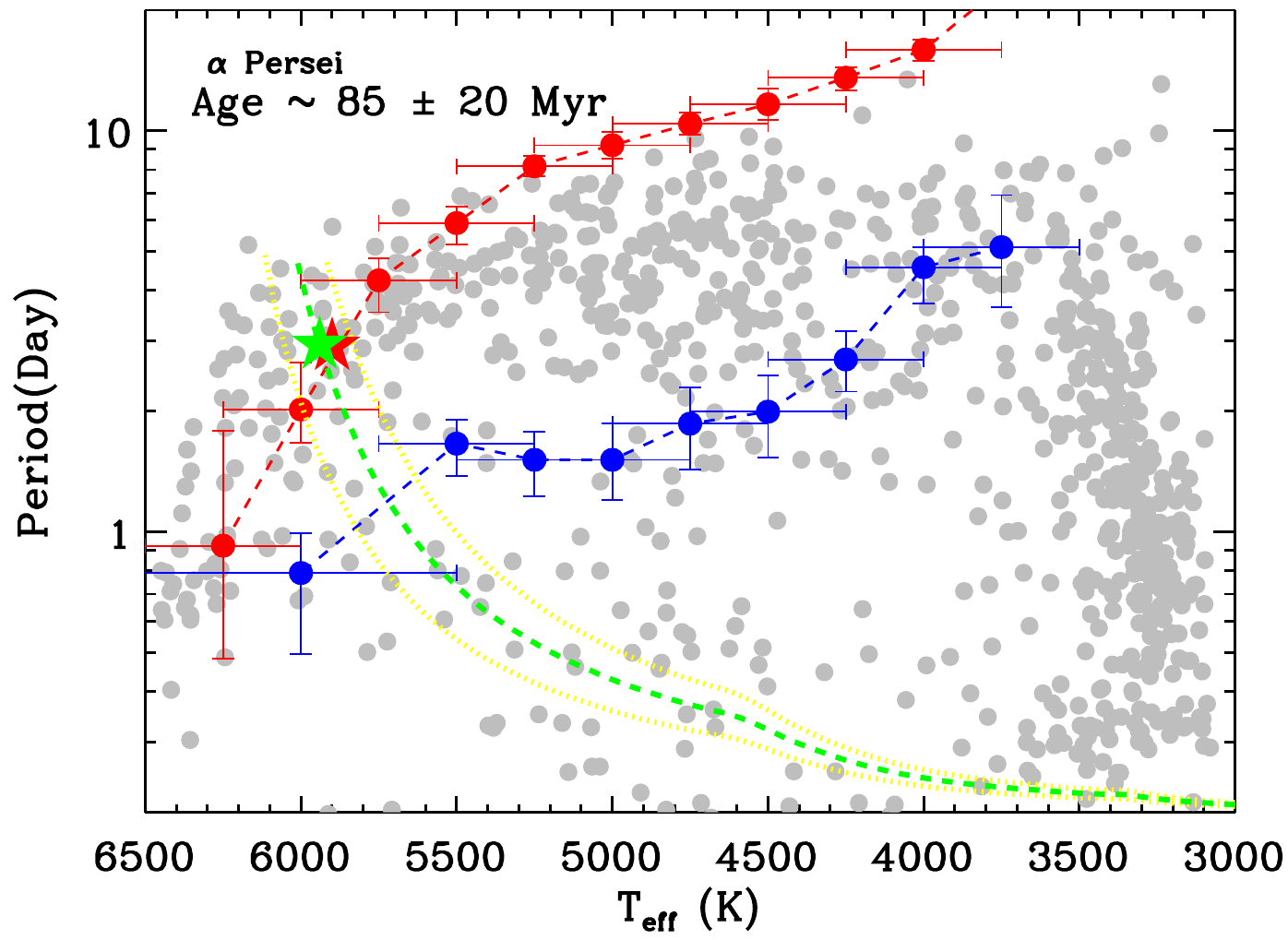}
\includegraphics[width=0.5\textwidth]{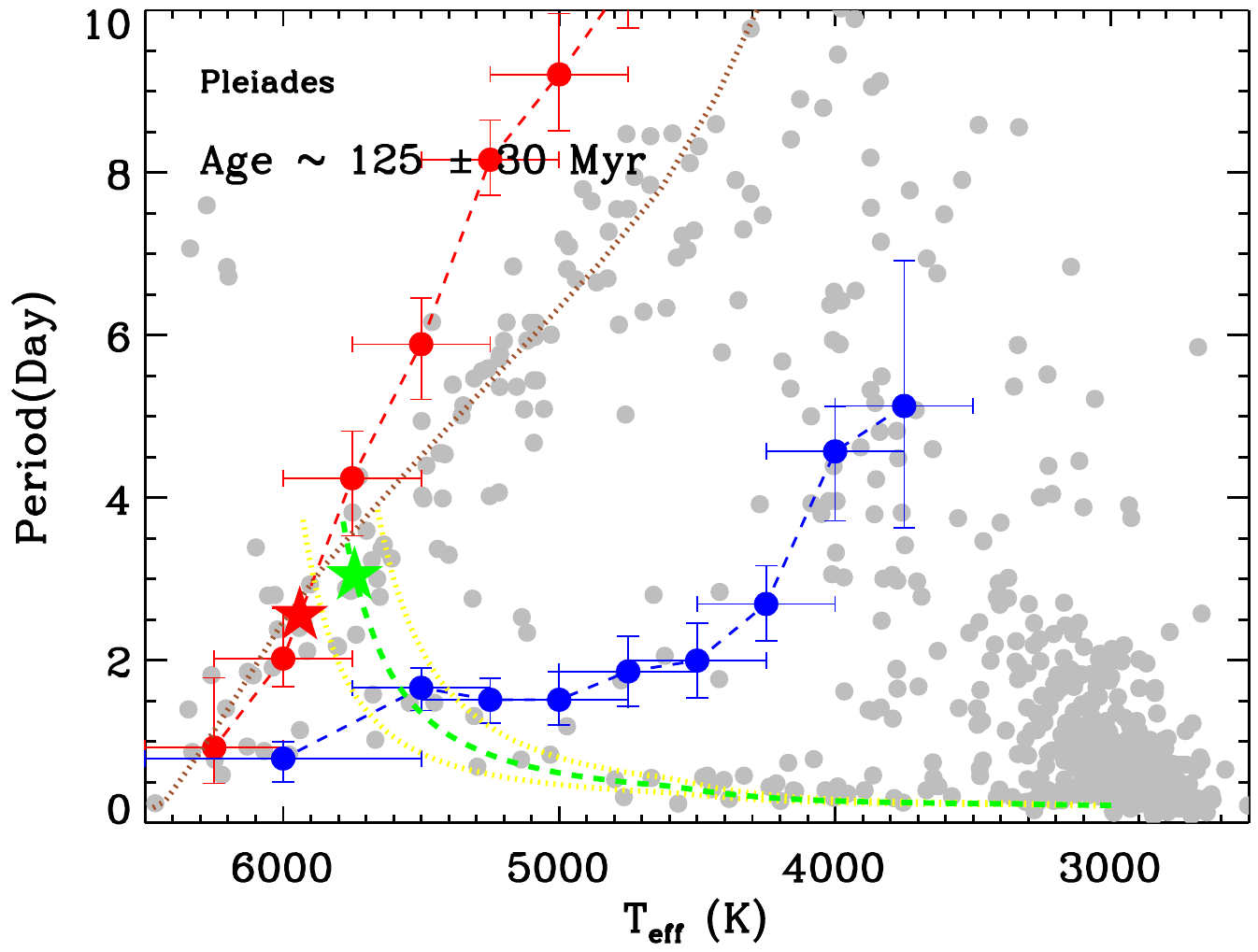}
\includegraphics[width=0.5\textwidth]{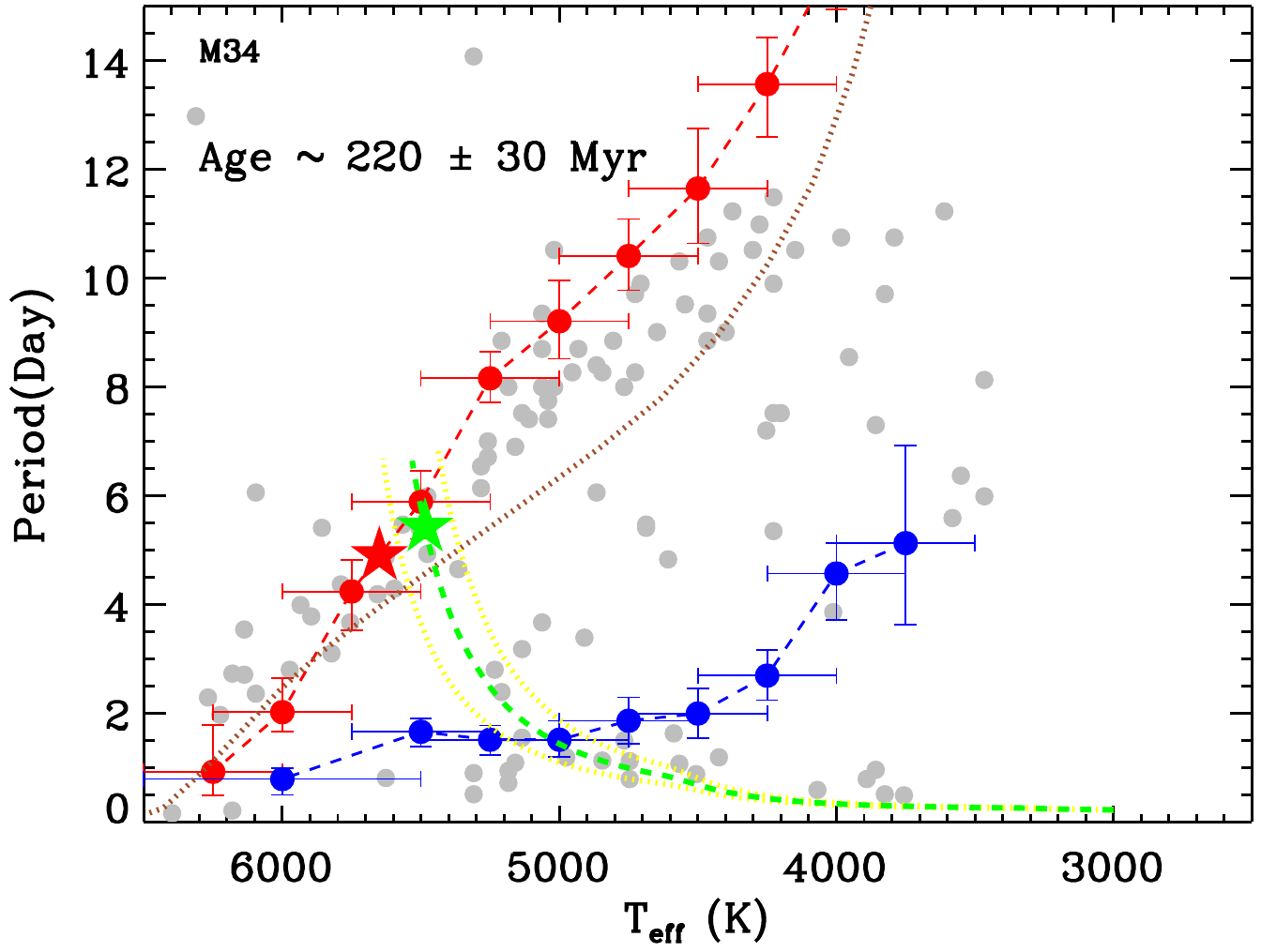}
\includegraphics[width=0.5\textwidth]{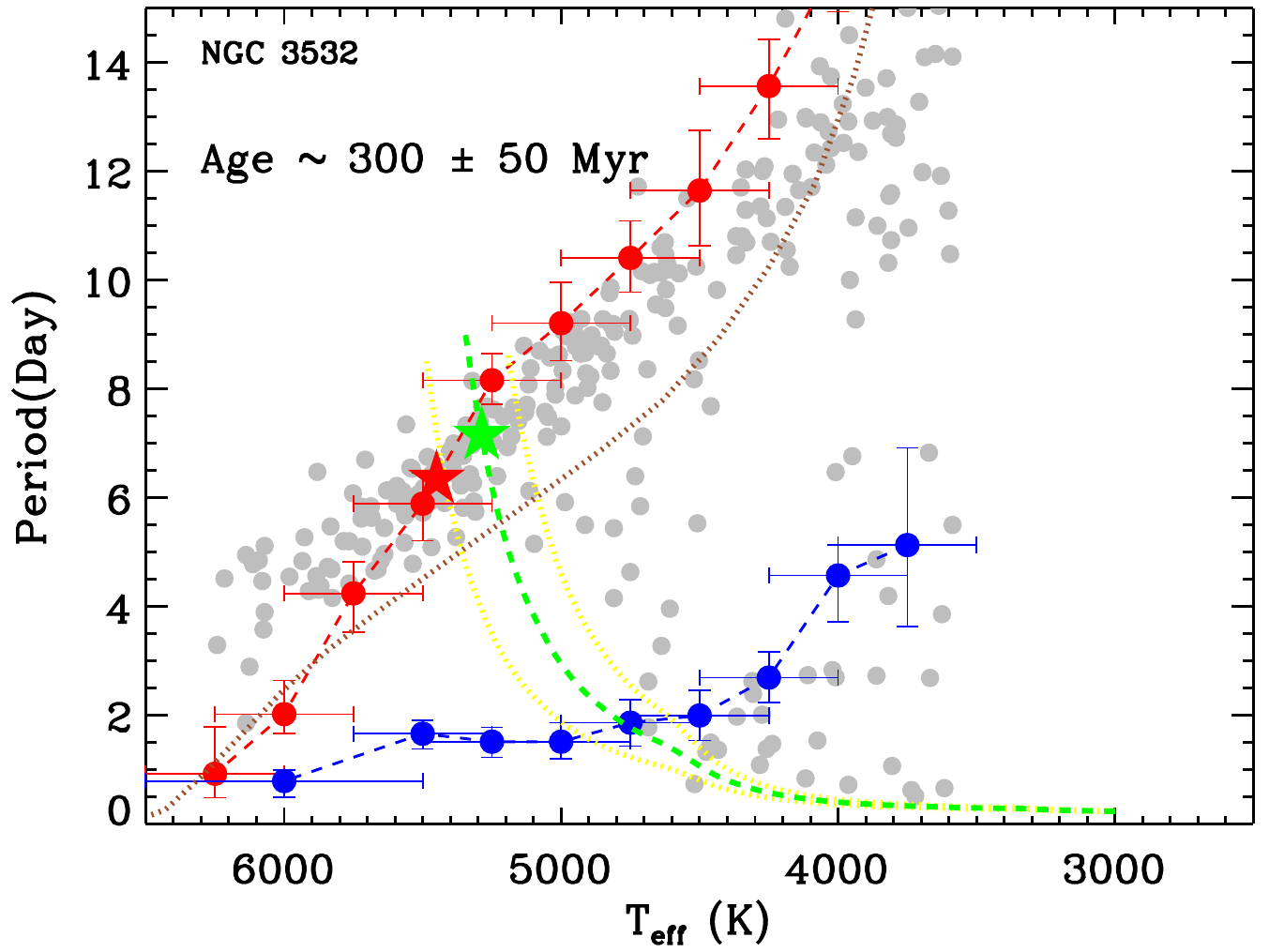}
\includegraphics[width=0.5\textwidth]{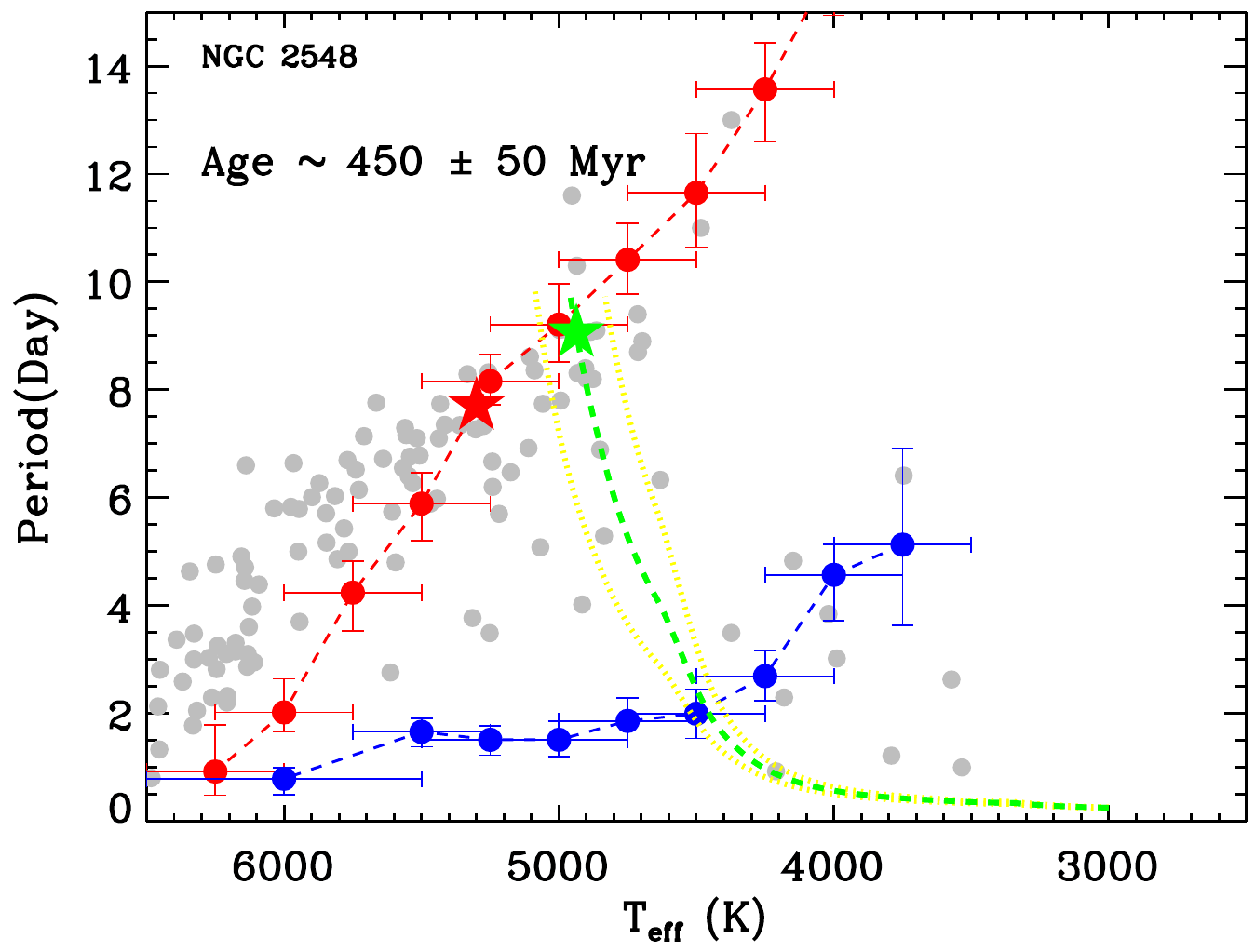}
\caption{Temperature--period diagrams for six open clusters. We have superimposed the C-to-g transition lines (the blue dashed lines), the g-to-I transition lines (the red dashed lines; the ZAIS lines), and the C-to-I evolution tracks of gyrochronology (the green dashed lines). The crossing points between the ZAIS lines and the slowly rotating sequence are marked with the red five-point stars. The convergent points between the C-to-I evolution tracks and the slowly rotating sequence are marked with the green five-point stars. The yellow dashed lines denote the upper and lower limit of the C-to-I evolution tracks from the uncertainty of the age in each panel. The brown dotted lines in Pleiades, M34 and NGC 3532 represent the lower envelope of the visually convergent sequence (Ro=0.11, also see Fig.~\ref{fig_teff_period}) to illustrate the pre-I sequence stars. Some references do not have the effective temperatures for their rotating stars. We convert their intrinsic color index ($\rho$ Ophiuchus and Upper Scorpius:$(V-K)_0$;  M34 and NGC 3532:$(B-V)_0$) to the effective temperature by using the tabulated data of \citet{Barnes2010b}. We convert the $(G_{\rm BP}-G_{\rm RP})_0$ of NGC 2548 to the effective temperature by using the relation of \citet{Mucci2021}, in which the $(G_{\rm BP}-G_{\rm RP})_0$ is obtained by distance and extinction (see Appendix~\ref{appendix_lum}).}
\label{fig_comp_cluster1}
\end{figure*}

\begin{figure*}
\includegraphics[width=0.5\textwidth]{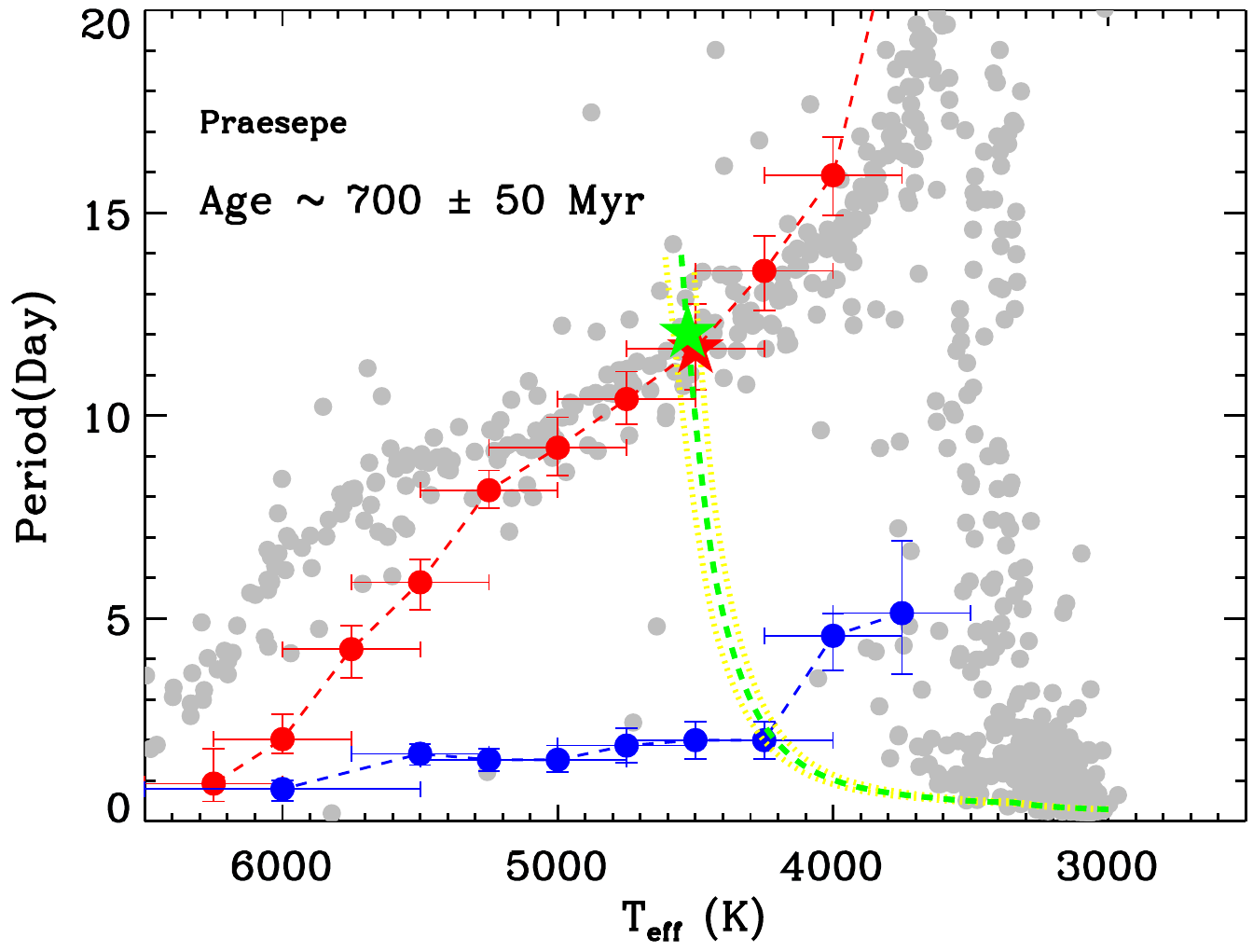}
\includegraphics[width=0.5\textwidth]{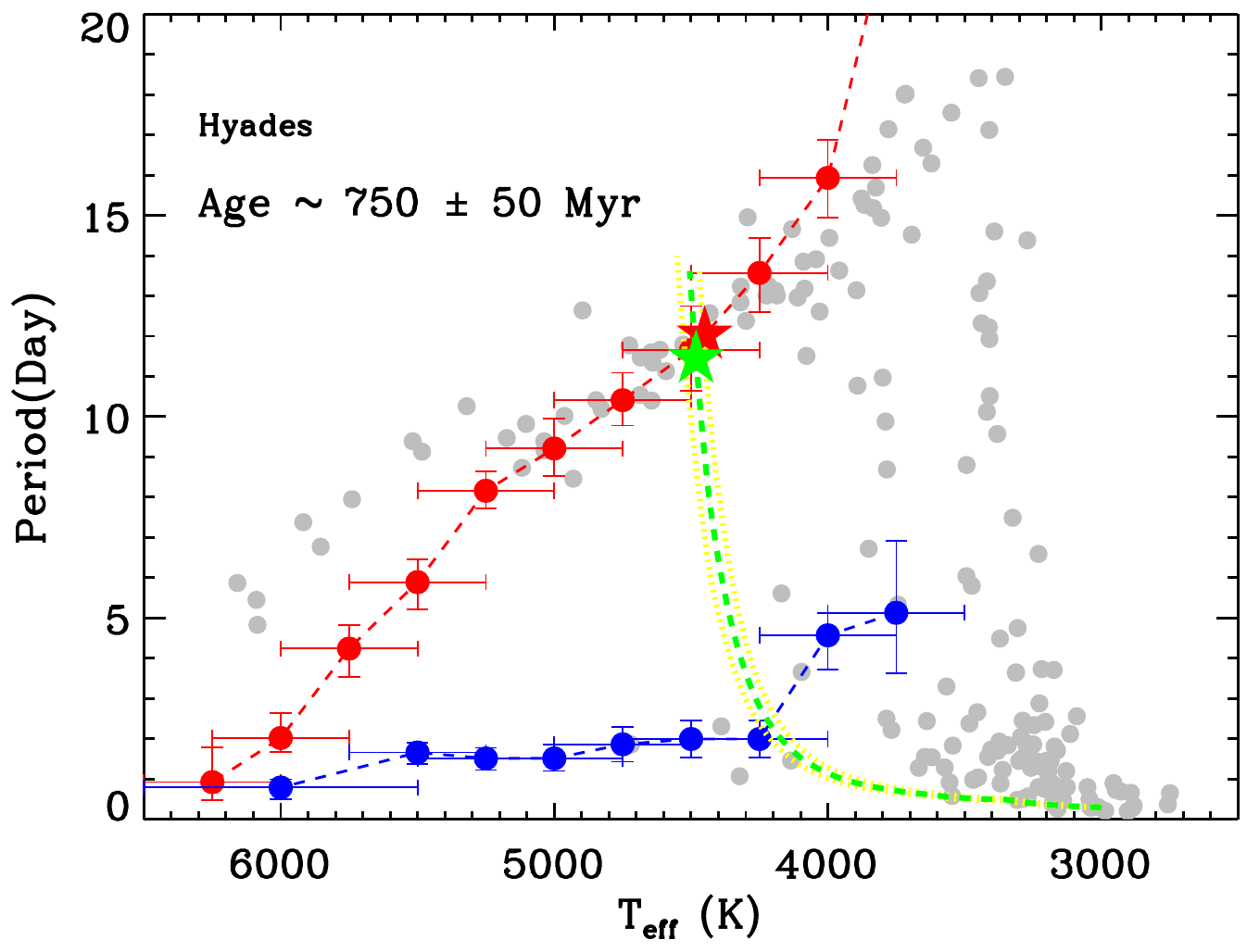}
\includegraphics[width=0.5\textwidth]{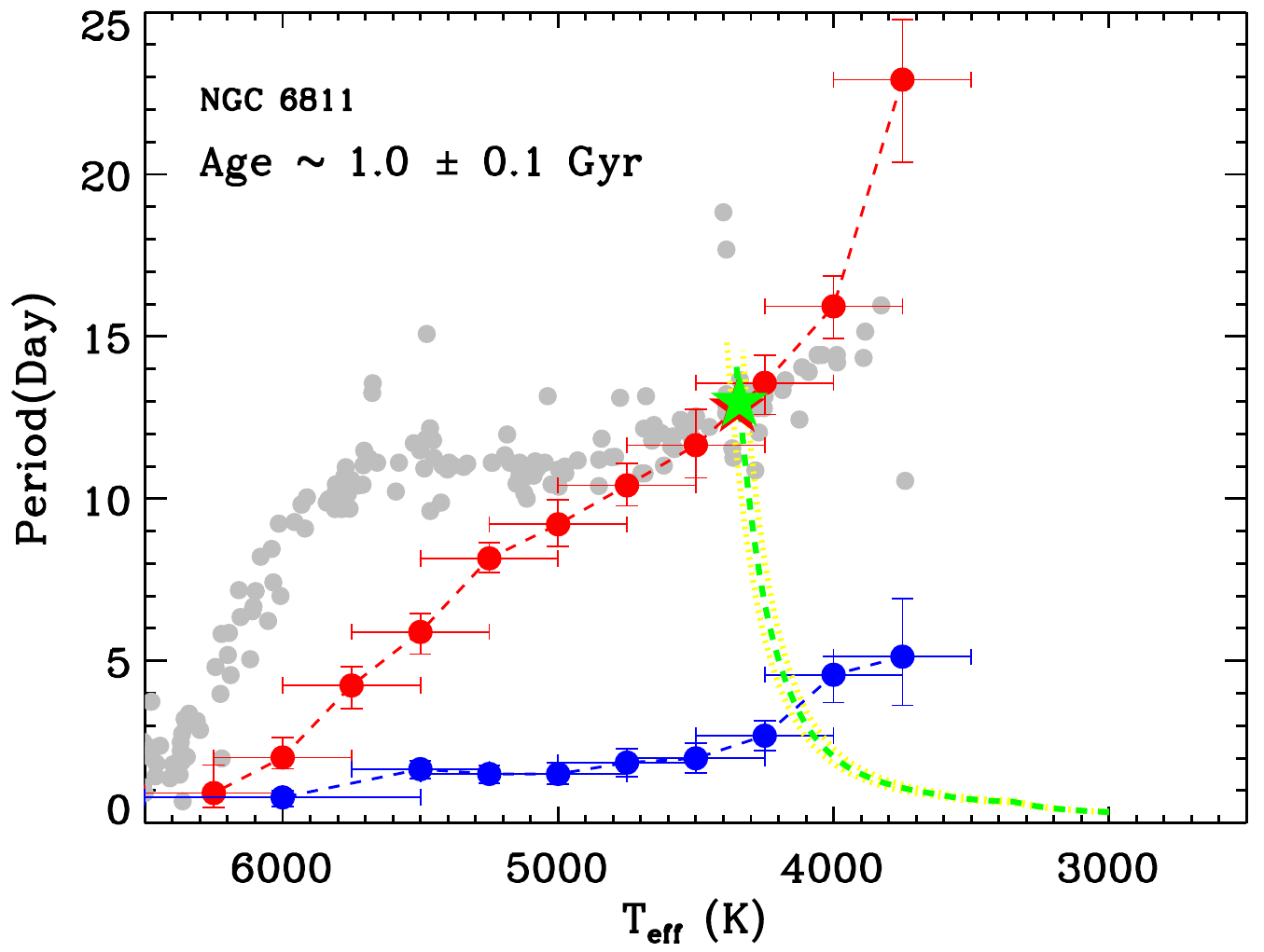}
\includegraphics[width=0.5\textwidth]{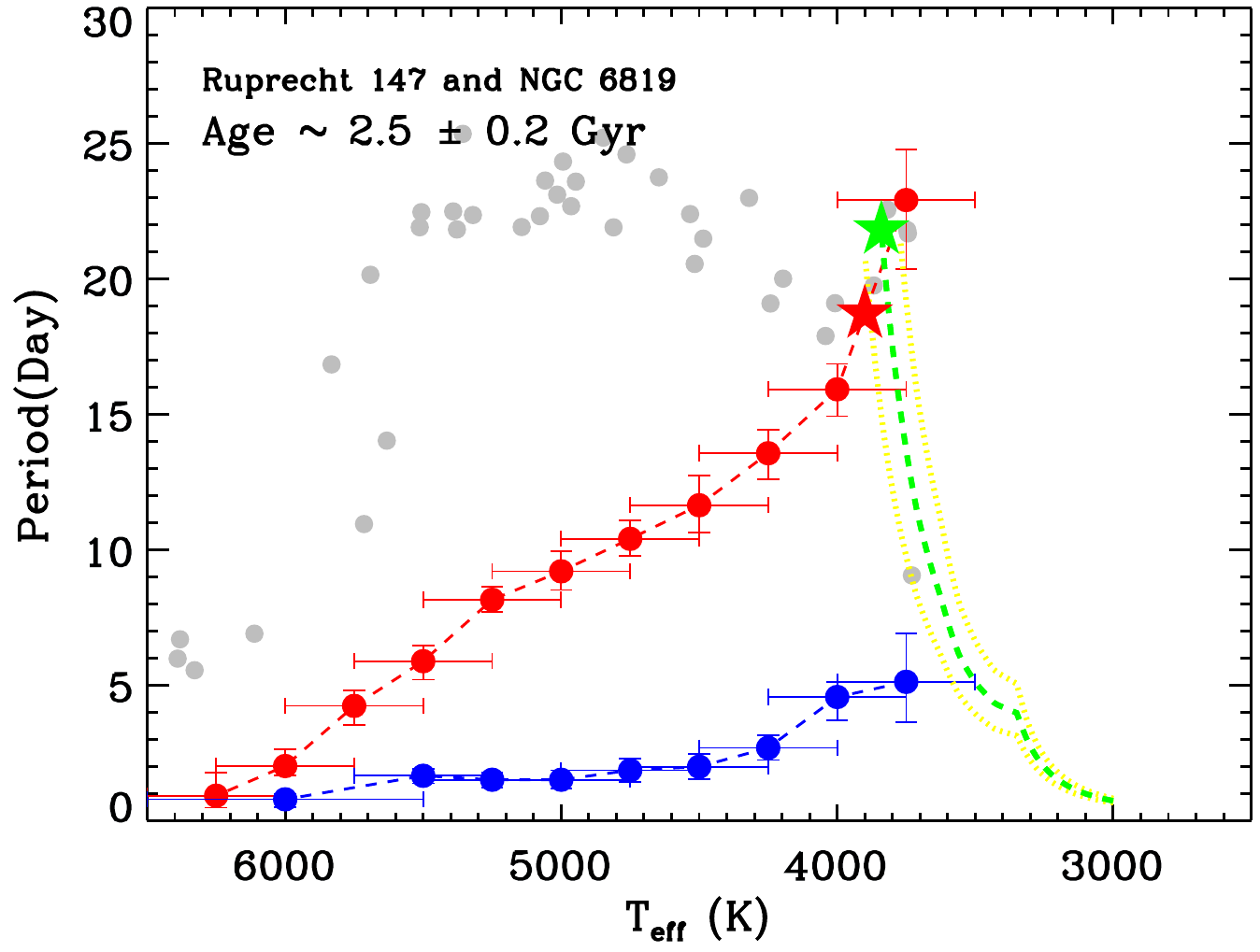}
\caption{Same as Fig.~\ref{fig_comp_cluster1} but for four older open clusters.}
\label{fig_comp_cluster2}
\end{figure*}

\begin{figure*}
\includegraphics[width=0.5\textwidth]{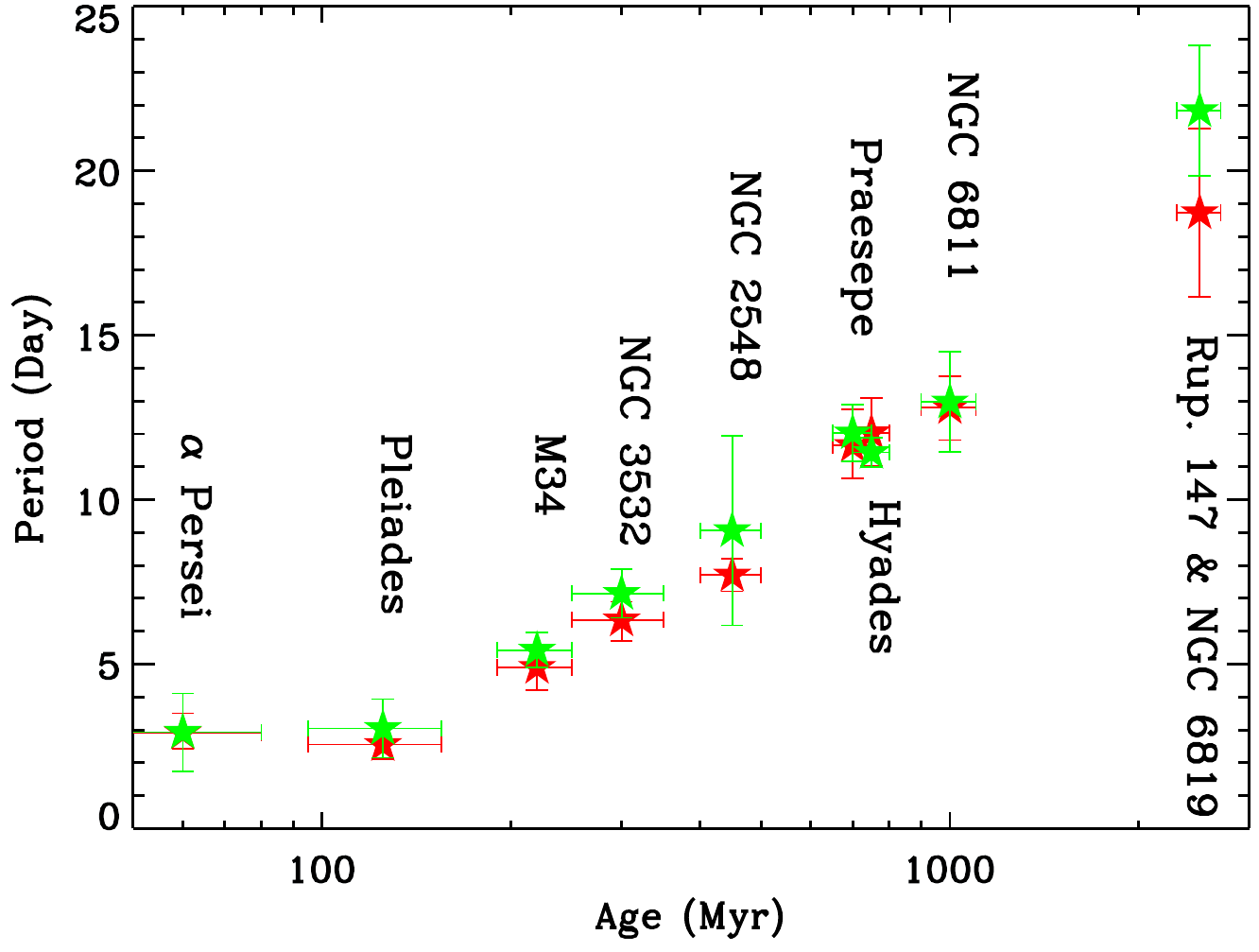}
\includegraphics[width=0.5\textwidth]{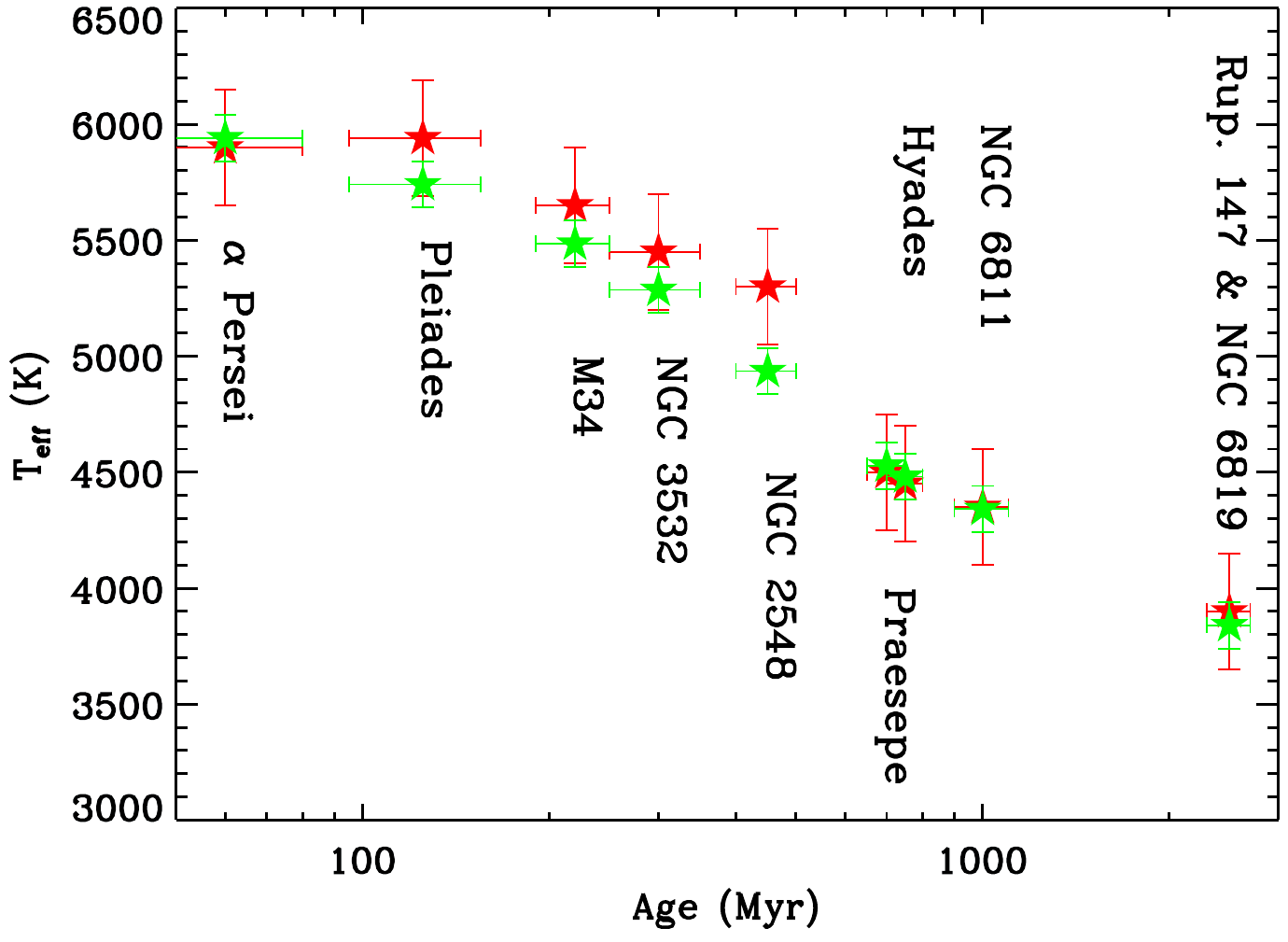}

\caption{Comparison between the crossing points from the ZAIS line and the convergent points from the evolution tracks in rotation period (left panel) and effective temperature (right panel). The mean difference of the rotation period is $\sim$ 0.39 days, and the mean difference of the effective temperature is $\sim$ 95 K. }
\label{fig_comp_cp}
\end{figure*}

 The color--period diagram shows that the duration from the C sequence to the I sequence depends on stellar mass \citep{Barnes2003b,Barnes2007}. Specifically, a larger stellar mass will have a shorter duration evolving onto the I sequence, and even F-type stars are born I sequence stars. Fig.~\ref{fig_cls} demonstrates this trend. The start point of each colorful line is the first point that Eq.~\ref{eq_age} is larger than 0. It indicates that the age and duration of Rossby number that a star enters the I sequence decreases along with increasing stellar mass until F-type stars are born I sequence stars. The physical understanding of this trend could be that a higher temperature corresponds to a thinner convective envelope, resulting in a shorter duration of the re-coupling process \citep{Spada2020}.

We show that the morphology of this trend can also be validated in the rotation--activity relationship. We repeated our fitting (with the same functional form) for groups of stars in separate temperature bins (see Fig.~\ref{fig_ro_act_teffbin1},Fig.~\ref{fig_ro_act_teffbin2} and Fig.~\ref{fig_ro_act_teffbin3}). Based on the motivation to compare with color--period diagram, we carry out the fitting by rotation period instead of Rossby number. The fitting results are listed in Table~\ref{table_pfit}. 

Fig.~\ref{fig_isp_r} and Fig.~\ref{fig_isp_age} illustrate the four regions in the temperature--period, temperature--Ro and temperature--age diagrams, whose data are from Table~\ref{table_pfit}.  The critical periods and ages of the g-to-I transition are larger than the traditional knowledge on the convergence of the slowly rotating sequence which is the lower envelope in Fig.\ref{fig_teff_period} (also see \citet{Curtis2020}). We argue that this is due to the ambiguous consensus on when gyrochronology starts to work \citep[e.g.,][]{Boyle2023} and how to define the I sequence. For example, in the temperature--period diagram of the open cluster $\alpha$ Persei (the top right panel of Fig.~\ref{fig_comp_cluster1}), a portion of stars cooler than the solar temperature tends to converge and form a slowly rotating sequence. However, there are still a considerable number of scattered stars, indicating that the initial rotation period still has a great impact on the rotation distribution. This phenomenon reaches its peak for late G-type and K-type stars from Pleiades to NGC 3532 (Fig.~\ref{fig_comp_cluster1}). We stress that the definition of the I sequence should be that all stars in a temperature bin have erased the imprint of the initial period, which is in line with the original intention of gyrochronology. Therefore, we conclude that those stars of $\alpha$ Persei cooler than the solar temperature have not completely evolved into the I phase forming a pure I sequence. In this sense, the critical points in the rotation--activity relationship defines the zero-age I sequence (ZAIS) that the vast majority of stars in a given temperature bin should be on the slowly rotating sequence (i.e., few fast rotating outliers are in the temperature bin).

We also argue that those ``visually" convergent stars are in a ``pre-I sequence" phase when the wind braking and rotational coupling are in a close contest \citep{Spada2020}, resulting in the decline or ``stall" of the spin-down \citep{Agueros2018,Curtis2019} (See the next subsection for the detailed discussion on the stall). In Pleiades, M34 and NGC 3532 of Fig.~\ref{fig_comp_cluster1}, the areas between the ZAIS lines (red dashed lines) and the lower envelope of the visually convergent sequence (Ro=0.11, brown dotted line, also see Fig.~\ref{fig_teff_period}) illustrate the region of the pre-I sequence. Although a data-based method may obtain an empirical rotation--age relationship for stars in this region, the uncertainty should be much larger than that of the ZAIS stars because of the fast rotating outliers that cannot be negligible and the stall of the spin-down.

\subsection{The comparison with the slowly rotating sequence in open clusters and the semi-empirical relation of gyrochronology}\label{sec_comp_cg}

 The red dashed line of Fig.~\ref{fig_isp_r} represents the ZAIS line based on the new rotation--activity relationship. It is important to validate it by comparing it with the slowly rotating sequences in open clusters and the semi-empirical relation of gyrochronology. We collect periods of 10 clusters (Table~\ref{table_cluster}) whose ages range from 1 Myr to 2.5 Gyr. Since the critical periods of the ZAIS increase with decreasing temperature, a crossing point between the ZAIS line and the slowly rotating sequence indicates the specific temperature and period of the ZAIS at the age of a cluster (red five-point stars in Fig.~\ref{fig_comp_cluster1} and Fig.~\ref{fig_comp_cluster2}, or $T_{\rm eff,ZAIS}$ and $P_{\rm ZAIS}$ in Table~\ref{table_cluster}). We excluded stars on the C sequence (fast rotating stars lower than the blue dashed lines in Fig.~\ref{fig_comp_cluster1} and Fig.~\ref{fig_comp_cluster2}), and compared the median period of each temperature bin with the ZAIS line. We take the closest point as the crossing point in each cluster. On the other hand, gyrochronology charts the roadmap of the transition from the C sequence to the I sequence. \citet{Barnes2003b} defines the rotational evolution track of the C sequence as

\begin{equation}
P=0.2 \times e^{t_{\rm age}/T(B-V)},
\label{eq_cp}
\end{equation}
where $T(B-V)$ is the time scale that a star evolve from the C sequence onto the I sequence \citep{Meibom2009}. We follow the choices of \citet{Barnes2010b} by adopting $T(B-V)$ = 0 Myr for $B-V$=0.47, $T(B-V)$ = 60 Myr for $B-V$=0.7, $T(B-V)$ = 140 Myr for $B-V$=1.05 and $T(B-V)$ = 500 Myr for $B-V$=1.40. \citet{Barnes2003b} pointed out that the convergent point between the evolution track and the slowly rotating sequence also indicate the temperature and period of the C-to-I transition (green five-point stars in Fig.~\ref{fig_comp_cluster1} and Fig.~\ref{fig_comp_cluster2}, or $T_{\rm eff,CI}$ and $P_{\rm CI}$ in Table~\ref{table_cluster}). We obtain the convergent point in each cluster by finding the closest point between the evolution track (the green dashed line in Fig.~\ref{fig_comp_cluster1} and Fig.~\ref{fig_comp_cluster2} ) and the median period of each temperature bin. 

It should be noted that $T(B-V)$ is smaller than what we find in Table~\ref{table_cluster} because: (1) The benchmarks they used are very limited which could result in large uncertainties on $B-V$. For example, the range of $B-V$ for $T(B-V)$ = 60 Myr is $0.6< B-V <0.8$ and the range of $B-V$ for $T(B-V)$ = 140 Myr is $0.8< B-V <1.3$. (2) The ages of clusters that they took as benchmarks are younger than that of the current knowledge. For example, the ages of $\alpha$ Persei, Pleiades and Hyades they adopted are $\sim $50, 100 and 600 Myr, respectively. By contrast, recent measurements on their ages are $\sim $ 85 \citep{Boyle2023}, 125 \citep{Rebull2016} and 750 \citep{Douglas2019} Myr, respectively, which are at least 25$\%$ older than them. However, we still use those evolution tracks and the corresponding convergent points for comparison because: (1) The influence of the discrepancy in $T(B-V)$ is limited in mathematics, especially for relatively old clusters as the influence decreases with increasing cluster ages. (2) Another important application of those tracks is to illustrate the lower envelope of fast rotating stars below which there should not be fast rotating outliers. The comparison could validate our results on the temperature and period of the ZAIS.  

Figure~\ref{fig_comp_cp} shows the comparison of the effective temperature and period of the ZAIS between our method and the semi-empirical relation of gyrochronology. We note that there is a systematical offset before NGC 2548 ($t_{\rm age} \sim 450$ Myr), which is due to the underestimate of $T(B-V)$ as mentioned above. For example, since gyrochronology has adopted several younger ages of clusters for the benchmark, it will result in the evolution tracks systematically shifting towards the lower temperature. This phenomenon can also be validated by Fig.~\ref{fig_comp_cluster1}, where the evolution tracks are systematically higher than the lower envelope of the fast rotating stars. By contrast, the ZAIS points derived by our rotation--activity relationship are near the edge of the lower envelopes, ensuring a pure I sequence at their temperature bins. We note that the mathematical influence of $T(B-V)$ decreases in Eq.~\ref{eq_cp} as $t_{\rm age}$ increases. This results in the two methods being basically consistent after Praesepe ($t_{\rm age} \sim 700$ Myr).

The popularity of gyrochronology is due to the fact that it is a dating method for I sequence stars rather than its classification on the evolutionary stages \citep[e.g.,][]{Meibom2015}. However, the ZAIS along effective temperature is important because it determines the application range of the dating method, which is only valid within the I sequence. For example, recent studies on gyrochronology find that its age prediction on K- and M-type stars deviates from the observation from the age of 700 Myr to 1.3 Gyr and the deviation increases with decreasing stellar mass \citep{Agueros2018,Curtis2019,Douglas2019}. To explain this discrepancy, several authors \citep{Agueros2018,Curtis2019} proposed that spin-down of those stars ``stalls'' for several hundred Myrs. \citet{Spada2020} proposed that the physical understanding of the stall is that the wind braking slows down the surface rotation of a star while the core-envelope re-coupling process speeds up the surface rotation because of the angular momentum transport from the fast rotating core. The stall of the surface rotation results from the competing effect of the two mechanisms, when the speed-up mechanism cancels out the magnetic braking. We propose that the efficiency of the angular momentum transport should monotonically increase with the proceeding of the re-coupling process. The transport would be the most efficient when the re-coupling is almost complete. This indicates that the stall can be associated with the pre-I sequence. Fig.~\ref{fig_comp_cluster2} shows that the stall region is near the ZAIS line (Praespe, Hyades and NGC 6811), which is in line with the physical understanding.

As the time scale of the re-coupling process is mass dependent, the ``cancel out" scenario \citep{Spada2020} also predicted that the stall should occur at other ages in other mass range. Interestingly,  the ZAIS points in $\alpha$ Persei and Pleiades (from 85 Myr to 125 Myr) are nearly unchanged, and the ZAIS points in NGC 3532 and NGC 2548 (from 300 Myr to 450 Myr) only have a small variation, implying that they are the counterpart of the stall at high-mass ranges. Given that the time scale of the stall should rapidly decrease with increasing temperature, more clusters near those age ranges are necessary to verify the prediction.

\section{Conclusion}\label{sec6}
In this study, we have combined LAMOST spectra with the Kepler mission and two open clusters to obtain the chromospheric activity, $R'_{\rm HK}$, of 6846 stars and their rotation periods. We used $R'_{\rm HK}$ and the rotation period to investigate the rotation--activity relationship. The rotation--activity relationship and gyrochronology are from the ensemble of the activity, rotation, and age, and both are used to depict evolutionary stages. However, their close relation has barely been discussed, and the current mapping between the two paradigms is in contradiction, which could raise many issues in physics and mathematics. The issues include the following (see Sect.~\ref{issue} for details): (1) The transition from the C phase to the I phase is unlikely to occur instantaneously, as it involves complicated physical changes. Much evidence indicates that the transition point is within an independent phase. The intermediate period gap found in the Kepler mission also indicates that stars in the unsaturated regime may have different physical properties regarding stellar structure and dynamos, which cannot be depicted by one mechanism. (2) The value of the transition point Ro$_\text{sat}$ has not yet been well determined mathematically. Different activity proxies could result in a large uncertainty of Ro$_\text{sat}$, while Ro$_\text{sat}$ is supposed to be independent of the activity proxy. Even for the same activity proxy, different samples give totally different values of Ro$_\text{sat}$. (3) The current mapping of evolutionary stages between the paradigms shows that the I sequence begins at Ro = 0.05, while studies on young open clusters find that the C sequence stars are dominated at that time.

We have proposed a tentative scheme based on the CgIW scenario to redefine the evolutionary phases of the rotation--activity relationship. Instead of the traditional two intervals, the scheme includes four intervals, namely, the C phase, the gap, the I phase, and the W phase. 

We propose that the new model can address the issues of the dichotomy. (1) The model ensures a pure saturated regime (see Appendix~\ref{app_newra}), avoiding the remnant dependence. (2) It enables the gap to be an independent transition phase rather than a critical point, whose physical nature is the re-coupling process between the radiative core and the convective envelope. (3) The critical point of the g-to-I transition separates the unsaturated regime near the intermediate period gap (Fig.\ref{fig_teff_period}), which can be associated with the transition from being spot dominated to facula dominated. It also defines a pure I sequence that excludes fast rotating outliers, which can be validated in open clusters (Fig.~\ref{fig_comp_cluster1} and Fig.~\ref{fig_comp_cluster2}). (4) The remapping of our new scheme shows that the C sequence stars and I sequence stars are dominated at the C phase and I phase (Fig.~\ref{fig_cls}), respectively, which makes sense in physics. (5) The new model sets the transition from emission line stars to absorption line stars within the gap (Fig.~\ref{fig_ro_act_teffbin1} and Fig.~\ref{fig_ro_act_teffbin2} ). (6) The transition point of the dichotomy could vary within the gap, and the variation can explain the large uncertainty of Ro$_{\rm sat}$ of the dichotomy.

Our $R'_{\rm HK}$ has a large uncertainty in the low-activity level, so we cannot directly gauge the rotation--activity relationship at a high Rossby number. However, evidence from previous studies indicates that there is a change of activity dependence on rotation near Ro = 0.7, which is in line with the W phase. The I-to-W transition may lead the Sun to a level of higher activity and variability in the future, and it also can be associated with the temperature rise on the Earth 500 Myr ago, which resulted in the diversification and evolution rates of life significantly accelerating.   

We further tested our scheme using a different activity proxy: the X-ray luminosity caused by the corona emission. We found the rotation--activity relationship can also be well classified by the four phases, which has a small improvement in reducing the scatter. However, compared with the re-kindling of chromospheric activity, the coronal activity has a significant drop in the W phase. We suggest that the weakening of the open magnetic field lines and large-scale magnetic field is responsible for the magnetic braking and corona activity, while the chromospheric activity is likely to be more strongly correlated with small-scale fields, as it is formed much closer to the photosphere. The W phase may experience a redistribution of magnetic field when the magnetic energy is concentrated in small-scale fields.

We repeated our fitting for groups of stars in separate temperature bins and found that our new scheme is mass dependent. Specifically, a larger stellar mass will have a shorter duration evolving onto the I sequence, and finally the transition points of C-to-g and g-to-I will converge at F-type stars, resulting in F-type stars being born as I sequence stars. This is consistent with gyrochronology, which has shown that the tracks of the C sequence and I sequence converge at F-type stars in the color--period diagram. 

However, the critical periods and ages of the g-to-I transition derived from our new rotation--activity relationship are larger than the traditional knowledge on the convergence of the slowly rotating sequence. We argue that this is due to the ambiguous consensus on when gyrochronology starts to work and on how to define the I sequence. Our findings define the pure I sequence, which we denote as the ZAIS. We compared our ZAIS line with the semi-empirical evolution tracks of gyrochronology in open clusters and found that our ZAIS line is more accurate in excluding fast rotating outliers and defining the pure I sequence. Moreover, we propose that the visually convergent stars cooler than the ZAIS line are pre-I sequence stars, which have not completely erased the imprint of the initial period (Fig.~\ref{fig_comp_cluster1} and Fig.~\ref{fig_comp_cluster2}). The pre-I sequence stars can be associated with the stall of the spin-down, which occurs near the ZAIS line for different temperature and age ranges (see Sect.~\ref{sec_comp_cg} for details)

 Gyrochronology classifies the evolutionary phase in terms of angular momentum loss so that the ``CgIW" scenario corresponds to a different spin-down rate in each phase. Equally, the activity dependence on rotation is a strong constraint and implication of the evolutionary phase. The different dependences in Fig.~\ref{fig_act_ro_color} demonstrate the CgIW scenario from the perspective of activity. In this sense, the CgIW scenario unifies the two paradigms and reveals their magnetic and structure evolution.

\section*{Data availability}
Data used in this study are available at the CDS via anonymous ftp to cdsarc.cds.unistra.fr (130.79.128.5) or via https://cdsarc.cds.unistra.fr/viz-bin/cat/J/A+A/vol/page

\begin{acknowledgements}
We sincerely thank the anonymous referee for the very careful and helpful comments, which have significantly improved this study. We sincerely thank A.R.G. Santos, J.J. Lehtinen, D. Foreman-Mackey, E. R. Newton,S.T. Douglas, N. Astudillo-Defru, M. Mittag, K. Vinay and Shu Wang for helpful discussion. JFL acknowledges support from the New Cornerstone Science Foundation through the New Cornerstone Investigator Program and the XPLORER PRIZE. This work is supported by the Joint Research Fund in Astronomy (U2031203) under cooperative agreement between the National Natural Science Foundation of China (NSFC) and Chinese Academy of Sciences (CAS). RS acknowledges support from the INAF grant number 1.05.23.04.04. FS is funded by the European Union – NextGenerationEU RRF M4C2 1.1 No.2022HY2NSX, "CHRONOS: adjusting the clock(s) to unveil the CHRONO-chemo-dynamical Structure of the Galaxy" (PI: S. Cassisi).
\end{acknowledgements}

%
%
\bibliographystyle{aa}
\bibliography{ref_aa}
\begin{appendix}

\section{The stellar bolometric luminosity}\label{appendix_lum}
The stellar bolometric luminosity was calculated from stellar distance, the Gaia magnitude, the extinction and stellar physical parameters. 
First, we obtained the distances to each source from the Gaia EDR3 geometric distance catalog \citep{2021AJ....161..147B}.The extinction was estimated using the 3D dust map from Pan-STARRS DR1 \citep{Green2019}, adopting $E(B-V) = 0.884 \times {\rm Bayestar19}$. We calculated the extinction in the Gaia $G$, $G_{\rm BP}$, and $G_{\rm RP}$ bands using $A_{\lambda} = R_{\lambda} \times E(B-V)$, where $R_{\lambda}$ represents the extinction coefficient in the corresponding Gaia band, following \citet{Cardelli1989}. Using stellar atmospheric parameters—effective temperature, surface gravity, and metallicity—we derived the bolometric corrections (BCs) from the MIST model via the {\it isochrones} Python module \citep{Morton2015}.
The bolometric magnitude for each band was then computed as 
\begin{equation} 
\label{luminosity.eq} 
M_{\rm bol} = m_{\lambda} - 5 \log d + 5 - A_{\lambda} + BC, 
\end{equation} 
where $m_{\lambda}$ is the apparent magnitude in the Gaia $G$, $G_{\rm BP}$, or $G_{\rm RP}$ band, $d$ is the distance in parsecs, $A_{\lambda}$ is the extinction in that band, and $BC$ is the corresponding bolometric correction.
The bolometric luminosity was calculated using 
\begin{equation} 
\label{lbol.eq} 
L_{\rm bol} = 10^{-0.4 (M_{\rm bol} - M_{\odot})} L_{\odot}, 
\end{equation} 
where $M_{\odot} = 4.74$ mag is the solar bolometric magnitude, and $L_{\odot} = 3.828 \times 10^{33}$ erg/s is the solar bolometric luminosity.
The final luminosity was adopted as the average of the values obtained from the three Gaia bands.

\section{The conversion of S to $R'_{\rm HK}$}\label{strhk}

The S-index includes contribution from chromospheric and photospheric radiation, and also depends on reference continua of different spectra types. In order to calculate the true contribution from chromosphere and normalize to the bolometric luminosity, the relation that converts {\it S} to $R'_{\rm HK}$ in terms of the color index $B-V$ was developed \citep{Noyes1984,Mid1982,Rut1984}:

\begin{equation}\label{eq_rhk}
 R'_{\rm HK} = R_{\rm HK} - R_{\rm phot} = K \cdot \sigma^{-1} \cdot 10^{-14} \cdot C_{\rm cf} (S - S_{\rm phot}).
 \end{equation}

Here, $R_{\rm phot}$ and $S_{\rm phto}$ represent the photospheric contribution, and $C_{\rm cf}$ is used to remove the color term from the S-index and thus is a conversion factor that corrects flux variations in the continuum passbands. The term $\sigma$ is the Stefan-Boltzman constant, and $10^{-14}$ is a scaling factor, while K converts the arbitrary flux units to physical flux units on the stellar surface. Previous studies investigated different instruments and spectra types and respectively found $K = 0.76 \times 10^6$ \citep{Mid1982}, $1.29 \times 10^6$ \citep{Rut1984}, $1.07 \times 10^6$ \citep{Hall2007} $~\rm erg~cm^{-2}~s^{-1}$ for a triangle filter of 1.09 \AA ~Mount Wilson passband. We adopt the first one, because taking the Sun as an anchor point with $S_{\rm MWO} \sim 0.17$ and ${\rm log}R'_{\rm HK} \sim -4.9$ \citep{Egeland2017}, we find it could obtain a consistent ${\rm log}R'_{\rm HK}$ for the Sun through the expressions we used.
\subsection{The conversion of S to $R'_{\rm HK}$ for F-,G-, and K-type stars}

The conversion factor $C_{\rm cf}$ differs between MS stars and post-MS stars. we adopt the third order polynomial fits of $C_{\rm cf}$ \citep{Rut1984} for MS stars and for post-MS stars respectively. The determination of the evolutionary stage of each star is introduced in Section~\ref{ms_pms}. We adopted the same $R_{\rm phot}$ \citep{Noyes1984} for the MS and post-MS stars.

\subsection{The conversion of S to $R'_{\rm HK}$ for M-type stars}

Although the most commonly used conversion factor $C_{\rm cf}$ is from two studies \citep{Mid1982,Rut1984}, a large fraction of their sample are F-,G-,K-type stars, having poor constraint on M-type stars($B-V > 1.4$). Besides, the stellar models they used probably have underestimated the contribution of $R_{\rm phot}$ for M-type stars \citep{Def2017,Boro2018}. Here, we use a new third order polynomial fits of $C_{\rm cf}$ and $R_{\rm phot}$ \citep{Def2017} for the conversion, which focus on the color index range of M-type stars.

\begin{figure}[ht]
\includegraphics[width=0.5\textwidth]{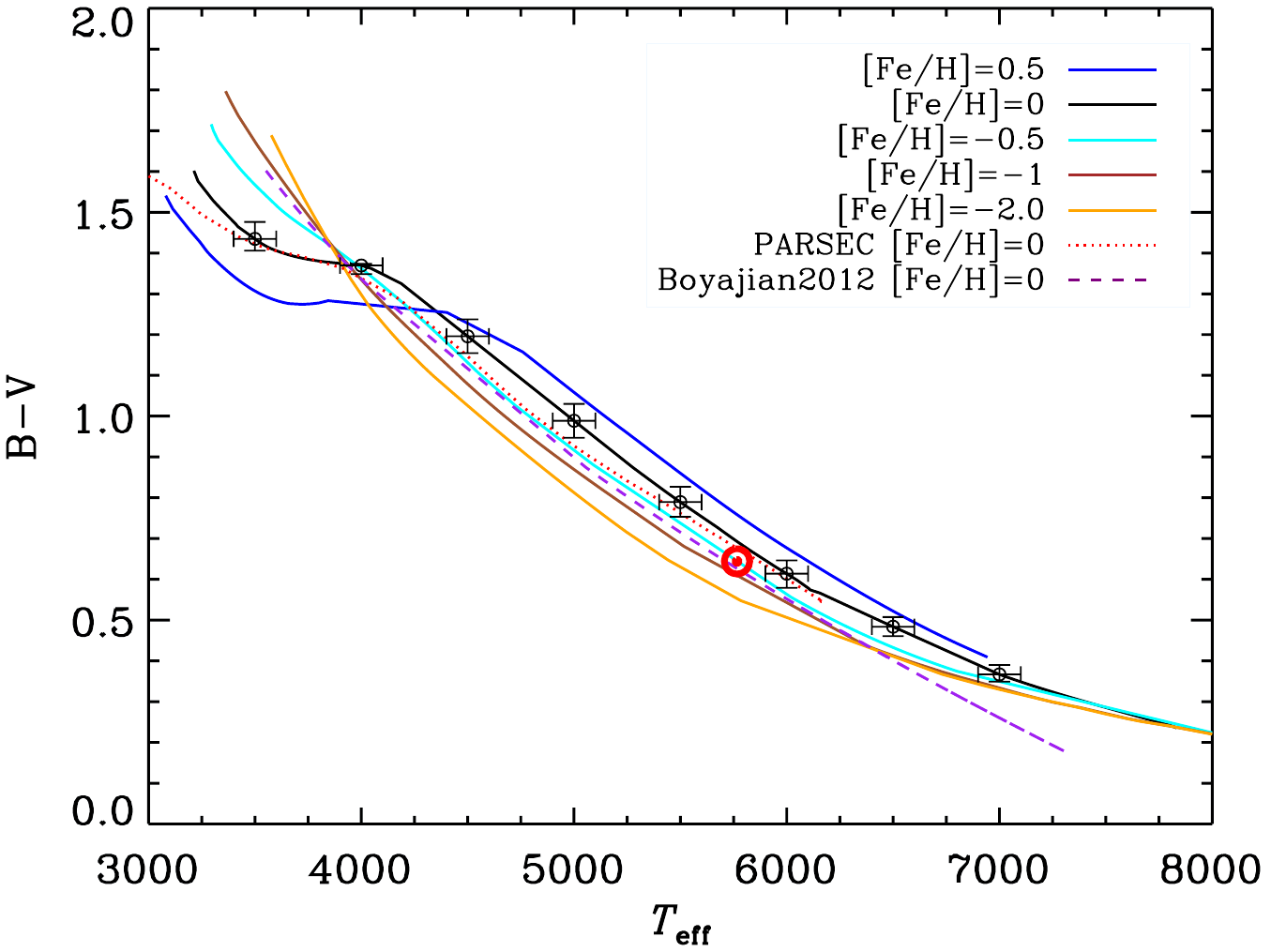}
\caption{Relationships of $T_{\rm eff}$ and color in different metallicities from the Dartmouth model. The Sun is marked as a red circle. The error bars of black circles indicates the uncertainty of the conversion for [Fe/H]=0, whose typical value is $\sim 0.03$ dex. For comparison, the dotted line is from PARSEC model atmosphere \citep{Chen2014} with [Fe/H]=0, the dashed line is an empirical relation from \citet{Boya2012} with [Fe/H]=0.}\label{figbvt}
\end{figure}

\subsection{The transformation of $T_{\rm eff}$ to $B-V$}

We need to perform the transformation from effective temperature to $B-V$~to complete the above calculations. Compared to empirical relations, synthetic photometries can provide a unified and extended transformation for effective temperature and metallicity \citep{Lejeune1997}. We use the Dartmouth Stellar Evolutionary Database \citep{Dotter2008}\footnote{Available online: URL http://stellar.dartmouth.edu/models/grid.html} whose synthetic $T_{\rm eff}$--color relations are derived from the PHOENIX model atmosphere grid for $T_{\rm eff} < 10000$ to obtain the color index $B-V$ of each star. The Dartmouth models provide synthetic photometry for a range of photometric systems by integrating the spectrum of each star over the relevant bandpass. The $T_{\rm eff}$--color relation is affected by metallicity \citep{Boya2012}. The synthetic photometry and spectra establish the $T_{\rm eff}$--color relation for different metallicities, and the influence of metallicity on the relation is demonstrated in Fig.~\ref{figbvt}. The empirical relation and the PARSEC model \citep{Chen2014} with solar metallicity are also plotted for comparison. We calculate $B-V$~for the full range of Dartmouth model metallicities (-2.5$\leq$ [Fe/H]$\leq$0.5) with steps of 0.5 dex at age $t_{\rm age} =1$ Gyr. We then obtain the final value of $B-V$~by linear interpolation in metallicity. We calculate the uncertainty of the conversion by using $T_{\rm eff}$+100K, and $T_{\rm eff}$-100K and their corresponding $B-V$ as the upper and lower limit for each star, where the error of $T_{\rm eff} \sim$ 100K is given by LASP. The typical uncertainties of $B-V$ are $\sim 0.03$ dex, which are shown in Fig.~\ref{figbvt}.

\section{Fit of the canonical rotation--activity relationship}\label{rare}
We fit the rotation--activity relationship by the Python package {\it emcee} \citep{Foreman2013}, which is an affine-invariant ensemble sampler for Markov chain Monte Carlo. As shown in Fig.~\ref{fig_fgkm_dic}, the scatter is large in the unsaturated regime, so we applied a Bayesian mixture model to the likelihood function, which contains two-dimensional uncertainties in both $x$ and $y$ direction \citep{Hogg2010}\footnote{ The mathematical derivation of the likelihood function is beyond the scope of this work. We refer to \citet{Hogg2010} for more details on the model. The likelihood function of this work is a combination of Equation 17 and 32.}. It is a probabilistic generative model that can prune outliers objectively with a uninformative prior. As suggested by \citet{Hogg2010}, it is strongly preferable to that rejecting points by hand or the so-called sigma clipping method, both of which are not the outcome of justifiable modeling.

The model needs three extra parameters ($P_b,Y_b,V_b$) to construct an objective function, where $P_b$ is the amplitude of the ``bad-data" distribution function in the mixture that indicates the probability that a point is ``bad". $P_b$ is sensitive to data error that also can be used to assess the validity of the model to the data. $Y_b$ and $V_b$ are the mean and variance of the ``bad-data" distribution in y axis. We set the correlation coefficient $\rho_{xy}=1$ in the uncertainty covariance matrix. Note that the correlation coefficient ranging from -1 to 1 could result in variation of 0.05 dex on $\beta$ and little influence on other parameters. 

The likelihood function also includes an intrinsic Gaussian variance $V_b$ that can be used to testify if our data error is underestimated or there is an intrinsic scatter in the relation \citep{Hogg2010}. The fit takes $V_b$ as a parameter in a logarithmic form to guarantee that $V_b$ is always a positive value. The total error is obtained by adding the data error and $V_b$ in quadrature. Note the parameter $V_b$ includes both the underestimated error of $R'_{\rm HK}$ and the intrinsic scatter of the relation. Its small value makes the fitting results be close to a weighted least-square solution and indicates the present errors on ${\rm log} R'_{\rm HK}$ are proper (Foreman-Mackey; private communication).

As suggested by the tutorials of emcee\footnote{ See URL https://emcee.readthedocs.io/en/stable/  for further details.} , we run 200 walkers for a total of 1000 steps, discarding the first 300 as burn-in. We calculate the mean acceptance fraction of the ensemble is $\sim$0.56, and the mean autocorrelation timescale is $\sim$48.2 steps.

\begin{figure}
\includegraphics[width=0.5\textwidth]{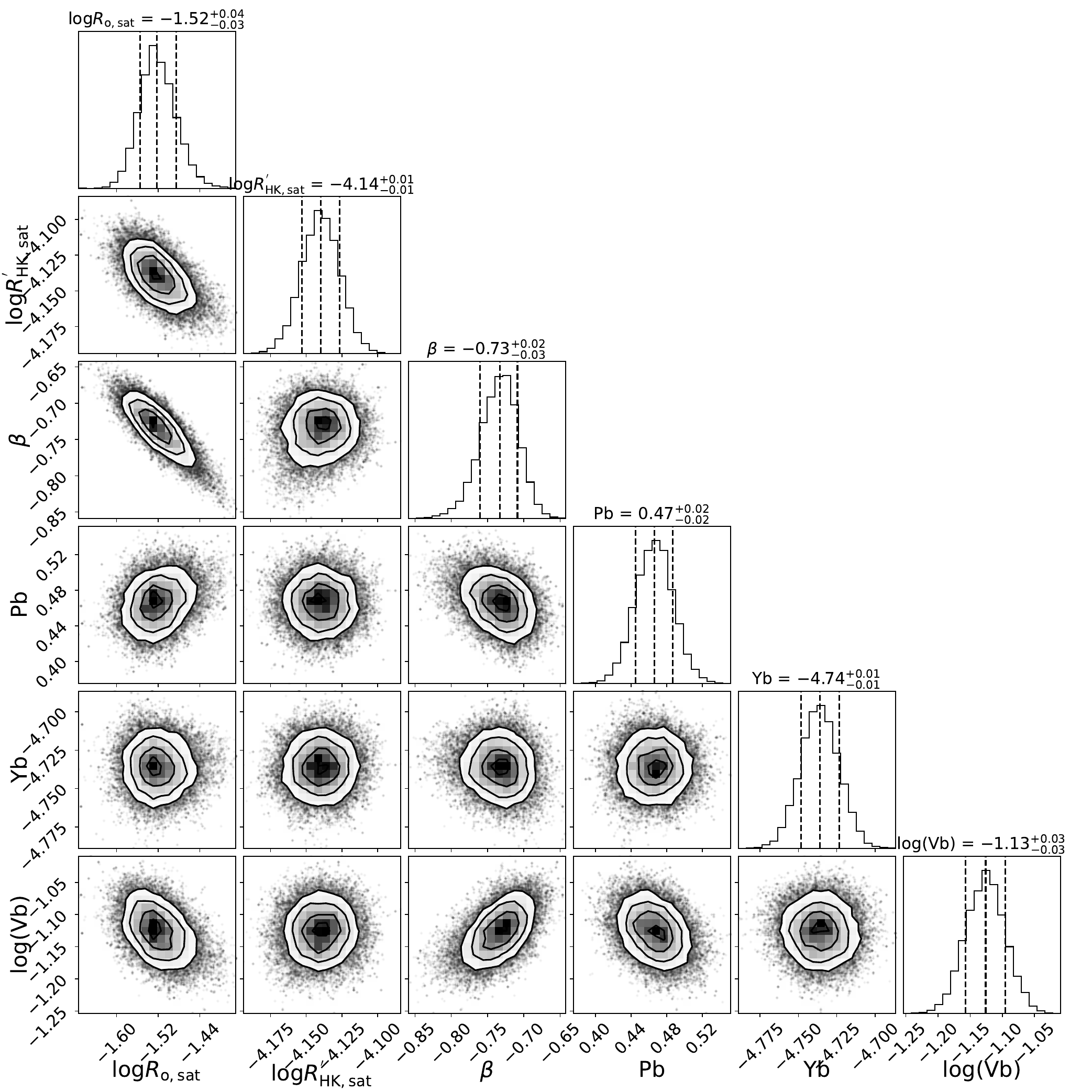}
\caption{Posterior probability distributions of the fitting parameters for the rotation--activity relationship of K-type stars. Each panel shows a two-dimensional projection of the distribution, above which the marginalized distributions indicate the 16th, 50th and 84th percentiles with dashed lines.}

\label{fig_corner_K}
\end{figure}
An example of the posterior probability distributions over each parameter is shown in Fig.~\ref{fig_corner_K} by using the {\it corner} package \citep{Foreman2016}. Each panel shows a two-dimensional projection of the distributions, above which the 16th, 50th, and 84th percentiles in the marginalized distributions are also marked. According to the suggestion of \citet{Hogg2010}, we take the median and 16th, 84th percentiles of the marginalized distributions as the best fit and errors.

\section{The conversion of various activity proxies}\label{app_cda}
\begin{figure}
\includegraphics[width=0.5\textwidth]{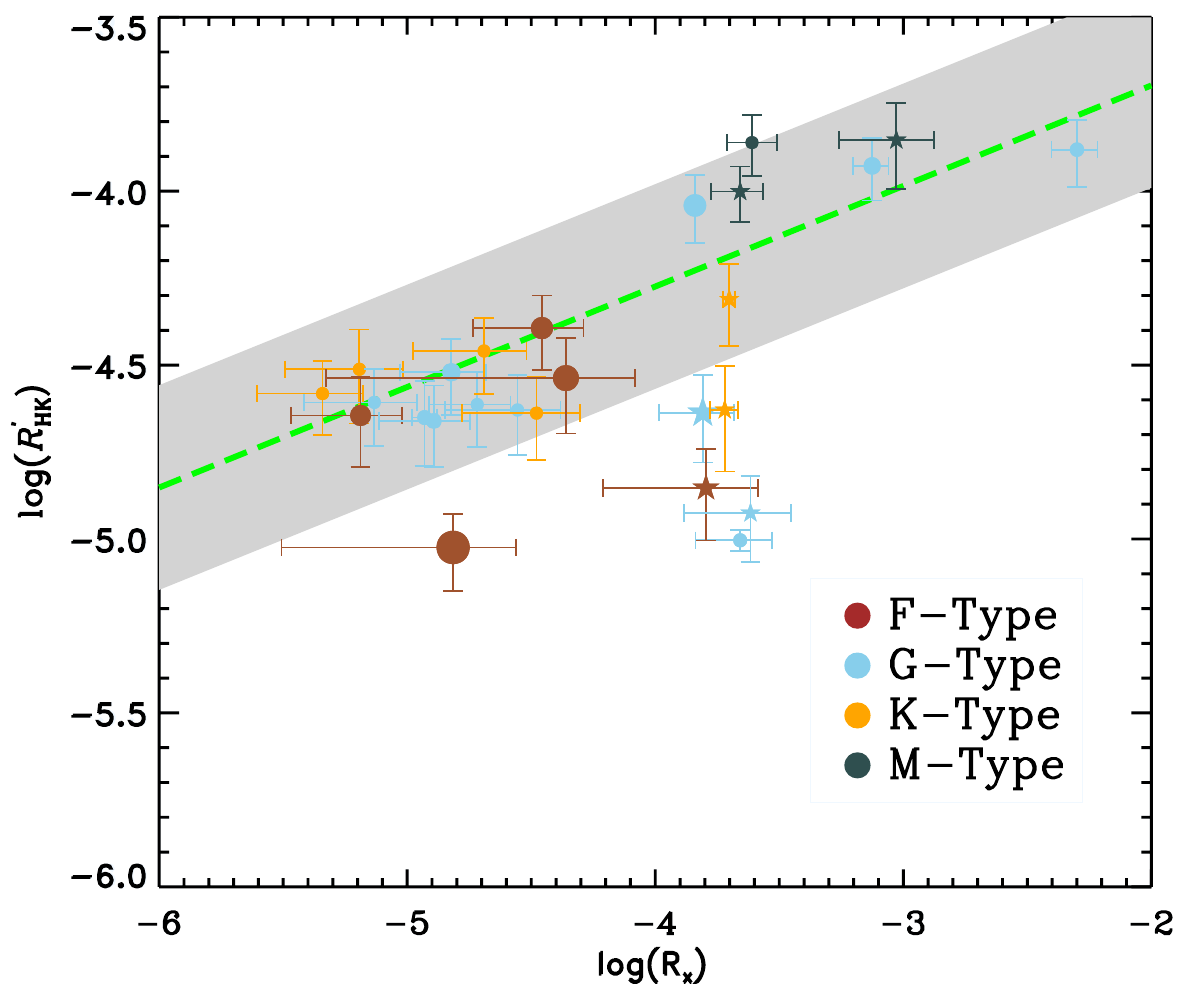}
\caption{Coronal activity $R_{\rm x}$ vs. chromospheric activity $R'_{\rm HK}$. Our sample have 26 stars with $R_{\rm x}$, three of them are from \citet{Wang2020} and the rest are from \citet{Pizz2019}. The green dashed line is given by \citet{Mama2008}, for which the boundary of the gray area indicates 1-$\sigma$ level of the residuals of all the points. The symbol of five-pointed star indicates common stars among $R'_{\rm HK}$, $R_{\rm X}$ and $R_{\rm flare}$. The size of symbols represents the surface gravity ${\rm log}g$ which is the same as Fig.~\ref{fig_fgkm_dic}}

\label{fig_hk_rx}
\end{figure}

\begin{figure}
\includegraphics[width=0.5\textwidth]{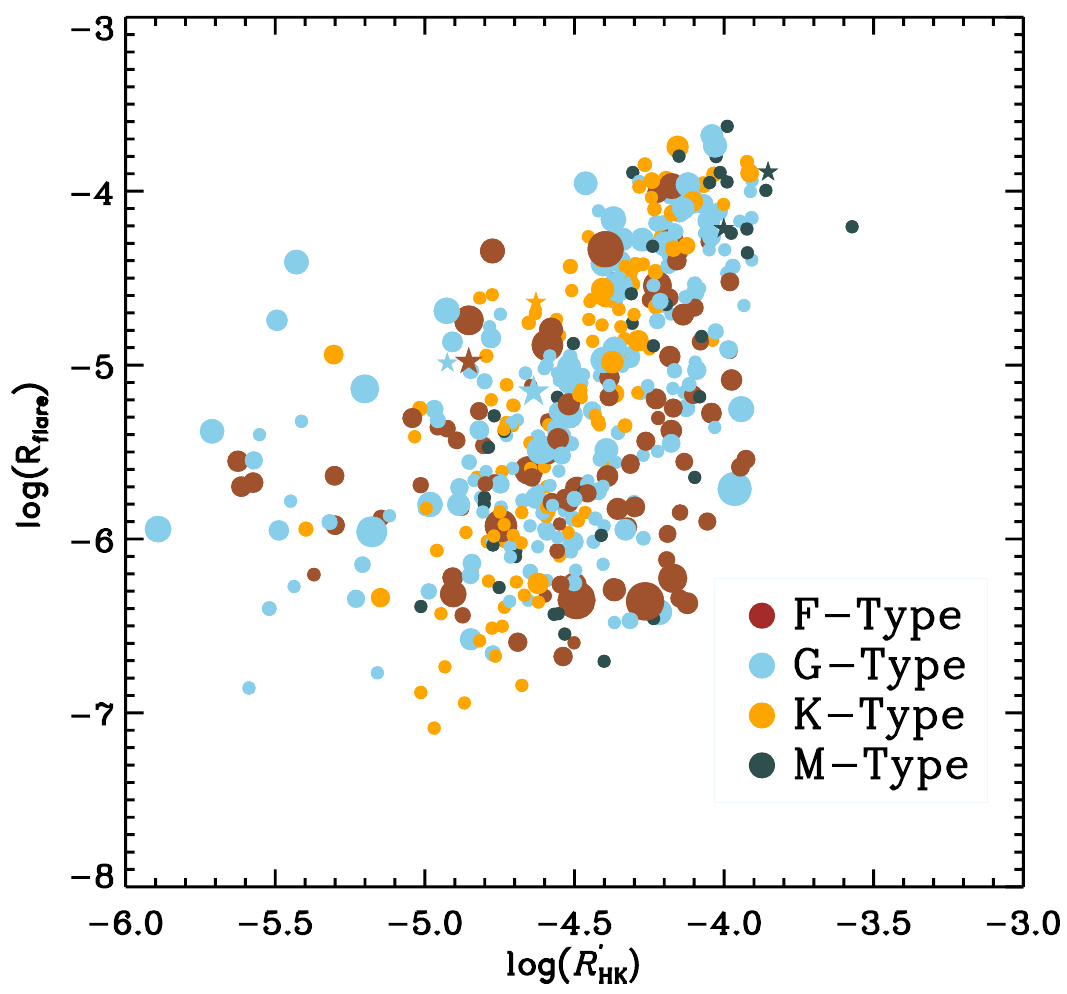}
\caption{Chromospheric activity $R'_{\rm HK}$ vs. the flaring activity $R_{\rm flare}$. Our sample have 515 flaring stars, where the values of the flaring activity are from \citet{Yang2019}. The meaning of symbols are as the same as Fig.~\ref{fig_hk_rx}.}

\label{fig_hk_flare}
\end{figure}

Different activity proxies could have empirical relationships that convert them to each other \citep[e.g.,][]{Li2024}. In this section, we compare $R'_{\rm HK}$ with other activity proxies including X-ray, flare, H$\alpha$ activity, and $R^{+}_{\rm HK}$ to verify if the fitting parameters of the canonical rotation--activity relationship from different proxies are consistent.

The coronal activity is represented by fractional X-ray luminosity: $R_{\rm x} = L_{\rm x}/L_{\rm bol}$. We find 26 stars in our sample with $R_{\rm x}$ from the X-ray studies in the Kepler field \citep{Pizz2019,Wang2020} and show them in Fig.~\ref{fig_hk_rx}. \citet{Mama2008} derived an improved $R_{\rm x}$--$R'_{\rm HK}$ relation based on $\sim$ 200 stars, which is plotted as the green dashed line(Equation~\ref{eq_hk_x}). Most of our stars obey this relation, whereas six of them are systematically lower by $\sim$ 0.5 dex in ${\rm log}R'_{\rm HK}$. This phenomenon is also reported by \citet{Mama2008}, who pointed out that the scatter of the relation increases substantially for ${\rm log}R_{\rm x} > -4$, which corresponded to the transition from active regimes to very active regimes.

\begin{equation}\label{eq_hk_x}
\begin{aligned}
 {\rm log}  R'_{\rm HK}= (-4.54\pm 0.01) + (0.289 \pm 0.015) ({\rm log} R_{\rm X}+4.92).
\end{aligned}
 \end{equation}

\begin{figure}
\includegraphics[width=0.5\textwidth]{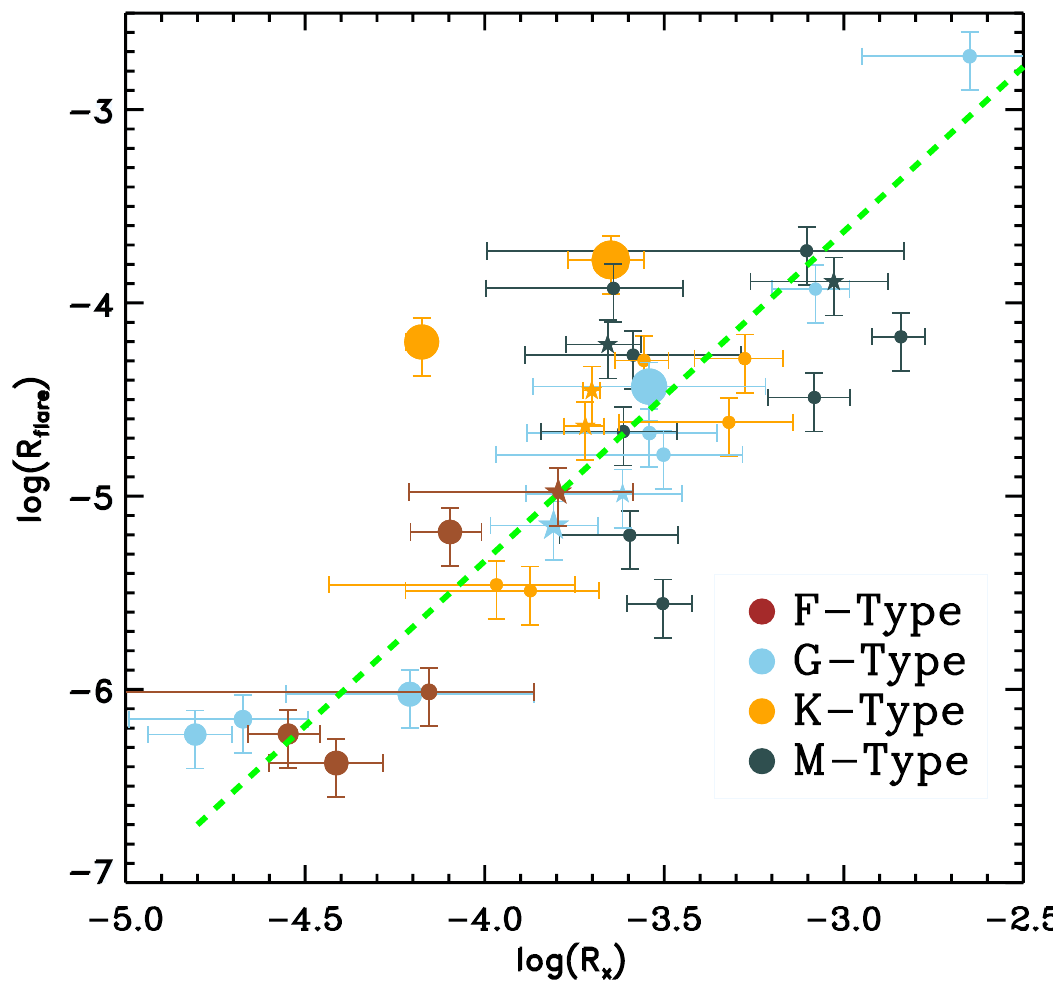}
\caption{Coronal activity $R_{\rm x}$ vs. the flaring activity $R_{\rm flare}$. There are 34 common stars with $R_{\rm x}$ and $R_{\rm flare}$ \citep{Pizz2019,Yang2019}. The green dashed line is the best fit given by a linear regression of the ordinary least-square bisector \citep{Isobe1990}. The meaning of symbols are as the same as Fig.~\ref{fig_hk_rx}.}
\label{fig_rx_flare}
\end{figure}

Our sample includes 515 flaring stars, whose flaring activity ($R_{\rm flare} = L_{\rm flare}/L_{\rm bol}$) is also a good indicator of activity \citep{Yang2017,Yang2019}. Fig.~\ref{fig_hk_flare} shows the relationship between chromospheric activity $R'_{\rm HK}$ and flaring activity $R_{\rm flare}$. There is a clearly positive correlation with a large scatter for ${\rm log} R'_{\rm HK} >-5$, below which the two parameters do not correlate each other. The dispersion is mainly caused by F- and G-type stars, whose internal uncertainties are greater in both $R'_{\rm HK}$ and $R_{\rm flare}$.  The correlation coefficient of the sample is r =0.60 for ${\rm log} R'_{\rm HK} >-5$.

We also plot the relationship between coronal activity $R_{\rm x}$ and flaring activity $R_{\rm flare}$ in Fig.~\ref{fig_rx_flare}, whose data are from \citet{Pizz2019} and \citet{Yang2019}. The correlation coefficient of the sample is r=0.82. Their best fit by the ordinary least-square bisector linear regression is

\begin{equation}\label{eq_flare_x}
\begin{aligned}
 {\rm log} R_{\rm flare}= (1.48\pm 0.49) + (1.71 \pm 0.13) {\rm log} R_{\rm x}
\end{aligned}
 \end{equation}

with an rms of 0.50 dex in ${\rm log}R_{\rm flare}$.

\begin{figure}
\includegraphics[width=0.5\textwidth]{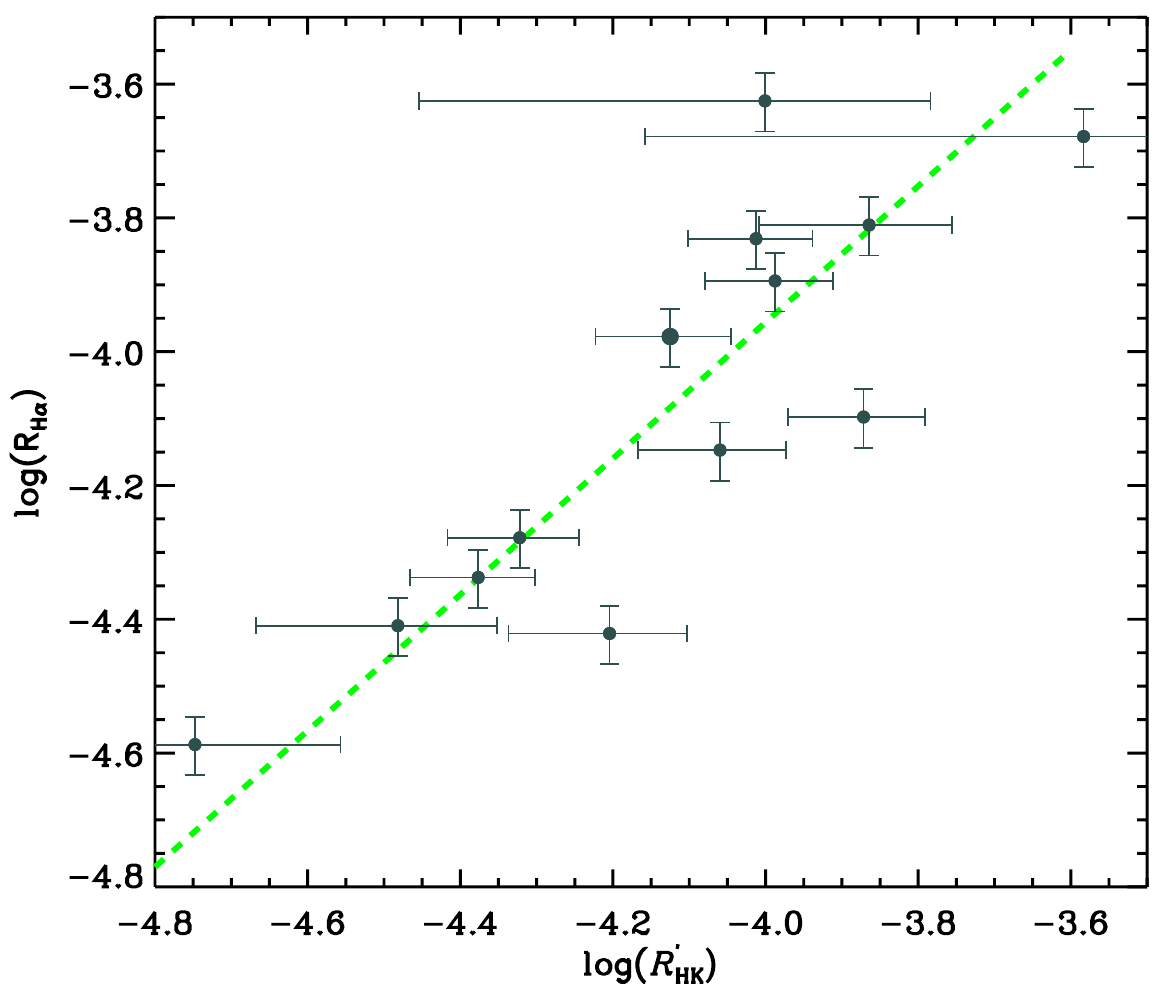}
\caption{Chromospheric activity $R'_{\rm HK}$ vs. the H$\alpha$ activity $R_{H\alpha}$. There are 13 stars with $R_{\rm H\alpha}$ \citep{Yang2017}, all of which are M-type emission stars. The green dashed line is the best fit given by a linear regression of the ordinary least-square bisector \citep{Isobe1990}. The meaning of symbols are as the same as Fig.~\ref{fig_hk_rx}.}
\label{fig_hk_ha}
\end{figure}

Our sample have 13 stars with H$\alpha$ activity ($R_{\rm H\alpha}=L_{\rm H\alpha} / L_{\rm bol}$) in \citet{Yang2017}, all of which are M-type emission line stars. We plot the chromospheric activity $R'_{\rm HK}$ vs. the H$\alpha$ activity $R_{\rm H\alpha}$ in Fig.~\ref{fig_hk_ha} and get the best fit of the ordinary least-square bisector linear regression as

\begin{equation}\label{eq_hk_ha}
\begin{aligned}
 {\rm log} R_{\rm H\alpha}= (0.11\pm 0.59) + (1.02 \pm 0.14) {\rm log} R'_{\rm HK}
\end{aligned}
 \end{equation}
with an rms of 0.16 dex in ${\rm log}R_{\rm H\alpha}$.
The relation of $R_{\rm X}$ vs. $R^{+}_{\rm HK}$ is fit by \citet{Mittag2018} as

\begin{equation}\label{eq_hkplus_x}
\begin{aligned}
 {\rm log}  R^{+}_{\rm HK}= (-2.46\pm 0.09) + (0.4 \pm 0.02) {\rm log} R_{\rm X}.
\end{aligned}
\end{equation}

Although the underlying physical trigger of all activity proxies is magnetism, the precise way to connect these proxies is complex. Their differences include different atmospheric layers, different wavelength bands, global or local events and the calculation method. However, from an observational point of view, we have shown the linear relations of different proxies in logarithmic form. It allowed us to establish a conversion of the rotation--activity relationship by different proxies and in turn gauge the applicability of the dichotomy. 


According to Equation~\ref{eq_hk_x},\ref{eq_flare_x}, \ref{eq_hk_ha} and \ref{eq_hkplus_x}, the activity proxies should obey the following conversion:

\begin{equation}\label{eq_sat}
\begin{aligned}
{\rm log}R'_{\rm HK}&=0.29 \times {\rm log}R_{\rm X}-3.11 \\
                        &=0.17 \times {\rm log}R_{\rm flare}-3.36\\
                        &=0.98 \times {\rm log}R_{\rm H\alpha}-0.11\\
                        &=0.73 \times {\rm log}R^{+}_{\rm HK}-1.33,
 \end{aligned}
 \end{equation}

and their corresponding $\beta$ in the rotation--activity relationship rely on the coefficient of the first-order terms of the conversion, whose ratios should be

\begin{equation}\label{eq_beta}
\begin{aligned}
 \beta'_{\rm HK} : \beta_{\rm X} :\beta_{\rm flare} :\beta_{\rm H\alpha}: \beta^{+}_{\rm HK}= 1:3.46:5.91:1.02:1.38.
 \end{aligned}
 \end{equation}

We took $R'_{\rm HK}$ as a benchmark and obtained

\begin{equation}\label{eq_value}
\begin{aligned}
\beta'_{\rm HK}&=-0.62 \pm 0.05, & {\rm log}R'_{\rm HK,sat}&=-4.06 \pm 0.10\\
\beta_{\rm X}&=-2.15 \pm 0.17, & {\rm log}R_{\rm X,sat}&=-3.28 \pm 0.35\\
\beta_{\rm flare}&=-3.66 \pm 0.30, & {\rm log}R_{\rm flare,sat}&=-4.12 \pm 0.59 \\
\beta_{\rm H\alpha}&=-0.63 \pm 0.05, & {\rm log}R_{\rm H\alpha,sat}&=-4.03 \pm 0.12\\
\beta^{+}_{\rm HK}&=-0.86 \pm 0.1, & {\rm log}R^{+}_{\rm HK,sat}&=-3.74 \pm 0.13
 \end{aligned}
 \end{equation}

The derived parameters in Equation~\ref{eq_value} are validated by Table~\ref{table_para_act_ro}, where the five proxies follow the above conversion within a uncertainty of 0.1 dex on $\beta'_{\rm HK}$ and ${\rm log}R'_{\rm HK}$. There is an exception in Table~\ref{table_para_act_ro} \citep{Newton2017}, whose $\beta$ significantly deviates from the expectation. \citet{Newton2017} and \cite{Douglas2014} used the same proxy H$\alpha$, but got totally different parameters. We present some potential reasons here: (1) One mathematic issue of the dichotomy is that $\beta$ and $Ro_{\rm sat}$ are degeneracy with an anti-correlation, which can be clearly seen from the projection of the posterior distribution in this study. As a result, two biased samples with different ages could influence the fitting parameters when they concentrate at the saturated or unsaturated regime. The sample of \citet{Douglas2014} is from open clusters: the Praesepe and the Hyades ($t_{\rm age} \sim 750$ Myr), while \citet{Newton2017} has more slowly rotating stars and fully convective stars in the unsaturated regime, which implies the mean age of this sample is much older than that of \citet{Douglas2014}. This makes the sample have a lot of inactive stars with large uncertainties in the unsaturated regime. We note that $\beta$ will drop to -1.2 when the points $R_{\rm H\alpha} <-5$ are neglected. For comparison, \citet{Def2017} used $R'_{\rm HK}$ to investigate the relationship ($N_{\rm star} \sim 40$) for M dwarfs by arbitrarily setting $P_{\rm sat} =10$ days and found $\beta \sim -1.5$. This reason also explains $\beta$ of \citet{Lehtinen2020} in Table~\ref{table_para_act_ro}, which do not have enough fast rotators to determine $Ro_{\rm sat}$. For comparison, \citet{Mas2015} used $R'_{\rm HK}$ to investigate the relation ($N_{\rm star} \sim 20$) in the range of $-6 <{\rm log}R'_{\rm HK}< -4.6$, which also does not have fast rotators. It found $\beta \sim -1.2$. (2) Another factor that could influence the conversion is the scale of $\tau_c$, not only for the difference between global and local $\tau_c$, but also for the potential internal factor that used to derive $\tau_c$. For example, \citet{Douglas2014} and \citet{Newton2017} adopted the same empirical Mass--$\tau_c$ relation but use different color--Mass relations to estimate their stellar mass (e.g., see Fig.~\ref{figcom_bv}). Given that $\tau_c$ is very sensitive to mass for M-type stars, a small discrepancy could have a much larger impact on parameters.

\section{Fit of the new rotation--activity relationship}\label{app_newra}
\begin{figure}
\includegraphics[width=0.5\textwidth]{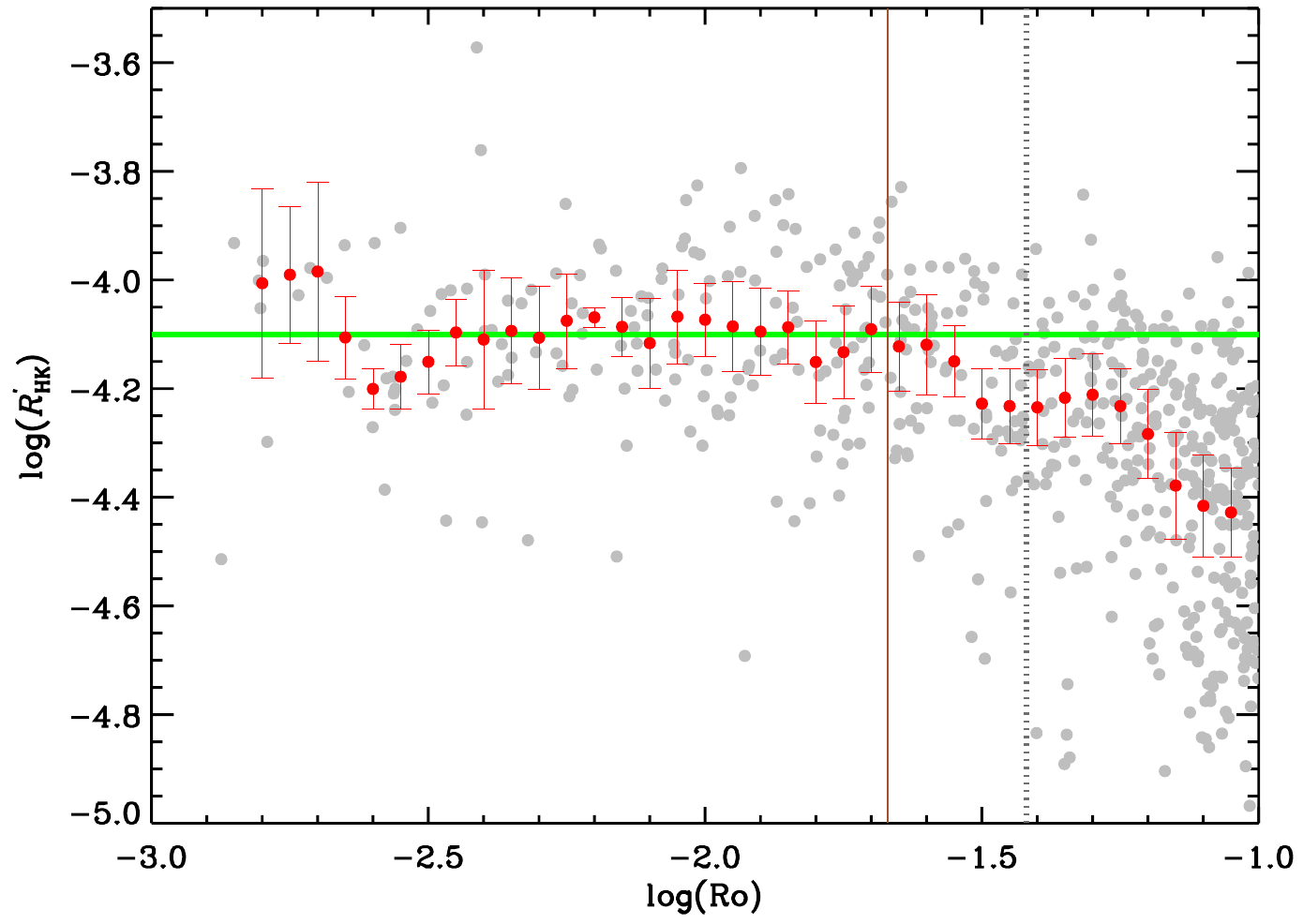}
\includegraphics[width=0.5\textwidth]{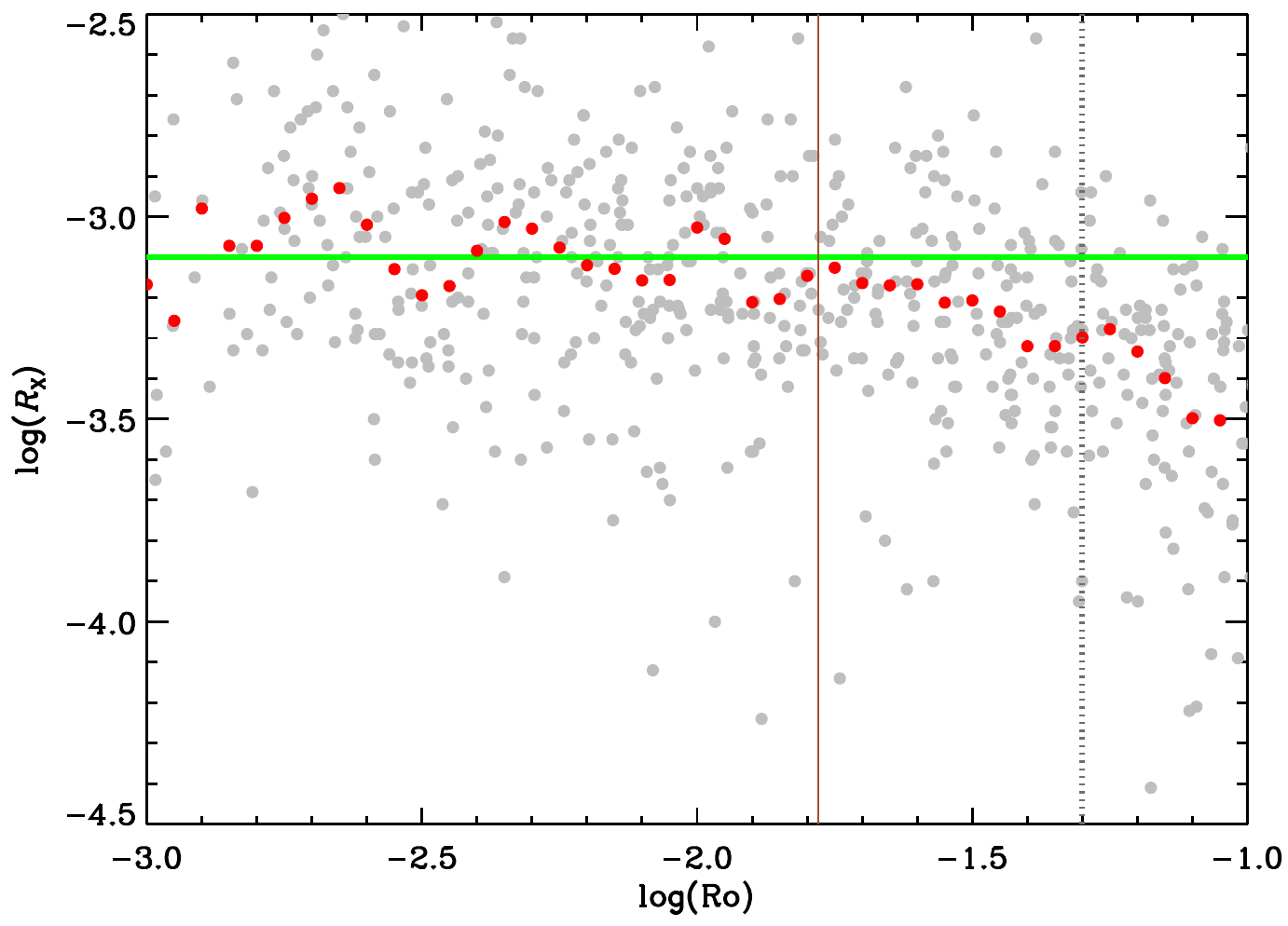}
\caption{Top panel: Rotation--activity relationship in the saturation regime for chromospheric data. The green line indicates the saturation value ${\rm log}R'_{\rm HK} = -4.10$. The red circles represent the mean ${\rm log}R'_{\rm HK}$ in a bin of 0.05 dex. The error bar is the mean error of ${\rm log}R'_{\rm HK}$ in a bin. The vertical line indicates the critical Rossby number Ro = 0.022, where the chromospheric activity reaches the saturation. The dotted vertical line indicates Ro=0.038 where the dichotomy separates the saturated and unsaturated regime. Bottom panel: Same as the top panel but for X-ray data. The vertical line indicates the critical Rossby number Ro = 0.017, where the X-ray activity reaches the saturation.The dotted vertical line indicates Ro=0.05 where the dichotomy separates the saturated and unsaturated regime.}
\label{fig_sat}
\end{figure}

We carry out the fit of the new rotation--activity relationship in three steps. First, we determine the activity level of the saturation ${\rm log}R'_{\rm HK,sat}$ by fitting the rotation--acitvity relationship with the dichotomy model and find ${\rm log} R'_{\rm HK,sat}=-4.10 \pm 0.1 $. Second, we determine the critical Rossby number $\rm Ro_\text{sat}$ that reaches saturation. As shown in Fig.~\ref{fig_sat}, we calculate the mean ${\rm log}R'_{\rm HK}$ in a bin of 0.05 dex along log(Ro). We decrease the value of Ro by steps of 0.05 dex until the mean ${\rm log}R'_{\rm HK}$ of a bin is greater than ${\rm log} R'_{\rm HK,sat}=-4.10$ and get $\rm Ro_\text{sat} =0.022 \pm 0.005$. In this way, we could find a ``pure" saturated regime and avoid the remnant dependence on rotation caused by the dichotomy of the model\citep{Reiners2014}.

Third, we fit the relation in the second and third intervals (gap and the I region). The model we used is

\begin{equation}
{\rm log} R'_{\rm HK} = \begin{cases} \beta_g({\rm Ro}-{\rm Ro_{sat}})+{\rm log} R'_{\rm HK,sat}, & {\rm Ro_\text{sat}} \le {\rm Ro} \le {\rm Ro_{g-to-I}}  \cr \beta_i{\rm Ro}+C, &{\rm Ro_{g-to-I}<Ro<0.7} \end{cases}.
\end{equation}

Here, $R'_{\rm HK,sat}$ is the saturation activity; $\beta_g$ and $\beta_i$ are the slope in the second (gap) and third intervals (the I region), respectively; and $\rm Ro_{g-to-I}$ is the critical Rossby number separating the gap and the I region. The fitting method is the same as Appendix~\ref{rare}. We obtained the best fit of the parameters as follows(the green line in Fig.~\ref{fig_act_ro_color}):

\begin{equation}\label{eq_fitM}
\begin{aligned}
                 \beta_g&= -3.83^{+0.33}_{-0.40}\\
                 {\rm Ro_{g-to-I}}&=0.15^{0.01}_{-0.01} \\
                \beta_i&= -0.68^{+0.03}_{-0.02}.\\
 \end{aligned}
 \end{equation}
We applied the same method to the X-ray data \citep{Wright2011}, and found the best fit is (Fig.~\ref{fig_ro_rx})

\begin{equation}\label{eq_fitx}
\begin{aligned}
                 {\rm log} R'_{\rm X,sat}&=-3.1^{+0.20}_{-0.20}\\
                 {\rm Ro_{sat}}&=0.017\\
                 \beta_g&= -8.59^{+0.27}_{-0.29}\\
                 {\rm Ro_{g-to-I}}&=0.19^{0.01}_{-0.01} \\
                \beta_i&= -0.90^{+0.27}_{-0.29}.\\
 \end{aligned}
 \end{equation}
 \section{The new rotation--activity relationship along the effective temperature}\label{app_arst}
 \begin{figure*}[ht]
\includegraphics[width=1\textwidth]{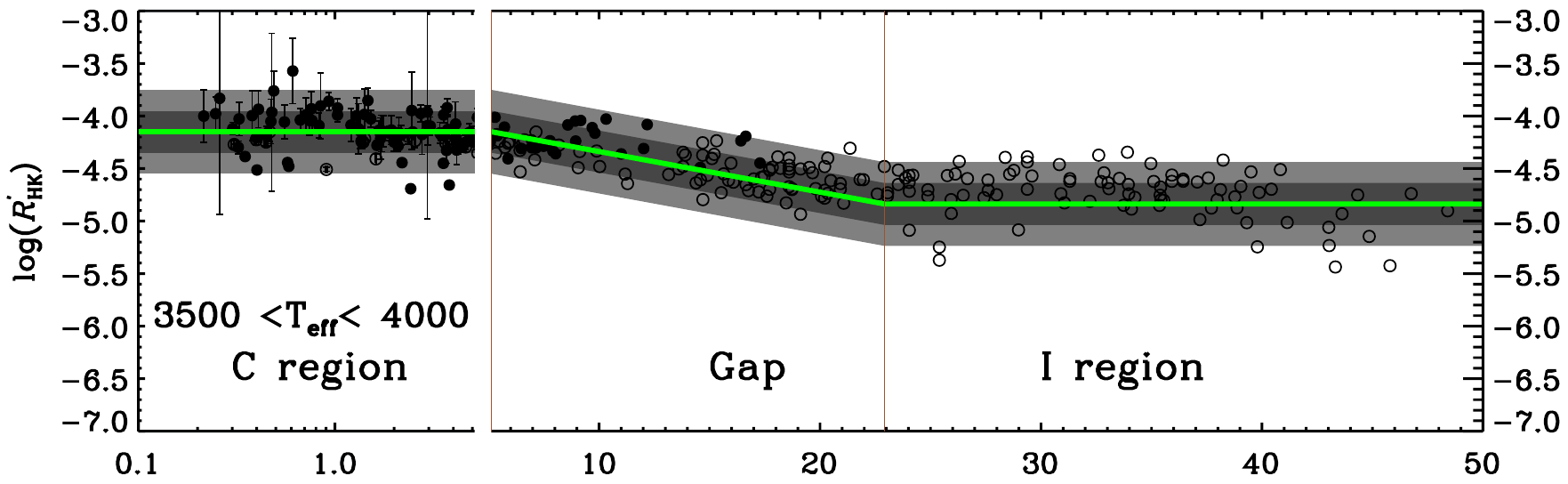}
\includegraphics[width=1\textwidth]{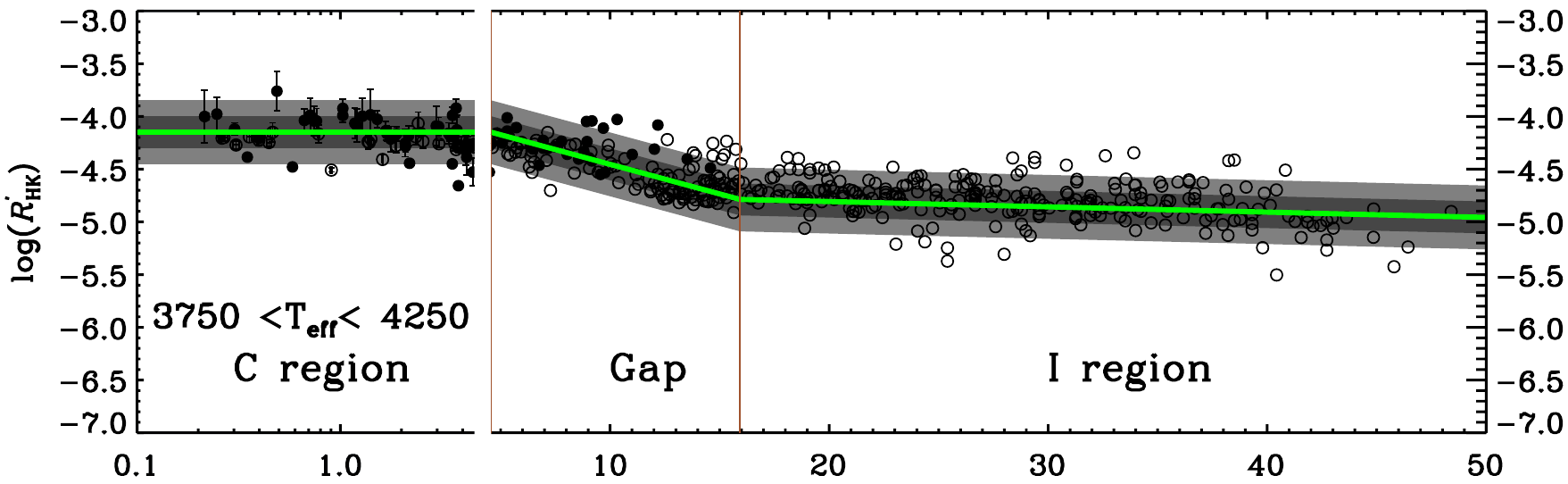}
\includegraphics[width=1\textwidth]{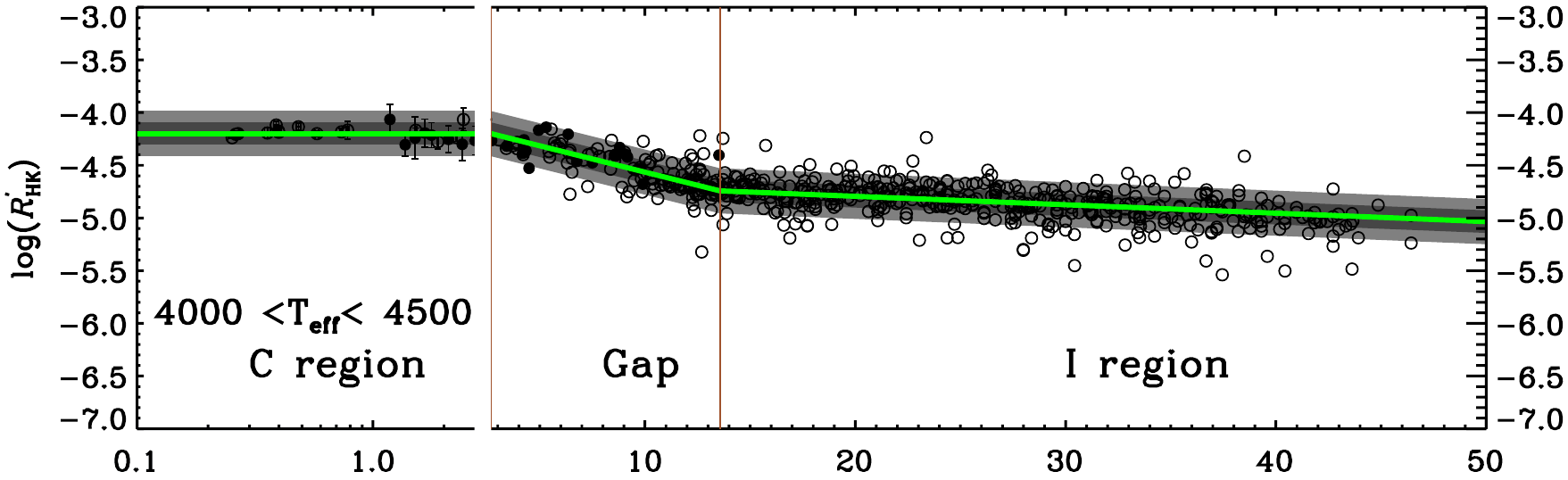}
\includegraphics[width=1\textwidth]{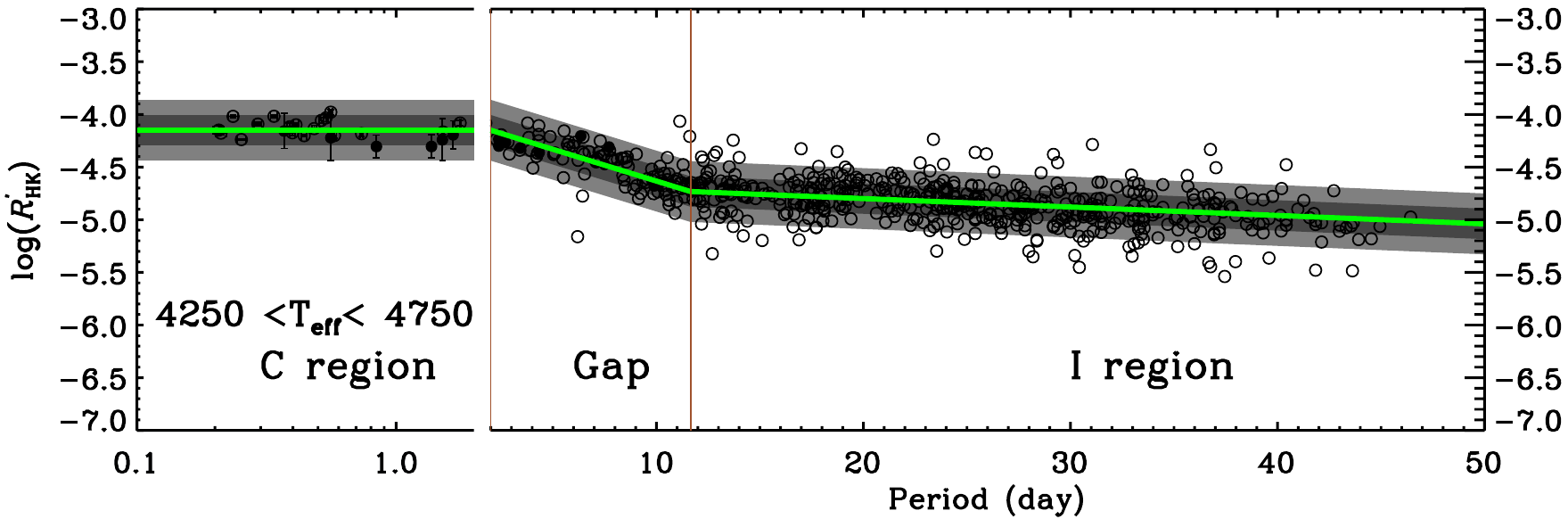}
\caption{New period–chromospheric activity relationship for temperature bins. Closed circles denote the emission line stars and open circles denote the absorption line stars. The fitting results are listed in Table~\ref{table_pfit}. The meaning of symbols are as the same as Fig.~\ref{fig_act_ro_color}.}
\label{fig_ro_act_teffbin1}
\end{figure*}

 \begin{figure*}[ht]
\includegraphics[width=1\textwidth]{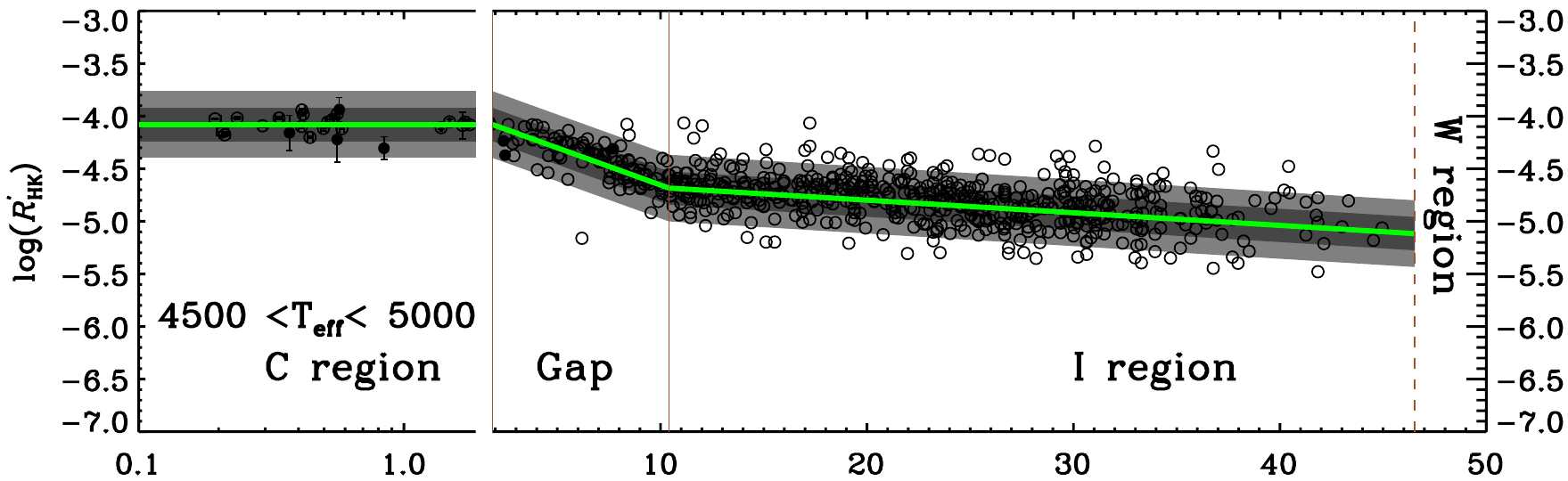}
\includegraphics[width=1\textwidth]{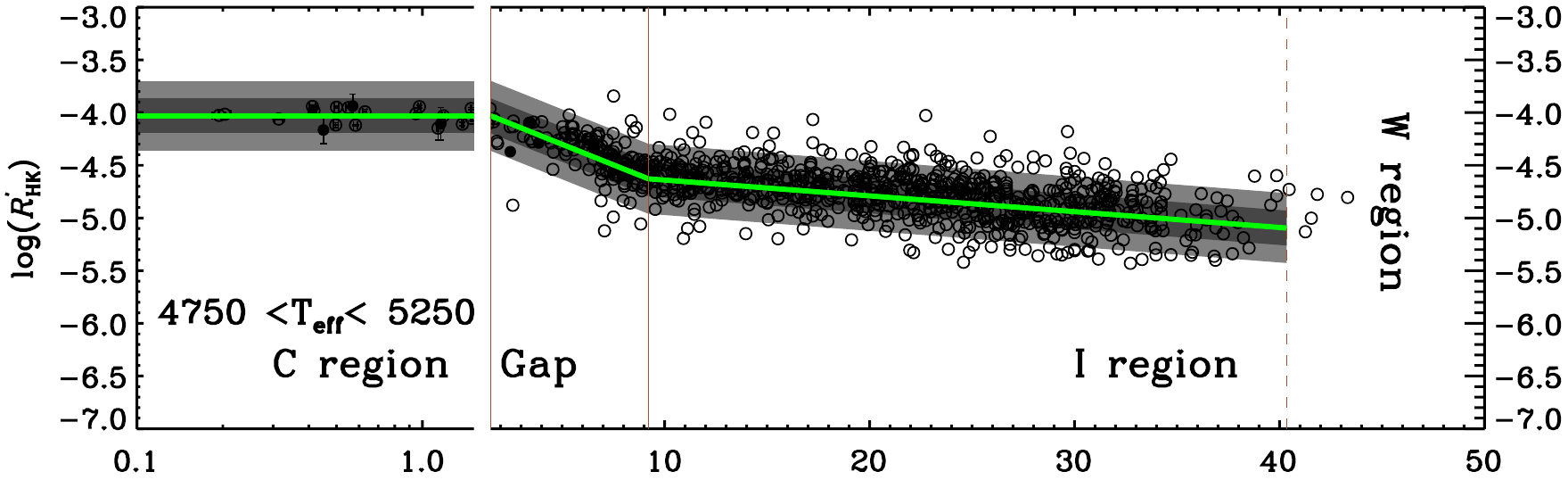}
\includegraphics[width=1\textwidth]{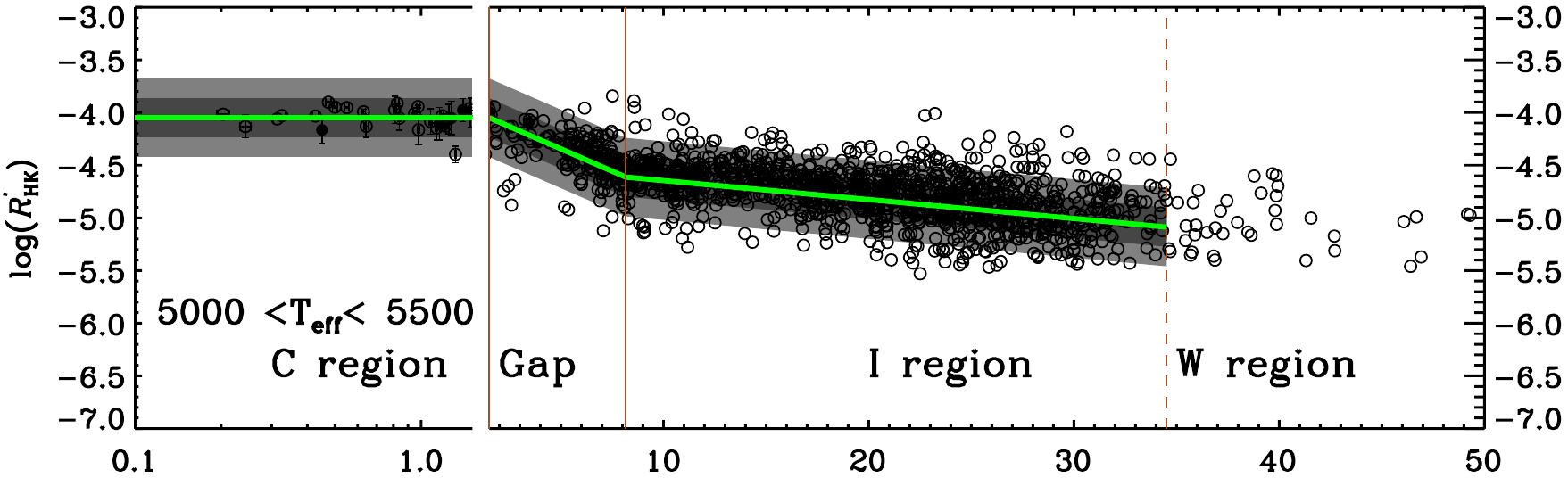}
\includegraphics[width=1\textwidth]{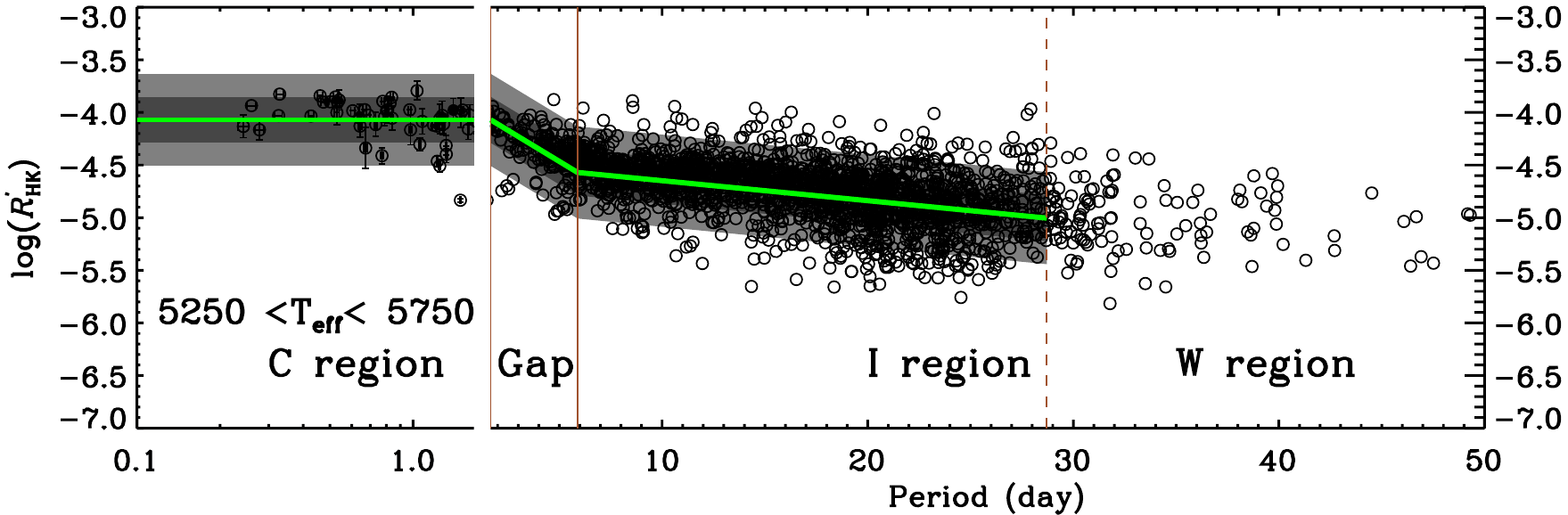}
\caption{New period–chromospheric activity relationship for temperature bins. Closed circles denote the emission line stars and open circles denote the absorption line stars. The fitting results are listed in Table~\ref{table_pfit}. The meaning of symbols are as the same as Fig.~\ref{fig_act_ro_color}.}
\label{fig_ro_act_teffbin2}
\end{figure*}

 \begin{figure*}[ht]
\includegraphics[width=1\textwidth]{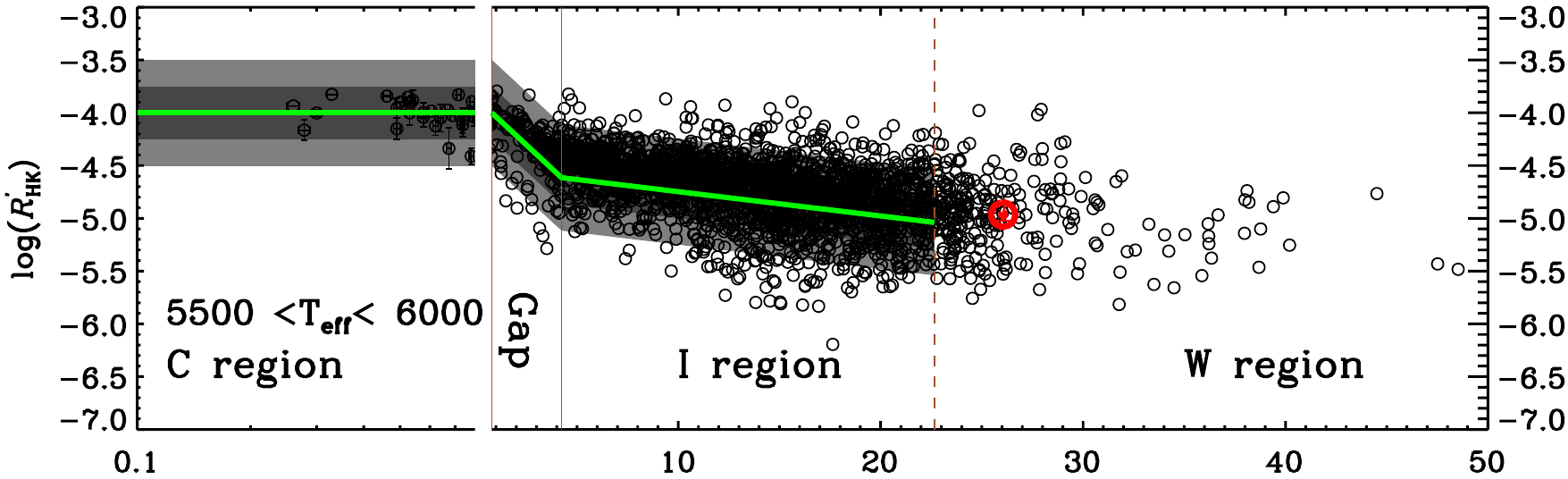}
\includegraphics[width=1\textwidth]{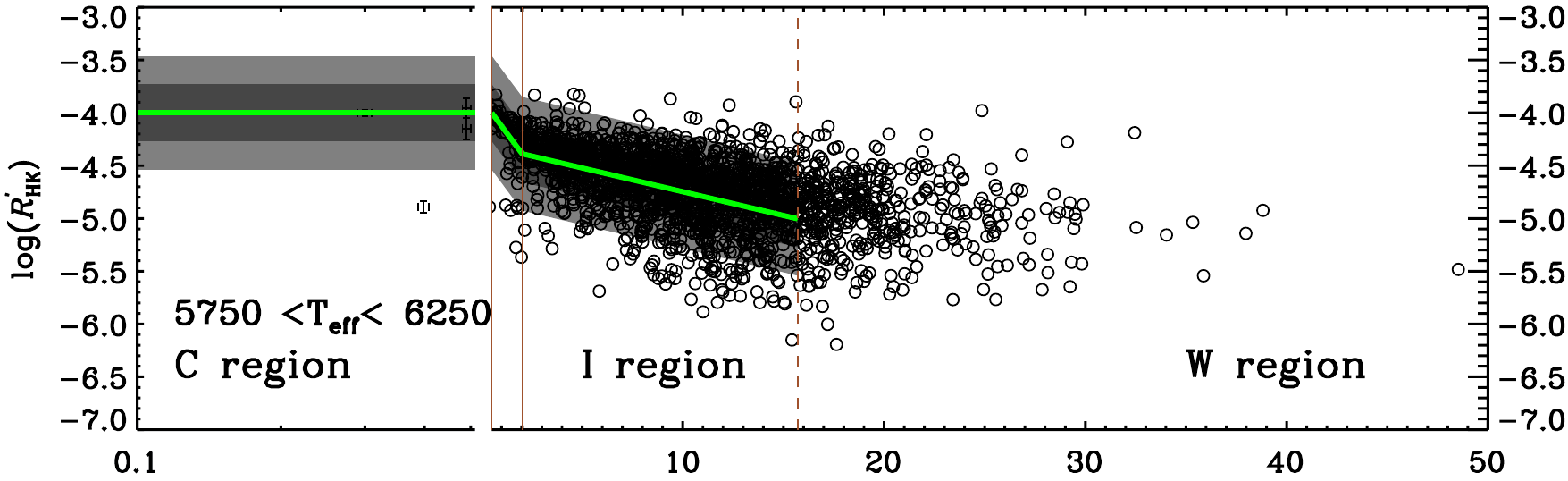}
\includegraphics[width=1\textwidth]{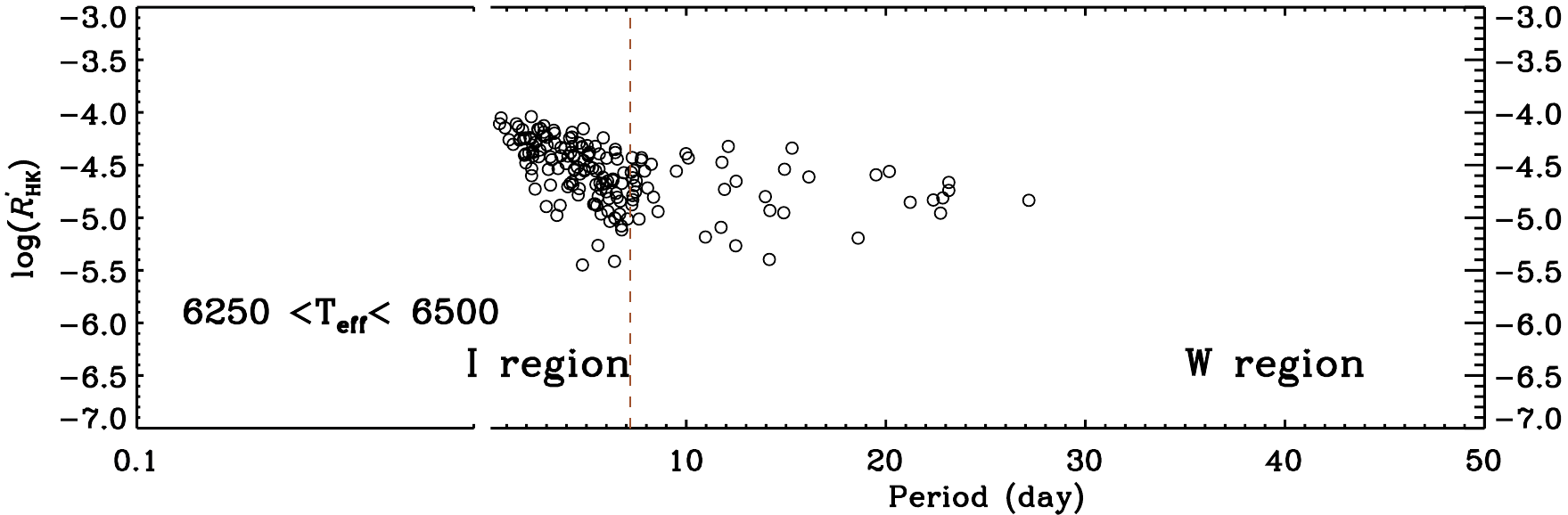}
\caption{The new period–chromospheric activity relationship for temperature bins. Closed circles denote the emission line stars and open circles denote the absorption line stars. The fitting results are listed in Table~\ref{table_pfit}. The meaning of symbols are as the same as Fig.~\ref{fig_act_ro_color}. The location of the Sun is marked with an $\odot$ symbol.}
\label{fig_ro_act_teffbin3}
\end{figure*}

\end{appendix}
\end{document}